\documentclass[twocolumn,aps,superscriptaddress,showpacs,nofootinbib,floatfix]{revtex4-1}

\usepackage{multirow}
\usepackage{graphicx}
\usepackage{amsmath}

\usepackage{latexsym}
\usepackage{feynmp}

\usepackage{epsfig,bm,color}
\usepackage[normalem]{ulem}  

\renewcommand\sout{\bgroup \color{red} \ULdepth=-.5ex \ULset}

\begin{document}

\title{The doubly-charmed pentaquark in a quark model with a complete set of harmonic oscillator bases}

\author{Woosung Park}
\email{diracdelta@daum.net}
\affiliation{Department of Physics and Institute of Physics and Applied Physics, Yonsei University, Seoul 03722, Korea}
\author{Sungsik Noh}
\email{sungsiknoh@yonsei.ac.kr}
\affiliation{Department of Physics and Institute of Physics and Applied Physics, Yonsei University, Seoul 03722, Korea}

\begin{abstract}
As our recent quark model calculation~\cite{Noh:2023fdy} suggests a strong possibility of a compact $T_{cc}$ that closely reproduces experimental mass, we have a strong incentive to extend our work to investigate the possible compact configuration of a pentaquark $udcc\bar{s}$, which is related to the structure of the doubly charmed tetraquark $T_{cc}$.
Since the introduction of a complete set of 3-dimensional harmonic oscillator bases to a spatial wave function in solving a quark model-based Hamiltonian with variational method leads to a more accurate value of the mass, it seems natural that future studies of the pentaquark should be treated with the same elaborate technical approach. To attain such precision for the ground state energy, we utilize  a complete set of 3-dimensional harmonic oscillator base up to 6th quanta.
Before carrying out this process, one important thing that has to be taken into account is to find out the color $\otimes$ spin states of the pentaquark for the evaluation of color and spin interaction most  essential to the quark model configuration.
To easily identify the suitable configuration, we make a systematic analysis of $SU(6)_{CS}$ irreducible representation of the pentaquark, from which we find that there is a correspondence between the color $\otimes$ spin states obtained from their coupling scheme  and the multiplet of the $SU(6)_{CS}$ irreducible representation of the pentaquark.
We find that the energy of the pentaquark configuration is +18.5 MeV above the lowest threshold for decay into $\Xi_{cc}$ and $K$, suggesting that this configuration is not stable against its decay. 
Nonetheless, while we used a Gaussian hyperfine potential, it was recently found that a Yukawa form leads to a stronger attraction for the $T_{cc}$ configuration.  Therefore it is important to study the same configuration using the latter potential.
\end{abstract}

\maketitle

\section{INTRODUCTION}
Since the discovery of X(3872)\cite{Choi:2003ue}, many theoretical works have been devoted to understand the structure of exotic hadrons. However, the results are rather controversial and vary depending on the models. Above all, it is an important fact that a fundamental understanding of the internal structure of X(3872) is as crucial as the controversies surrounding it. This will provide a way to advance our physical investigation of the properties of other exotic hadrons.

On the other hand, amidst such controversies, the recent discovery of $T_{cc}$\cite{LHCbTcc} which has an open charm quark in its flavor structure distinct as the X(3872) with a hidden charm quark  may provide a good opportunity to understand other exotic hadrons, including X(3872).

In particular, for the case of the $T_{cc}$,  even  before the experimental  discovery of $T_{cc}$ there have been many theoretical studies for  the $T_{cc}$ as another candidate of  exotic particle in several branches of  hadron physics\cite{Ballot:1983iv,Zouzou:1986qh,R1,R2,R3,R4,R5,R6,R7,R9,R11,R13,R14,R15,R16,R17,R18,R19,R20,R21,R22,R23,R24,R25,R27,R28,Woosung:NPA2019,Noh:Prd2021,Park:2013fda}.
However, most of these results have shown significant discrepancies not only in the mass of the discovered particles but also in describing their binding structures.
  
Recently, we have improved our quark model calculation for the mass of $T_{cc}$ given in Ref \cite{Woosung:NPA2019,Park:2013fda} to take into account a
complete set of harmonic oscillator basis in variational method.  We found in Ref.~\cite{Noh:Prd2021} that with such a complete set, the mass is found to be almost near that of the experiment, but not  a bound state  in contrast to the experiment. 
Furthermore, according to our recent calculation\cite{Noh:2023fdy}, we find a stable compact $T_{cc}$ configuration with a positive slightly binding energy below the lowest threshold.  Such binding was possible due to the 
strength of  Yukawa type of hyperfine potential, which provides more attraction at short range than the Gaussian type as used in the previous work\cite{Noh:Prd2021}. 

Motivated  by the most probable account of $T_{cc}$\cite{Noh:2023fdy},  we further attempt to investigate the possibility of the existence of a bound state in a pentaquark configuration whose constituent flavors consist of $udcc\bar{s}$, including $ud$ with $I=0$  and two open charms similar to $T_{cc}$. 
To obtain convergence in solving a quark model-based Hamiltonian for the ground state of the pentaquark, it is essential to introduce a complete set of 3-dimensional harmonic oscillator bases to the spatial wave function using the variational method. To attain such precision for the ground state energy, we utilize  a complete set of 3-dimensional harmonic oscillator base up to 6th quanta. The configuration we are studying here is calculated by considering Gaussian type of hyperfine potential rather than Yukawa type of hyperfine potential. Due to its higher degree of freedom compared to tetraquarks, calculating the pentaquark is a highly challenging task. Furthermore, the presence of a Yukawa-type potential further adds to greatly the complexity of such calculations, owing to its intricate form.

The main purpose of our present work is to calculate the mass of the pentaquark with a high level of precision for the first time, in order to understand its structure and investigate its stability, using a complete set of harmonic oscillator basis in variational method.

In performing this, one of the difficulties which has to be taken into account is that for the evaluation of the color and spin interaction most essential to the quark model, the color $\otimes$ spin states of the pentaquark has to  be classified according to $SU(6)_{CS)}$ to better understand mathematically than in the previous work\cite{Park:2017jbn}.

For this purpose, we first find the color $\otimes$ spin states of the pentaquark directly from a coupling scheme between a color and a spin state, and next illuminate the subject extensively with the help of the related formula, by categorizing those states into the multiplet of $SU_{CS}(6)$ irreducible representation containing $ SU_{C}(3)$ $\times$ $ SU_{S}(2)$ as its subgroup. This makes it possible ultimately to build a full wave function which satisfy the Pauli principle, by combining these states with a spatial function.

Our paper is organized as follows;
in section II, we present  the color, and spin states of pentaquark; in section III, we deal with the color $\otimes$ spin states, from the coupling of color and spin state; in section IV, we analyze the $SU_{CS}(6)$ irreducible representation of pentaquark; in section V, we present the Hamiltonian of the pentaquark and discuss about the effect of a complete set of harmonic oscillator basis in variational method upon the practical calculation of the Hamiltonian; in section VI, we deal with the wave function of pentaquark required to describe the Hamiltonian; in section VII, we discuss about a numerical analysis.

\section{The flavor, color, and spin states of pentaquark}
In this section, we discuss about the property of the configurations of the pentaquark which is based on the constituent quark model, involving chromomagnetic one gluon interaction.
Since the chromomagnetic interaction is determined by the color $\otimes$ spin states of the pentaquark which comprises parts of its wave functions, it is obviously essential to make a systematic analysis of those states.  In analyzing   both the characteristics and structure of the color $\otimes$ spin states of the pentaquark,
we  adopt the $SU(6)_{CS}$ representation which is useful to perceive  the 
property of  the chromomagnetic interaction. 

We will first present the structure of $q^4\bar{q}$ consisting of $SU(3)_F$ symmetry for $q^4$ and an antiquark for $\bar{q}$, for the time being, assuming that the spatial wave function for all the quarks is in the S wave, because
this assumption later allows us to find out the color $\otimes$ spin states of the pentaquark  from the $SU(6)_{CS}$ representation point of view, using the related formula which is very useful to a certain full symmetric wave function.
Then, we must consider the flavor $\otimes$ color $\otimes$ spin states of the pentaquark which are allowed to the Pauli principle. We now investigate the flavor, color, and spin states of the pentaquark system which have to be considered as a preliminary step.

\subsection{The flavor states for $q^4$}
In order to describe the flavor states for $q^4$, we present the $SU(3)_F$ multiplets, which are classified by the decomposition of $\mathbf{3}_F \otimes \mathbf{3}_F \otimes \mathbf{3}_F \otimes \mathbf{3}_F$,  given as the corresponding Young
diagram:
\begin{itemize}
\item $\mathbf{15}$ multiplets : One basis function with Young diagram [4] 
\begin{align}
\begin{tabular}{c}
$\vert F \rangle$=
\end{tabular}
\begin{tabular}{|c|c|c|c|}
\hline
1 & 2 & 3& 4  \\
\hline
\end{tabular}
\label{F15}
\end{align}
\item $\mathbf{{15}^\prime}$ multiplets : Three basis functions with Young diagram [31] 
\begin{align}
\begin{tabular}{c}
$\vert F_1 \rangle$=
\end{tabular}
\begin{tabular}{|c|c|c|}
\hline
        1   & 2   & 3    \\
\cline{1-3} 
\multicolumn{1}{|c|}{4}  \\
\cline{1-1} 
\end{tabular}
, \quad \quad 
\begin{tabular}{c}
$\vert F_2 \rangle$=
\end{tabular}
\begin{tabular}{|c|c|c|}
\hline
        1   & 2   & 4    \\
\cline{1-3} 
\multicolumn{1}{|c|}{3}  \\
\cline{1-1} 
\end{tabular}
, \quad \quad 
\begin{tabular}{c}
$\vert F_3 \rangle$=
\end{tabular}
\begin{tabular}{|c|c|c|}
\hline
        1   & 3   & 4    \\
\cline{1-3} 
\multicolumn{1}{|c|}{2}  \\
\cline{1-1} 
\end{tabular}
.
\label{F15-1}
\end{align}
\item $\mathbf{3}$ multiplets : Three basis functions with Young diagram $[21^2]$ 
\begin{align}
\begin{tabular}{c}
$\vert F_1 \rangle$=
\end{tabular}
\begin{tabular}{|c|c|}
\hline
        1   & 2     \\
\cline{1-2} 
\multicolumn{1}{|c|}{3}  \\
\cline{1-1} 
\multicolumn{1}{|c|}{4}  \\
\cline{1-1} 
\end{tabular}
, \quad \quad \quad
\begin{tabular}{c}
$\vert F_2 \rangle$=
\end{tabular}
\begin{tabular}{|c|c|}
\hline
        1   & 3     \\
\cline{1-2} 
\multicolumn{1}{|c|}{2}  \\
\cline{1-1} 
\multicolumn{1}{|c|}{4}  \\
\cline{1-1} 
\end{tabular}
, \quad \quad \quad
\begin{tabular}{c}
$\vert F_3 \rangle$=
\end{tabular}
\begin{tabular}{|c|c|}
\hline
        1   & 4     \\
\cline{1-2} 
\multicolumn{1}{|c|}{2}  \\
\cline{1-1} 
\multicolumn{1}{|c|}{3}  \\
\cline{1-1} 
\end{tabular}
.
\label{F3}
\end{align}
\item $\mathbf{\bar{6}}$ multiplets : Two basis functions with Young diagram $[2^2]$ 
\begin{align}
\begin{tabular}{c}
$\vert F_1 \rangle$=
\end{tabular}
\begin{tabular}{|c|c|}
\hline
        1   & 2     \\
\hline
        3   & 4     \\
\hline        
\end{tabular}
, \quad \quad \quad \quad \quad 
\begin{tabular}{c}
$\vert F_2 \rangle$=
\end{tabular}
\begin{tabular}{|c|c|}
\hline
        1   & 3     \\
\hline
        2   & 4     \\
\hline        
\end{tabular}
.
\label{F6}
\end{align}
\end{itemize}

\subsection{The color singlets}

Since the observable hadron states must be in a color singlet, we have to choose only the  color singlet states of the pentaquark system. One way of constructing the color singlet is to combine the color antitriplet of the antiquark $\bar{q}$ with the color triplet of $q^4$, which can be obtained from the decomposition of $\mathbf{3}_C \otimes \mathbf{3}_C \otimes \mathbf{3}_C \otimes \mathbf{3}_C$ for the four quarks. The color triplets of $q^4$ which correspond to Young diagram $[21^2]$ are given as follows:

\begin{align}
&\begin{tabular}{|c|c|}
\hline
           1 &  2    \\
\cline{1-2}
\multicolumn{1}{|c|}{3} \\
\cline{1-1}
\multicolumn{1}{|c|}{4}  \\
\cline{1-1}
\end{tabular}
=\{(12)_6(34)_{\bar{3}}\}_3,
\quad
\begin{tabular}{|c|c|}
\hline
           1 &  3    \\
\cline{1-2}
\multicolumn{1}{|c|}{2} \\
\cline{1-1}
\multicolumn{1}{|c|}{4}  \\
\cline{1-1}
\end{tabular}
=\{(12)_{\bar{3}}34\}_3,
\nonumber \\
 &\begin{tabular}{|c|c|}
\hline
           1 &  4    \\
\cline{1-2}
\multicolumn{1}{|c|}{2} \\
\cline{1-1}
\multicolumn{1}{|c|}{3}  \\
\cline{1-1}
\end{tabular}
=\{(123)_14\}_3.
\label{color1} 
\end{align}
Here, the subscript on the right hand side of Eq.~(\ref{color1}) indicates the irreducible representation of $SU(3)_C$. Then, one can construct three orthogonal color singlets by combining the color antitriplet of the antiquark with the color triplet of $q^4$,  given as follows:

\begin{align}
&\vert C_1 \rangle=
\begin{tabular}{|c|c|}
\hline
           1 &  2    \\
\cline{1-2}
\multicolumn{1}{|c|}{3} \\
\cline{1-1}
\multicolumn{1}{|c|}{4}  \\
\cline{1-1}
\end{tabular}_C
\otimes(5)_{\bar{3}},
\quad
\vert C_2 \rangle=
\begin{tabular}{|c|c|}
\hline
           1 &  3    \\
\cline{1-2}
\multicolumn{1}{|c|}{2} \\
\cline{1-1}
\multicolumn{1}{|c|}{4}  \\
\cline{1-1}
\end{tabular}_C
\otimes(5)_{\bar{3}},
\nonumber \\
&\vert C_3 \rangle=
\begin{tabular}{|c|c|}
\hline
           1 &  4    \\
\cline{1-2}
\multicolumn{1}{|c|}{2} \\
\cline{1-1}
\multicolumn{1}{|c|}{3}  \\
\cline{1-1}
\end{tabular}_C
\otimes(5)_{\bar{3}}.
\label{color2}
\end{align}

\subsection{The spin states}

The spin states of the pentaquark system consisting of $S=1/2$, $S=3/2$, and $S=5/2$ are also obtained from the decomposition of  $\mathbf{2}_S \otimes \mathbf{2}_S \otimes \mathbf{2}_S \otimes \mathbf{2}_S  \otimes \mathbf{2}_S$ into   the irreducible representation of $SU(2)_S$, as follows:

\begin{itemize}
\item $S=1/2$ states : 5 basis functions with Young diagram [32] 
\begin{align}
&\vert S^{1/2}_1 \rangle=
\begin{tabular}{|c|c|c|}
\hline
1& 2 &  3  \\
\cline{1-3}
\multicolumn{1}{|c|}{4} & \multicolumn{1}{c|}{5}  \\
\cline{1-2}
\end{tabular},
\quad
\vert S^{1/2}_2 \rangle=
\begin{tabular}{|c|c|c|}
\hline
1& 2 &  4  \\
\cline{1-3}
\multicolumn{1}{|c|}{3} & \multicolumn{1}{c|}{5}  \\
\cline{1-2}
\end{tabular},
\quad
\vert S^{1/2}_3 \rangle=
\begin{tabular}{|c|c|c|}
\hline
1& 3 &  4  \\
\cline{1-3}
\multicolumn{1}{|c|}{2} & \multicolumn{1}{c|}{5}  \\
\cline{1-2}
\end{tabular},
\nonumber \\
&\vert S^{1/2}_4 \rangle=
\begin{tabular}{|c|c|c|}
\hline
1& 2 &  5  \\
\cline{1-3}
\multicolumn{1}{|c|}{3} & \multicolumn{1}{c|}{4}  \\
\cline{1-2}
\end{tabular},
\quad
\vert S^{1/2}_5 \rangle=
\begin{tabular}{|c|c|c|}
\hline
1& 3 &  5  \\
\cline{1-3}
\multicolumn{1}{|c|}{2} & \multicolumn{1}{c|}{4}  \\
\cline{1-2}
\end{tabular}.
\label{spin1/2}
\end{align}
\end{itemize}

When we investigate the properties of the pentaquark against the
strong decay into a baryon and a meson, it is very useful to
describe the spin 1/2 states associated with the decay mode. We denote the five spin states by,

\begin{align}
&\vert{\chi}_1 \rangle=[\{(12)_13\}_{\frac{3}{2}}(45)_1]_{\frac{1}{2}}, \quad
\vert{\chi}_2 \rangle=[\{(12)_13\}_{\frac{1}{2}}(45)_1]_{\frac{1}{2}}, \nonumber \\
&\vert{\chi}_3 \rangle=[\{(12)_03\}_{\frac{1}{2}}(45)_1]_{\frac{1}{2}}, \quad
\vert{\chi}_4 \rangle=[\{(12)_13\}_{\frac{1}{2}}(45)_0]_{\frac{1}{2}},\nonumber \\
&\vert{\chi}_5 \rangle=[\{(12)_03\}_{\frac{1}{2}}(45)_0]_{\frac{1}{2}},
\label{spin1/2-1}
\end{align}
where the subscript indicates the spin state.
The bases set of Eq.~(\ref{spin1/2-1}) is transformed into the bases set of Eq.~(\ref{spin1/2})
through the following orthogonal transformation:
\begin{align}
 \left(\begin{array}{c}
 \vert{\chi}_1 \rangle    \\
 \vert{\chi}_2 \rangle    \\
   \vert{\chi}_3 \rangle \\ 
    \vert{\chi}_4 \rangle \\
     \vert{\chi}_5 \rangle \\
\end{array} \right)
=
  \left(\begin{array}{ccccc}
1    &  0     &  0    &   0  &  0  \\
0  &  \frac{1}{2}  &  0 &   \frac{\sqrt{3}}{2}   &   0   \\
0  &   0   & \frac{1}{2}  &  0 &   \frac{\sqrt{3}}{2}       \\
 0  &  \frac{\sqrt{3}}{2}  &  0 &   -\frac{1}{2}   &   0    \\
0  & 0 &    \frac{\sqrt{3}}{2}  &  0 &   -\frac{1}{2}       \\
\end{array} \right)
 \left(\begin{array}{c}
\vert S^{1/2}_1 \rangle  \\
\vert S^{1/2}_2 \rangle    \\
 \vert S^{1/2}_3 \rangle     \\
 \vert S^{1/2}_4 \rangle    \\
 \vert S^{1/2}_5 \rangle     \\
\end{array} \right).
 \label{S-1/2-orthogonal-matrix}  
\end{align}

\begin{itemize}
\item $S=3/2$ states : 4 basis functions with Young diagram [41] 
\begin{align}
&\vert S^{3/2}_1 \rangle=
\begin{tabular}{|c|c|c|c|}
\hline
1& 2 &  3& 4  \\
\cline{1-4}
\multicolumn{1}{|c|}{5} \\
\cline{1-1}
\end{tabular},
\vert S^{3/2}_2 \rangle=
\begin{tabular}{|c|c|c|c|}
\hline
1& 2 &  3& 5  \\
\cline{1-4}
\multicolumn{1}{|c|}{4} \\
\cline{1-1}
\end{tabular},
\vert S^{3/2}_3 \rangle=
\begin{tabular}{|c|c|c|c|}
\hline
1& 2 &  4& 5  \\
\cline{1-4}
\multicolumn{1}{|c|}{3} \\
\cline{1-1}
\end{tabular},
\nonumber \\
&\vert S^{3/2}_4 \rangle=
\begin{tabular}{|c|c|c|c|}
\hline
1& 3 &  4& 5  \\
\cline{1-4}
\multicolumn{1}{|c|}{2} \\
\cline{1-1}
\end{tabular}.
\label{spin3/2}
\end{align}
\end{itemize}
The spin 3/2 states related to the decay mode for the separation of a baryon and a meson are denoted by the following:

\begin{align}
&\vert{\phi}_1 \rangle=[\{(12)_13\}_{\frac{3}{2}}(45)_0]_{\frac{3}{2}},
\quad \vert{\phi}_2 \rangle=[\{(12)_13\}_{\frac{3}{2}}(45)_1]_{\frac{3}{2}}, \nonumber \\
&\vert{\phi}_3 \rangle=[\{(12)_13\}_{\frac{1}{2}}(45)_1]_{\frac{3}{2}},
\quad  \vert{\phi}_4 \rangle=[\{(12)_03\}_{\frac{1}{2}}(45)_1]_{\frac{3}{2}}.
\label{spin3/2-1}
\end{align}
Then, the orthogonal matrix which transform the bases set of Eq.~(\ref{spin3/2-1}) into the bases set of Eq.~(\ref{spin3/2}) is given as follows.

\begin{align}
 \left(\begin{array}{c}
 \vert{\phi}_1 \rangle    \\
 \vert{\phi}_2 \rangle    \\
   \vert{\phi}_3 \rangle \\ 
    \vert{\phi}_4 \rangle \\
   \end{array} \right)
=
\left(\begin{array}{cccc} \sqrt{\frac{5}{8}}  &   -\sqrt{\frac{3}{8}} & 0 & 0 \\
                         \sqrt{\frac{3}{8}}  &   \sqrt{\frac{5}{8}} & 0 & 0         \\
                                          0 &  0 & 1 & 0         \\
                            0 &  0 & 0 & 1    \end{array} \right)
 \left(\begin{array}{c}
\vert S^{3/2}_1 \rangle  \\
\vert S^{3/2}_2 \rangle    \\
 \vert S^{3/2}_3 \rangle     \\
 \vert S^{3/2}_4 \rangle    \\
 \end{array} \right).
 \label{S-3/2-orthogonal-matrix}  
\end{align}

\begin{itemize}
\item $S=5/2$ states : 1 basis function with Young diagram [5] 
\begin{align}
&\vert S^{5/2} \rangle=
\begin{tabular}{|c|c|c|c|c|}
\hline
1& 2 &  3& 4 & 5  \\
\hline
\end{tabular}.
\label{spin5/2}
\end{align}
\end{itemize}

\section{The coupling of color and spin states}

Given a flavor multiplet for $q^4$ with a certain symmetry in terms of the corresponding Young diagram, both the symmetry and structure of color $\otimes$ spin states are of great importance when we choose a certain kind of symmetry of the full wave function, including a spatial wave function. In this section, for this purpose, we discuss about the method of finding out the color $\otimes$ spin states represented by the Young diagram for $q^4$; These states can be constructed by coupling the color and spin states using permutation group theory.

In order to construct the  color $\otimes$ spin states with a certain symmetry among particles 1-4, it is necessary to look into the inner product between the Young diagram $[21^2]$ of the color singlets in Eq.~(\ref{color2}) and the corresponding Young diagram for particles 1-4 with respect to  each spin state.

When particles 1-4 have spin 0, one can obtain the color $\otimes$ spin states using the following procedure:
\begin{align}
\begin{tabular}{|c|c|}
\hline
   $\quad$         &  $\quad$   \\
\cline{1-2}
\multicolumn{1}{|c|}{$\quad$} \\
\cline{1-1}
\multicolumn{1}{|c|}{$\quad$}  \\
\cline{1-1}
\end{tabular}_C
 \times
 \begin{tabular}{|c|c|}
\hline
 $\quad$   &  $\quad$       \\
\hline
 $\quad$   &  $\quad$  \\
\hline
\end{tabular}_S
=
\begin{tabular}{|c|c|}
\hline
   $\quad$         &  $\quad$   \\
\cline{1-2}
\multicolumn{1}{|c|}{$\quad$} \\
\cline{1-1}
\multicolumn{1}{|c|}{$\quad$}  \\
\cline{1-1}
\end{tabular}_{CS}
\oplus
 \begin{tabular}{|c|c|c|}
\hline
 $\quad$   &  $\quad$    &   $\quad$    \\
\cline{1-3}
\multicolumn{1}{|c|}{ $\quad$  } \\
\cline{1-1}
\end{tabular}_{CS}
\label{cs1}
\end{align}
This inner product between the color and the spin state for four quarks is only applied to $\vert S^{1/2}_4 \rangle$ and $\vert S^{1/2}_5 \rangle$ in Eq.~(\ref{spin1/2}) because the spin state among particles 1-4 for these two is 0, as indicated by the Young diagram of spin state in Eq.~(\ref{cs1}).

When particles 1-4 have spin 1:
\begin{align}
&\begin{tabular}{|c|c|}
\hline
   $\quad$         &  $\quad$   \\
\cline{1-2}
\multicolumn{1}{|c|}{$\quad$} \\
\cline{1-1}
\multicolumn{1}{|c|}{$\quad$}  \\
\cline{1-1}
\end{tabular}_C
 \times
 \begin{tabular}{|c|c|c|}
\hline
 $\quad$   &  $\quad$    &   $\quad$    \\
\cline{1-3}
\multicolumn{1}{|c|}{ $\quad$  } \\
\cline{1-1}
\end{tabular}_S
=
\begin{tabular}{|c|}
\hline
 $\quad$      \\
\hline
 $\quad$      \\
 \hline
 $\quad$      \\
 \hline
 $\quad$      \\
\hline
\end{tabular}_{CS}
\oplus
\begin{tabular}{|c|c|}
\hline
   $\quad$         &  $\quad$   \\
\cline{1-2}
\multicolumn{1}{|c|}{$\quad$} \\
\cline{1-1}
\multicolumn{1}{|c|}{$\quad$}  \\
\cline{1-1}
\end{tabular}_{CS}
\oplus
 \begin{tabular}{|c|c|}
\hline
 $\quad$   &  $\quad$       \\
\hline
 $\quad$   &  $\quad$       \\
\hline
\end{tabular}_{CS}
\nonumber \\
&\oplus
\begin{tabular}{|c|c|c|}
\hline
 $\quad$   &  $\quad$    &   $\quad$    \\
\cline{1-3}
\multicolumn{1}{|c|}{ $\quad$  } \\
\cline{1-1}
\end{tabular}_{CS}
\label{cs2}
\end{align}
This holds good for the case of $S=1/2$ and $S=3/2$ states in Eq.~(\ref{spin1/2}) and Eq.~(\ref{spin3/2}), respectively,
where the spin state among particles 1-4 is 1.

When particles 1-4 have spin 2:
\begin{align}
\begin{tabular}{|c|c|}
\hline
   $\quad$         &  $\quad$   \\
\cline{1-2}
\multicolumn{1}{|c|}{$\quad$} \\
\cline{1-1}
\multicolumn{1}{|c|}{$\quad$}  \\
\cline{1-1}
\end{tabular}_C
 \times
 \begin{tabular}{|c|c|c|c|}
\hline
 $\quad$   &  $\quad$    &   $\quad$ &  $\quad$ \\
\hline
\end{tabular}_S
=
\begin{tabular}{|c|c|}
\hline
   $\quad$         &  $\quad$   \\
\cline{1-2}
\multicolumn{1}{|c|}{$\quad$} \\
\cline{1-1}
\multicolumn{1}{|c|}{$\quad$}  \\
\cline{1-1}
\end{tabular}_{CS}
\label{cs3}
\end{align}
The last is applied to the case of $S=3/2$ and $S=5/2$ states in Eq.~(\ref{spin3/2}) and Eq.~(\ref{spin5/2}), respectively, 
where the spin state among  particles 1-4 is 2.

To obtain the combinative coefficients in the coupling scheme of color and spin states, we calculate the Clebsch-Gordan (CG) coefficients of the permutation group $S_n$, where $n$ is the number of particles. The CG coefficients can be factorized into the K matrix and the CG coefficients of $S_{n-1}$, which can be found using the reduction property. This approach allows us to determine the proper combination of color and spin states that satisfy the desired symmetry of the full wave function. The reduction property is given by Ref.\cite{Stancu:1999qr}:
\begin{align}
&S([f^{\prime}]p^{\prime}q^{\prime}y^{\prime}[f^{\prime\prime}]p^{\prime\prime}q^{\prime\prime}y^{\prime\prime}\vert[f]pqy)=
K([f^{\prime}]p^{\prime}[f^{\prime\prime}]p^{\prime\prime}\vert[f]p)  \times
\nonumber \\
&S([f^{\prime}_{p^{\prime}}]q^{\prime}y^{\prime}[f^{\prime\prime}_{p^{\prime\prime}}]q^{\prime\prime}y^{\prime\prime}\vert[f_p]qy),
\label{K-matrix}
\end{align}
where $S$ in the left-hand (right-hand) side is a CG coefficient of $S_n$ ($S_{n-1}$), and $n$ is the number of the participant particles.

Before completing the coupling scheme of the color and spin state, it is necessary to investigate the color $\otimes$ spin states of $q^4$. This can be done by considering the direct product of the fundamental representation of  $SU(6)_{CS}$ for $q^4$, which is decomposed into the direct sum of the irreducible representations of $SU(6)_{CS}$, as the following:

\begin{align}
&\mathbf{6}_{CS}\otimes\mathbf{6}_{CS}\otimes\mathbf{6}_{CS}\otimes\mathbf{6}_{CS}
=
\begin{tabular}{|c|c|c|c|}
\hline
$\quad$ &$\quad$  &$\quad$ & $\quad$   \\
\hline
\end{tabular}
\oplus
3\times
\begin{tabular}{|c|c|c|}
\hline
    $\quad$       &  $\quad$  & $\quad$    \\
\cline{1-3} 
\multicolumn{1}{|c|}{$\quad$}  \\
\cline{1-1} 
\end{tabular}
\nonumber \\
&\oplus
3\times
\begin{tabular}{|c|c|}
\hline
      $\quad$        &    $\quad$      \\
\cline{1-2} 
\multicolumn{1}{|c|}{  $\quad$  }  \\
\cline{1-1} 
\multicolumn{1}{|c|}{  $\quad$  }  \\
\cline{1-1} 
\end{tabular}
\oplus
2\times
\begin{tabular}{|c|c|}
\hline
   $\quad$    & $\quad$         \\
\hline
   $\quad$    & $\quad$         \\
\hline
\end{tabular}
\oplus
\begin{tabular}{|c|}
\hline
     $\quad$          \\
\hline
    $\quad$          \\
  \hline
     $\quad$          \\
\hline
    $\quad$          \\
\hline         
\end{tabular}
\label{cs-decom}      
\end{align}

We further divide the SU(6)$_{CS}$ representation for $q^4$ in  Eq.~(\ref{cs-decom}) into the direct sum of SU(3)$_C \otimes {\rm SU(2})_S$ multiplets, represented by both a color and a spin state. This allows us to establish the correspondence between the $SU(6)_{CS}$ representation and the color $\otimes$ spin state obtained from the coupling scheme, and thereby obtain a better understanding of the symmetry properties of the system.
\begin{table}[htp]
\caption{ The composition of $SU(3)_C$ and $SU(2)_S$ concerning the $SU(6)_{CS}$ representation
for $q^4$    }
\begin{center}
\begin{tabular}{c|c|c|c}
\hline \hline
$SU(6)_{CS}$      & $SU(3)_C$ $\otimes$  $SU(2)_S$ & Dimension  &  Eigenvalue \\
\hline
 $[4]$  &    $(\mathbf{15},\mathbf{5})$, $(\mathbf{{15}^{\prime}},\mathbf{3})$,  $(\mathbf{\bar{6}},\mathbf{1})$            &    126         &  50/3        \\
   \hline  
 $[1^4]$  &  $(\mathbf{\bar{6}},\mathbf{1})$, $(\mathbf{3},\mathbf{3})$                            &   15         &  14/3         \\
   \hline  
  $[31]$  &   $(\mathbf{15},\mathbf{3})$,  $(\mathbf{{15}^{\prime}},\mathbf{5})$, $(\mathbf{{15}^{\prime}},\mathbf{3})$,      &  210          &  38/3        \\
     &  $(\mathbf{{15}^{\prime}},\mathbf{1})$, $(\mathbf{\bar{6}},\mathbf{3})$,  $(\mathbf{3},\mathbf{3})$,                                                         &                 &         \\
  &          $(\mathbf{3},\mathbf{1})$                                                      &                 &         \\   
   \hline  
    $[21^2]$  &   $(\mathbf{{15}^{\prime}},\mathbf{3})$,  $(\mathbf{{15}^{\prime}},\mathbf{1})$, 
    $(\mathbf{\bar{6}},\mathbf{3})$,                       &   105         &      26/3    \\
    &  $(\mathbf{3},\mathbf{5})$,   $(\mathbf{3},\mathbf{3})$,     $(\mathbf{3},\mathbf{1})$                                              &                            &  \\
   \hline  
    $[2^2]$  &    $(\mathbf{15},\mathbf{1})$,  $(\mathbf{{15}^{\prime}},\mathbf{3})$,   
    $(\mathbf{\bar{6}},\mathbf{5})$,    & 105           &  32/3        \\
    &    $(\mathbf{\bar{6}},\mathbf{1})$,                      $(\mathbf{3},\mathbf{3})$                                                                                      &                           &  \\
   \hline \hline        
\end{tabular}
\end{center}
\label{su6-2}
\end{table}

In Table~\ref{su6-2}, the fourth column indicates the eigenvalues of the quadratic Casimir operator of the $SU(6)_{CS}$ representation for $q^4$, and the first column the Young diagram of the irreducible representation of $SU(6)_{CS}$.
The eigenvalues of the quadratic Casimir operator for the $SU(6)_{CS}$ representation in Table~\ref{su6-2} can be easily obtained using a formula~\cite{Jaffe:1976ih} given by:
\begin{align}
\frac{280}{72D(6)}{\sum}_{j}D(3)_j(2S_j+1)S_j(S_j+1),
\label{SU(6)-formula}  
 \end{align}
where $D(6)$ is the dimension of the $SU(6)_{CS}$ representation, and $D(3)$ is the dimension of the $SU(3)_{C}$ representation contained in the composition of $SU(3)_C$ and  $SU(2)_S$ multiplets. For a given $SU(6)_{CS}$ irreducible representation, the subscript $j$ indicates the different $SU(3)_C \otimes SU(2)_S$ composite states that compose the  irreducible representation of  $SU(6)_{CS}$. Our overall normalization is different from that used in Ref.\cite{Jaffe:1976ih}.

Moreover, since the symmetry property for three quarks is apparently given to the color $\otimes$ spin states of the coupling scheme from a Young diagram for four quarks, we also examine the $SU(6)_{CS}$ representation for three quarks, which can be expressed by the corresponding Young diagram for three quarks in the same method. Similarly, by decomposing the $SU(6)_{CS}$ representation into the sum of $SU(3)_C$ $\otimes$  $SU(2)_S$, we find the composition of $SU(3)_C$ and $SU(2)_S$ for the $SU(6)_{CS}$ representation for three quarks, and the eigenvalue of the quadratic Casimir operator of $SU(6)_{CS}$ for $q^3$ using Eq.~(\ref{SU(6)-formula}). The results are summarized in Table\ref{su6-3}. It should be noted that the symmetry property for three quarks is apparent in the color $\otimes$ spin states of the coupling scheme, as it is derived from the Young diagram for four quarks.
 
\begin{table}[htp]
\caption{ The composition of $SU(3)_C$ and $SU(2)_S$ concerning the $SU(6)_{CS}$ representation
for $q^3$    }
\begin{center}
\begin{tabular}{c|c|c|c}
\hline \hline
$SU(6)_{CS}$    & $SU(3)_C$ $\otimes$  $SU(2)_S$ & Dimesion  &  Eigenvalue \\
\hline
 $[3]$  &    $(\mathbf{10},\mathbf{4})$,    $(\mathbf{8},\mathbf{2})$        &    56         &  45/4        \\
   \hline  
 $[21]$  &     $(\mathbf{10},\mathbf{2})$,   $(\mathbf{8},\mathbf{4})$,     
 $(\mathbf{8},\mathbf{2})$,     $(\mathbf{1},\mathbf{2})$                     &   70         &  33/4         \\
   \hline  
  $[1^3]$  &      $(\mathbf{8},\mathbf{2})$,     $(\mathbf{1},\mathbf{4})$                         &  20         &  21/4        \\
 \hline \hline        
\end{tabular}
\end{center}
\label{su6-3}
\end{table}

\subsection{The Young-Yamanouchi bases of coupling scheme for $S=1/2$}

Now, we present the Young-Yamanouchi bases from the coupling scheme for the total $S=1/2$ state in detail. The coupling scheme in Eq.(\ref{cs1}) results in the Young diagram $[21^2]$ with the total $S=1/2$ and $S=0$ among particles 1-4.

\begin{align}
\begin{tabular}{|c|c|}
\hline
           1 &  2    \\
\cline{1-2}
\multicolumn{1}{|c|}{3} \\
\cline{1-1}
\multicolumn{1}{|c|}{4}  \\
\cline{1-1}
\end{tabular}_{CS^{1/2};S^0}
=&-\frac{1}{2}\vert C_1 \rangle \otimes \vert S^{1/2}_4 \rangle
+\frac{1}{2}\vert C_2 \rangle \otimes \vert S^{1/2}_5 \rangle+
\nonumber \\
&\frac{1}{\sqrt{2}}\vert C_3 \rangle \otimes \vert S^{1/2}_5 \rangle,
\nonumber
\end{align}
\begin{align}
\begin{tabular}{|c|c|}
\hline
           1 &  3    \\
\cline{1-2}
\multicolumn{1}{|c|}{2} \\
\cline{1-1}
\multicolumn{1}{|c|}{4}  \\
\cline{1-1}
\end{tabular}_{CS^{1/2};S^0}
=&\frac{1}{2}\vert C_1 \rangle \otimes \vert S^{1/2}_5 \rangle
+\frac{1}{2}\vert C_2 \rangle \otimes \vert S^{1/2}_4 \rangle-
\nonumber \\
&\frac{1}{\sqrt{2}}\vert C_3 \rangle \otimes \vert S^{1/2}_4 \rangle,
\nonumber
\end{align}
\begin{align}
\begin{tabular}{|c|c|}
\hline
           1 &  4   \\
\cline{1-2}
\multicolumn{1}{|c|}{2} \\
\cline{1-1}
\multicolumn{1}{|c|}{3}  \\
\cline{1-1}
\end{tabular}_{CS^{1/2};S^0}
=\frac{1}{\sqrt{2}}\vert C_1 \rangle \otimes \vert S^{1/2}_5 \rangle
-\frac{1}{\sqrt{2}}\vert C_2 \rangle \otimes \vert S^{1/2}_4 \rangle.
\label{cs-1/2-0-211}
\end{align}
In the left hand side of Eq.(\ref{cs-1/2-0-211}), the superscript of $S$ after $CS^{1/2}$ in the Young-Yamanouchi bases indicates the total spin of particles 1-4. Also, the explicit product of the antiquark state is neglected as the Young-Yamanouchi bases represent a certain symmetry properties for $q^4$ but the total spin of particles 1-5 is given in the superscript of $CS$. It should be noted that these states in Eq.~(\ref{cs-1/2-0-211}) belong to the multiplet of the $SU(6)_{CS}$ representation $[21^2]$ for $q^4$, as can be seen in Table~\ref{su6-2}. This is so because first, it is obvious that the color state for $q^4$ in Eq.~(\ref{color2}) is in a triplet state; second, the spin state for $q^4$ in Eq.~(\ref{spin1/2}) is in a singlet state; third, these Young-Yamanouchi bases in the left hand side of Eq.~(\ref{cs-1/2-0-211}) should be represented in the form of Young diagram $[21^2]$ for $q^4$. It follows that the states in Eq.~(\ref{cs-1/2-0-211}) are eigenstates of the quadratic Casimir operator of $SU(6)_{CS}$ for $q^4$, with an eigenvalue of 26/3, as shown in Table~\ref{su6-2}.

In addition, since the Young-Yamanouchi basis of the coupling scheme in Eq.~(\ref{cs-1/2-0-211}) also involves the Young tableau corresponding to the irreducible representation of the permutation group, $S_3$ among three particles, we can infer that there is another correspondence between these states and the multiplet of $SU(6)_{CS}$ representation for $q^3$. From the standpoint of the Young tableau for $q^3$ in Eq.~(\ref{cs-1/2-0-211}), we can see that both the first and second states belong to the multiplet of $SU(6)_{CS}$ representation for $q^3$ represented by the Young diagram [21], while the third state belongs to the multiplet of $SU(6)_{CS}$ representation for $q^3$ represented by the Young diagram $[1^3]$. Table~\ref{su6-3} confirms that this is the case because the $SU(6)_{CS}$ representation  [21] for $q^3$ contains (octet, doublet) and  (singlet, doublet) states, and the $SU(6)_{CS}$ representation $[1^3]$ for $q^3$ contains (octet, doublet) states in the form of $SU(3)_C$ $\otimes$  $SU(2)_S$.

Next, we show the Young-Yamanouchi bases of the coupling scheme corresponding to Young diagram $[21^2]$ with the total  $S=1/2$ and at the same time $S=1$ among particles 1-4 which come from the coupling scheme in Eq.~(\ref{cs2}):

\begin{align}
\begin{tabular}{|c|c|}
\hline
           1 &  2    \\
\cline{1-2}
\multicolumn{1}{|c|}{3} \\
\cline{1-1}
\multicolumn{1}{|c|}{4}  \\
\cline{1-1}
\end{tabular}_{CS^{1/2};S^1}
=&-\frac{1}{\sqrt{6}}\vert C_1 \rangle \otimes \vert S^{1/2}_1 \rangle
-\frac{1}{\sqrt{3}}\vert C_1 \rangle \otimes \vert S^{1/2}_2 \rangle+
\nonumber \\
&\frac{1}{\sqrt{3}}\vert C_2 \rangle \otimes \vert S^{1/2}_3 \rangle
-\frac{1}{\sqrt{6}}\vert C_3 \rangle \otimes \vert S^{1/2}_3 \rangle,
\nonumber
\end{align}
\begin{align}
\begin{tabular}{|c|c|}
\hline
           1 &  3    \\
\cline{1-2}
\multicolumn{1}{|c|}{2} \\
\cline{1-1}
\multicolumn{1}{|c|}{4}  \\
\cline{1-1}
\end{tabular}_{CS^{1/2};S^1}
=&\frac{1}{\sqrt{3}}\vert C_1 \rangle \otimes \vert S^{1/2}_3 \rangle
-\frac{1}{\sqrt{6}}\vert C_2 \rangle \otimes \vert S^{1/2}_1 \rangle+
\nonumber \\
&\frac{1}{\sqrt{3}}\vert C_2 \rangle \otimes \vert S^{1/2}_2 \rangle
+\frac{1}{\sqrt{6}}\vert C_3 \rangle \otimes \vert S^{1/2}_2 \rangle,
\nonumber
\end{align}
\begin{align}
\begin{tabular}{|c|c|}
\hline
           1 &  4   \\
\cline{1-2}
\multicolumn{1}{|c|}{2} \\
\cline{1-1}
\multicolumn{1}{|c|}{3}  \\
\cline{1-1}
\end{tabular}_{CS^{1/2};S^1}
=&-\frac{1}{\sqrt{6}}\vert C_1 \rangle \otimes \vert S^{1/2}_3 \rangle
+\frac{1}{\sqrt{6}}\vert C_2 \rangle \otimes \vert S^{1/2}_2 \rangle+
\nonumber \\
&\sqrt{\frac{2}{3}}\vert C_3 \rangle \otimes \vert S^{1/2}_1 \rangle.
\label{cs-1/2-1-211}
\end{align}

It should be noted that the states in Eq.~(\ref{cs-1/2-1-211}) correspond to states of the multiplet of the  $SU(6)_{CS}$ representation for $q^4$, $[21^2]$, not only because it is obvious from Table~\ref{su6-2} that the color state for $q^4$ is in a triplet state, and at the same time the spin state  for $q^4$ is in a triplet state, but also because these Young-Yamanouchi bases should be represented in the form of Young diagram $[21^2]$ for $q^4$. Thus, the states in Eq.(\ref{cs-1/2-1-211}) are eigenstates of the quadratic Casimir operator of $SU(6)_{CS}$ for $q^4$, with 26/3 as their eigenvalue, as shown in Table~\ref{su6-2}.

Moreover, we can see from the Young tableau for $q^3$ in Eq.~(\ref{cs-1/2-1-211}) that both the first and the second states belong to the multiplet of $SU(6)_{CS}$ representation for $q^3$ represented by Young diagram [21], while the third belongs to the multiplet of $SU(6)_{CS}$ representation for $q^3$ represented by  Young diagram $[1^3]$.

Next, we show the Young-Yamanouchi bases for the coupling scheme corresponding to the Young diagram [31] with the total spin $S=1/2$ and at the same time $S=0$ among particles 1-4, which come from the coupling scheme in Eq.~(\ref{cs1}).

\begin{align}
\begin{tabular}{|c|c|c|}
\hline
           1 &  2 & 3   \\
\cline{1-3}
\multicolumn{1}{|c|}{4} \\
\cline{1-1}
\end{tabular}_{CS^{1/2};S^0}
=-\frac{1}{\sqrt{2}}\vert C_1 \rangle \otimes \vert S^{1/2}_4 \rangle
-\frac{1}{\sqrt{2}}\vert C_2 \rangle \otimes \vert S^{1/2}_5 \rangle,
\nonumber
\end{align}
\begin{align}
\begin{tabular}{|c|c|c|}
\hline
           1 &  2 & 4   \\
\cline{1-3}
\multicolumn{1}{|c|}{3} \\
\cline{1-1}
\end{tabular}_{CS^{1/2};S^0}
=&\frac{1}{2}\vert C_1 \rangle \otimes \vert S^{1/2}_4 \rangle
-\frac{1}{2}\vert C_2 \rangle \otimes \vert S^{1/2}_5 \rangle+
\nonumber \\
&\frac{1}{\sqrt{2}}\vert C_3 \rangle \otimes \vert S^{1/2}_5 \rangle,
\nonumber
\end{align}
\begin{align}
\begin{tabular}{|c|c|c|}
\hline
           1 &  3 & 4   \\
\cline{1-3}
\multicolumn{1}{|c|}{2} \\
\cline{1-1}
\end{tabular}_{CS^{1/2};S^0}
=&-\frac{1}{2}\vert C_1 \rangle \otimes \vert S^{1/2}_5 \rangle
-\frac{1}{2}\vert C_2 \rangle \otimes \vert S^{1/2}_4 \rangle-
\nonumber \\
&\frac{1}{\sqrt{2}}\vert C_3 \rangle \otimes \vert S^{1/2}_4 \rangle.
\label{cs-1/2-0-31}
\end{align}
 
The states in Eq.~(\ref{cs-1/2-0-31}) belong to the multiplet of the $SU(6)_{CS}$ representation for $q^4$, $[31]$, not only because it is obvious from Table~\ref{su6-2} that the color state for $q^4$ is in a triplet state, and at the same time  the spin state  for $q^4$ is in a singlet state, but also because these Young-Yamanouchi bases should be represented in the form of Young diagram $[31]$ for $q^4$. Therefore, the states in Eq.~(\ref{cs-1/2-0-31}) are eigenstates of the quadratic Casimir operator of $SU(6)_{CS}$ for $q^4$, with an eigenvalue of 38/3, as can be seen in Table~\ref{su6-2}.

On the other hand, we can see from the Young tableau for $q^3$ in Eq.~(\ref{cs-1/2-0-31}) that the first belongs to the multiplet of $SU(6)_{CS}$ representation for $q^3$ represented by the Young diagram [3], while both the second and the third states belong to the multiplet of $SU(6)_{CS}$ representation for $q^3$ represented by the Young diagram $[21]$.

Next, we show the Young-Yamanouchi bases of the coupling scheme corresponding to the Young diagram [31] with the total spin $S=1/2$ and at the same time $S=1$ among particles 1-4, which come from the coupling scheme in Eq.~(\ref{cs2}):

\begin{align}
\begin{tabular}{|c|c|c|}
\hline
           1 &  2 & 3   \\
\cline{1-3}
\multicolumn{1}{|c|}{4} \\
\cline{1-1}
\end{tabular}_{CS^{1/2};S^1}
=\frac{1}{\sqrt{2}}\vert C_1 \rangle \otimes \vert S^{1/2}_2 \rangle
+\frac{1}{\sqrt{2}}\vert C_2 \rangle \otimes \vert S^{1/2}_3 \rangle,
\nonumber
\end{align}
\begin{align}
\begin{tabular}{|c|c|c|}
\hline
           1 &  2 & 4   \\
\cline{1-3}
\multicolumn{1}{|c|}{3} \\
\cline{1-1}
\end{tabular}_{CS^{1/2};S^1}
=-\frac{1}{\sqrt{2}}\vert C_1 \rangle \otimes \vert S^{1/2}_1 \rangle
+\frac{1}{\sqrt{2}}\vert C_3 \rangle \otimes \vert S^{1/2}_3 \rangle,
\nonumber
\end{align}
\begin{align}
\begin{tabular}{|c|c|c|}
\hline
           1 &  3 & 4   \\
\cline{1-3}
\multicolumn{1}{|c|}{2} \\
\cline{1-1}
\end{tabular}_{CS^{1/2};S^1}
=-\frac{1}{\sqrt{2}}\vert C_2 \rangle \otimes \vert S^{1/2}_1 \rangle
-\frac{1}{\sqrt{2}}\vert C_3 \rangle \otimes \vert S^{1/2}_2 \rangle.
\label{cs-1/2-1-31}
\end{align}

The states in Eq.~(\ref{cs-1/2-1-31}) belong to the multiplet of the  $SU(6)_{CS}$ representation for $q^4$, $[31]$, not only because it is obvious from  Table~\ref{su6-2} that the color state for $q^4$ is in a triplet state, and at the same time the spin state for $q^4$ is in a triplet state, but also because these Young-Yamanouchi bases should be represented in the form of the Young diagram $[31]$ for $q^4$. Thus, the states in Eq.~(\ref{cs-1/2-1-31}) are eigenstates of the quadratic Casimir operator of $SU(6)_{CS}$ for $q^4$, with an eigenvalue of 38/3, as can be seen in  Table~\ref{su6-2}.

Moreover, we can see from the Young tableau for $q^3$ in Eq.~(\ref{cs-1/2-1-31}) that the first state belongs to the multiplet of $SU(6)_{CS}$ representation for $q^3$ represented by  Young diagram [3], while both the second and the third states belong to the multiplet of $SU(6)_{CS}$ representation for $q^3$ represented by Young diagram $[21]$.

Next, we show the Young-Yamanouchi bases of the coupling scheme corresponding to the Young diagram $[2^2]$ with the total spin $S=1/2$ and at the same time $S=1$ among particles 1-4 which come from the coupling scheme in Eq.~(\ref{cs2}):

\begin{align}
\begin{tabular}{|c|c|}
\hline
           1 &  2    \\
\hline
          3  &  4 \\
\hline
\end{tabular}_{CS^{1/2};S^1}
=&-\frac{1}{\sqrt{3}}\vert C_1 \rangle \otimes \vert S^{1/2}_1 \rangle
+\frac{1}{\sqrt{6}}\vert C_1 \rangle \otimes \vert S^{1/2}_2 \rangle-
\nonumber \\
&\frac{1}{\sqrt{6}}\vert C_2 \rangle \otimes \vert S^{1/2}_3 \rangle
-\frac{1}{\sqrt{3}}\vert C_3 \rangle \otimes \vert S^{1/2}_3 \rangle,
\nonumber
\end{align}
\begin{align}
\begin{tabular}{|c|c|}
\hline
        1 &  3    \\
\hline
          2  &  4 \\
\hline
\end{tabular}_{CS^{1/2};S^1}
=&-\frac{1}{\sqrt{6}}\vert C_1 \rangle \otimes \vert S^{1/2}_3 \rangle
-\frac{1}{\sqrt{3}}\vert C_2 \rangle \otimes \vert S^{1/2}_1 \rangle-
\nonumber \\
&\frac{1}{\sqrt{6}}\vert C_2 \rangle \otimes \vert S^{1/2}_2 \rangle
+\frac{1}{\sqrt{3}}\vert C_3 \rangle \otimes \vert S^{1/2}_2 \rangle.
\label{cs-1/2-1-22}
\end{align}

The states in Eq.~(\ref{cs-1/2-1-22}) belong to the multiplet of the $SU(6)_{CS}$ representation for $q^4$, $[2^2]$, not only because it is obvious from Table~\ref{su6-2} that the color state for $q^4$ is in a triplet state, and at the same time the spin state for $q^4$ is in a triplet state, but also because these Young-Yamanouchi bases should be represented in the form of Young diagram $[2^2]$ for $q^4$. Therefore, the states in Eq.~(\ref{cs-1/2-1-22}) are eigenstates of the quadratic Casimir operator of  $SU(6)_{CS}$ for $q^4$, with an eigenvalue 32/3, which can be seen in  Table~\ref{su6-2}.

Moreover, we can see from the Young tableau for  $q^3$ in Eq.~(\ref{cs-1/2-1-22}) that both the states belong to the multiplet of $SU(6)_{CS}$ representation for $q^3$ represented by the Young diagram [21].

Next, we show the Young-Yamanouchi bases of the coupling scheme corresponding to the Young diagram $[1^4]$ with the total  $S=1/2$ and at the same time $S=1$ among particles 1-4 which come from the coupling scheme in Eq.~(\ref{cs2}):
\begin{align}
\begin{tabular}{|c|}
\hline
           1     \\
\hline
          2   \\
 \hline 
        3  \\
 \hline 
 4     \\        
\hline
\end{tabular}_{CS^{1/2};S^1}
=\frac{1}{\sqrt{3}}(&\vert C_1 \rangle \otimes \vert S^{1/2}_3 \rangle
-\vert C_2 \rangle \otimes \vert S^{1/2}_2 \rangle+
\nonumber \\
&\vert C_3 \rangle \otimes \vert S^{1/2}_1 \rangle).
\label{cs-1/2-1-1111}
\end{align}

The states in Eq.~(\ref{cs-1/2-1-1111}) belong to the multiplet of the  $SU(6)_{CS}$ representation for $q^4$, $[1^4]$, not only because it is obvious from  Table~\ref{su6-2} that the color state for $q^4$ is in a triplet state, and at the same time the spin state  for $q^4$ is in a triplet state, but also because these Young-Yamanouchi bases should be represented in the form of Young diagram $[1^4]$ for $q^4$. Therefore, the state in Eq.~(\ref{cs-1/2-1-1111}) is eigenstate of the quadratic Casimir operator of  $SU(6)_{CS}$ for $q^4$, with an eigenvalue of 14/3, as can be seen in  Table~\ref{su6-2}. It should be noted that since the Young-Yamanouchi bases in the left hand side of Eq.~(\ref{cs-1/2-1-1111}) from the coupling scheme means a fully antisymmetric property, each term in the right hand side should be conjugated to each other in terms of the color state and spin state.

Moreover, we can see from the Young tableau for $q^3$ in Eq.~(\ref{cs-1/2-1-1111}) that the state belongs to the multiplet of $SU(6)_{CS}$ representation for $q^3$ represented by the Young diagram $[1^3]$.

\subsection{The Young-Yamanouchi bases of coupling scheme for $S=3/2$}
Next, we present the Young-Yamanouchi bases from the coupling scheme with regard to the total spin $S=3/2$ in detail. We show the Young-Yamanouchi bases of the coupling scheme corresponding to the Young diagram $[21^2]$ with the total spin $S=3/2$ and at the same time $S=1$ among particles 1-4 which come from the coupling scheme in Eq.~(\ref{cs2}):

\begin{align}
\begin{tabular}{|c|c|}
\hline
           1 &  2    \\
\cline{1-2}
\multicolumn{1}{|c|}{3} \\
\cline{1-1}
\multicolumn{1}{|c|}{4}  \\
\cline{1-1}
\end{tabular}_{CS^{3/2};S^1}
=&-\frac{1}{\sqrt{6}}\vert C_1 \rangle \otimes \vert S^{3/2}_2 \rangle
-\frac{1}{\sqrt{3}}\vert C_1 \rangle \otimes \vert S^{3/2}_3 \rangle+
\nonumber \\
&\frac{1}{\sqrt{3}}\vert C_2 \rangle \otimes \vert S^{3/2}_4 \rangle
-\frac{1}{\sqrt{6}}\vert C_3 \rangle \otimes \vert S^{3/2}_4 \rangle,
\nonumber
\end{align}
\begin{align}
\begin{tabular}{|c|c|}
\hline
           1 &  3    \\
\cline{1-2}
\multicolumn{1}{|c|}{2} \\
\cline{1-1}
\multicolumn{1}{|c|}{4}  \\
\cline{1-1}
\end{tabular}_{CS^{3/2};S^1}
=&\frac{1}{\sqrt{3}}\vert C_1 \rangle \otimes \vert S^{3/2}_4 \rangle
-\frac{1}{\sqrt{6}}\vert C_2 \rangle \otimes \vert S^{3/2}_2 \rangle+
\nonumber \\
&\frac{1}{\sqrt{3}}\vert C_2 \rangle \otimes \vert S^{3/2}_3 \rangle
+\frac{1}{\sqrt{6}}\vert C_3 \rangle \otimes \vert S^{3/2}_3 \rangle,
\nonumber
\end{align}
\begin{align}
\begin{tabular}{|c|c|}
\hline
           1 &  4   \\
\cline{1-2}
\multicolumn{1}{|c|}{2} \\
\cline{1-1}
\multicolumn{1}{|c|}{3}  \\
\cline{1-1}
\end{tabular}_{CS^{3/2};S^1}
=&-\frac{1}{\sqrt{6}}\vert C_1 \rangle \otimes \vert S^{3/2}_4 \rangle
+\frac{1}{\sqrt{6}}\vert C_2 \rangle \otimes \vert S^{3/2}_3 \rangle+
\nonumber \\
&\sqrt{\frac{2}{3}}\vert C_3 \rangle \otimes \vert S^{3/2}_2 \rangle.
\label{cs-3/2-1-211}
\end{align}

The states in Eq.~(\ref{cs-3/2-1-211}) correspond to the multiplet of the  $SU(6)_{CS}$ representation for $q^4$, $[21^2]$, not only because it is obvious from  Table~\ref{su6-2} that the color state for $q^4$ is in a triplet state, and at the same time  the spin state  for $q^4$ is in a triplet state but also because these Young-Yamanouchi bases should be represented in the form of Young diagram $[21^2]$ for $q^4$. Therefore, the states in Eq.~(\ref{cs-3/2-1-211}) are eigenstates of the quadratic Casimir operator of  $SU(6)_{CS}$ for $q^4$, with an eigenvalue of 26/3, as shown in Table\ref{su6-2}.

Furthermore, it is worth noting that from the standpoint of the permutation group, $S_4$, the states in Eq.~(\ref{cs-3/2-1-211}) are exactly equivalent to the states in Eq.~(\ref{cs-1/2-1-211}), since the Young tableau of the spin states in Eq.~(\ref{cs-3/2-1-211}) are the same as those of the spin states in Eq.~(\ref{cs-1/2-1-211}) for the four quarks. As the result, we can infer that the values of  the color and spin operator acted upon either the states in Eq.~(\ref{cs-3/2-1-211}) or the states in Eq.(\ref{cs-1/2-1-211}) are identical for four quarks.

For the three quarks, we can see from the Young tableau in Eq.~(\ref{cs-3/2-1-211}) that both the first and second states belong to the multiplet of $SU(6)_{CS}$ representation for $q^3$ represented by the Young diagram $[21]$, while the third belongs to the multiplet of $SU(6)_{CS}$ representation for $q^3$ represented by the Young diagram $[1^3]$.

Next, we show the Young-Yamanouchi bases of the coupling scheme corresponding to Young diagram $[21^2]$ with the total spin $S=3/2$ and at the same time $S=2$ among particles 1-4, which come from the coupling scheme in Eq.~(\ref{cs3}):

\begin{align}
&\begin{tabular}{|c|c|}
\hline
           1 &  2    \\
\cline{1-2}
\multicolumn{1}{|c|}{3} \\
\cline{1-1}
\multicolumn{1}{|c|}{4}  \\
\cline{1-1}
\end{tabular}_{CS^{3/2};S^2}
=\vert C_1 \rangle \otimes \vert S^{3/2}_1 \rangle,
\begin{tabular}{|c|c|}
\hline
           1 &  3    \\
\cline{1-2}
\multicolumn{1}{|c|}{2} \\
\cline{1-1}
\multicolumn{1}{|c|}{4}  \\
\cline{1-1}
\end{tabular}_{CS^{3/2};S^2}
=\vert C_2 \rangle \otimes \vert S^{3/2}_1 \rangle,
\nonumber \\
&\begin{tabular}{|c|c|}
\hline
           1 &  4   \\
\cline{1-2}
\multicolumn{1}{|c|}{2} \\
\cline{1-1}
\multicolumn{1}{|c|}{3}  \\
\cline{1-1}
\end{tabular}_{CS^{3/2};S^2}
=\vert C_3 \rangle \otimes \vert S^{3/2}_1 \rangle.
\label{cs-3/2-2-211}
\end{align}

The states in Eq.~(\ref{cs-3/2-2-211}) correspond to the multiplet of the  $SU(6)_{CS}$ representation for $q^4$, $[21^2]$, not only because it is obvious from  Table~\ref{su6-2} that the color state for $q^4$ is in a triplet state, and at the same time the spin state for $q^4$ is in a quintet state, but also because these Young-Yamanouchi bases should be represented in the form of the Young diagam $[21^2]$ for $q^4$. Therefore, the states in Eq.~(\ref{cs-3/2-2-211}) are eigenstates of the quadratic Casimir operator of  $SU(6)_{CS}$ for $q^4$, with an eigenvalue of 26/3, as shown in  Table~\ref{su6-2}. 

Moreover, we can see from the Young tableau for $q^3$ in Eq.~(\ref{cs-3/2-2-211}) that both the first and the second states belong to the multiplet of $SU(6)_{CS}$ representation for $q^3$ represented by Young diagram $[21]$, while the third state belongs to the multiplet of $SU(6)_{CS}$ representation for $q^3$ represented by Young diagram $[1^3]$.

Next, we show the Young-Yamanouchi bases of the coupling scheme corresponding to the Young diagram $[31]$ with the total spin $S=3/2$ and at the same time $S=1$ among particles 1-4 which come from the coupling scheme in Eq.~(\ref{cs2}):

\begin{align}
\begin{tabular}{|c|c|c|}
\hline
           1 &  2 & 3   \\
\cline{1-3}
\multicolumn{1}{|c|}{4} \\
\cline{1-1}
\end{tabular}_{CS^{3/2};S^1}
=\frac{1}{\sqrt{2}}\vert C_1 \rangle \otimes \vert S^{3/2}_3 \rangle
+\frac{1}{\sqrt{2}}\vert C_2 \rangle \otimes \vert S^{3/2}_4 \rangle,
\nonumber
\end{align}
\begin{align}
\begin{tabular}{|c|c|c|}
\hline
           1 &  2 & 4   \\
\cline{1-3}
\multicolumn{1}{|c|}{3} \\
\cline{1-1}
\end{tabular}_{CS^{3/2};S^1}
=-\frac{1}{\sqrt{2}}\vert C_1 \rangle \otimes \vert S^{3/2}_2 \rangle
+\frac{1}{\sqrt{2}}\vert C_3 \rangle \otimes \vert S^{3/2}_4 \rangle,
\nonumber
\end{align}
\begin{align}
\begin{tabular}{|c|c|c|}
\hline
           1 &  3 & 4   \\
\cline{1-3}
\multicolumn{1}{|c|}{2} \\
\cline{1-1}
\end{tabular}_{CS^{3/2};S^1}
=-\frac{1}{\sqrt{2}}\vert C_2 \rangle \otimes \vert S^{3/2}_2 \rangle
-\frac{1}{\sqrt{2}}\vert C_3 \rangle \otimes \vert S^{3/2}_3 \rangle.
\label{cs-3/2-1-31}
\end{align}
The states in Eq.~(\ref{cs-3/2-1-31}) belong to the multiplet of the  $SU(6)_{CS}$ representation for $q^4$, $[31]$, not only because it is obvious from  Table~\ref{su6-2} that the color state for $q^4$ is in a triplet state, and at the same time the spin state for $q^4$ is in a triplet state, but also because these Young-Yamanouchi bases states should be represented in the form of the Young diagram $[31]$ for $q^4$. Thus, the states in Eq.~(\ref{cs-3/2-1-31}) are eigenstates of the quadratic Casimir operator of $SU(6)_{CS}$ for $q^4$, with an eigenvalue of 38/3, as shown in Table\ref{su6-2}.

Furthermore, it is worth noting that from the standpoint of the permutation group, $S_4$, the states in Eq.~(\ref{cs-3/2-1-31}) are exactly equivalent to the states in Eq.~(\ref{cs-1/2-1-31}), since the Young tableau of the spin states in Eq.~(\ref{cs-3/2-1-31}) are the same as those of the spin states in Eq.~(\ref{cs-1/2-1-31}) for the four quarks. As the result, we can infer that the values of  the color and spin operator acted upon either the states in Eq.~(\ref{cs-3/2-1-31}) or the states in Eq.(\ref{cs-1/2-1-31}) are identical for four quarks.

In addition, we can see from the Young tableau for $q^3$ in Eq.~(\ref{cs-3/2-1-31}) that the first state belongs to the multiplet of $SU(6)_{CS}$ representation for $q^3$ represented by  Young diagram $[3]$, while both the second and the third states belong to the multiplet of $SU(6)_{CS}$ representation for $q^3$ represented by the Young diagram $[21]$.

Next, we show the Young-Yamanouchi bases of the coupling scheme corresponding to the Young diagram $[2^2]$ with the total $S=3/2$ and at the same time $S=1$ among particles 1-4, which come from the coupling scheme in Eq.~(\ref{cs2}):

\begin{align}
\begin{tabular}{|c|c|}
\hline
           1 &  2    \\
\hline
          3  &  4 \\
\hline
\end{tabular}_{CS^{3/2};S^1}
=&-\frac{1}{\sqrt{3}}\vert C_1 \rangle \otimes \vert S^{3/2}_2 \rangle
+\frac{1}{\sqrt{6}}\vert C_1 \rangle \otimes \vert S^{3/2}_3 \rangle-
\nonumber \\
&\frac{1}{\sqrt{6}}\vert C_2 \rangle \otimes \vert S^{3/2}_4 \rangle
-\frac{1}{\sqrt{3}}\vert C_3 \rangle \otimes \vert S^{3/2}_4 \rangle,
\nonumber
\end{align}
\begin{align}
\begin{tabular}{|c|c|}
\hline
        1 &  3    \\
\hline
          2  &  4 \\
\hline
\end{tabular}_{CS^{3/2};S^1}
=&-\frac{1}{\sqrt{6}}\vert C_1 \rangle \otimes \vert S^{3/2}_4 \rangle
-\frac{1}{\sqrt{3}}\vert C_2 \rangle \otimes \vert S^{3/2}_2 \rangle-
\nonumber \\
&\frac{1}{\sqrt{6}}\vert C_2 \rangle \otimes \vert S^{3/2}_3 \rangle
+\frac{1}{\sqrt{3}}\vert C_3 \rangle \otimes \vert S^{3/2}_3 \rangle.
\label{cs-3/2-1-22}
\end{align}
The states in Eq.~(\ref{cs-3/2-1-22}) correspond to the multiplet of the  $SU(6)_{CS}$ representation for $q^4$, $[2^2]$, not only because it is obvious from  Table~\ref{su6-2} that the color state for $q^4$ is in a triplet state, and at the same time  the spin state  for $q^4$ is in a triplet state, but also because these Young-Yamanouchi bases should be represented in the form of Young diagam $[2^2]$ for $q^4$. Therefore, the states in Eq.~(\ref{cs-3/2-1-22}) are eigenstates of the quadratic Casimir operator of  $SU(6)_{CS}$ for $q^4$, with an eigenvalue of 32/3, as can be seen in  Table~\ref{su6-2}.

From the standpoint of the permutation group, $S_4$, the states in Eq.~(\ref{cs-3/2-1-22}) are exactly equivalent to the states in Eq.~(\ref{cs-1/2-1-22}), since the Young tableau of the spin states in Eq.~(\ref{cs-3/2-1-22}) are the same as those of the spin states in Eq.~(\ref{cs-1/2-1-22}) for the four quarks.

We also can see from the  Young tableau for  $q^3$ in Eq.~(\ref{cs-3/2-1-22}) that both the first and the second   belong to the multiplet of $SU(6)_{CS}$ representation for $q^3$ represented by  Young diagram $[21]$.

Next, we show the Young-Yamanouchi bases of the coupling scheme corresponding to Young diagram $[1^4]$ with the total  $S=3/2$ and at the same time $S=1$ among particles 1-4 which come from the coupling scheme in Eq.~(\ref{cs2}):

\begin{align}
\begin{tabular}{|c|}
\hline
           1     \\
\hline
          2   \\
 \hline 
        3  \\
 \hline 
 4     \\        
\hline
\end{tabular}_{CS^{3/2};S^1}
=\frac{1}{\sqrt{3}}(&\vert C_1 \rangle \otimes \vert S^{3/2}_4 \rangle
-\vert C_2 \rangle \otimes \vert S^{3/2}_3 \rangle+
\nonumber \\
&\vert C_3 \rangle \otimes \vert S^{3/2}_2 \rangle).
\label{cs-3/2-1-1111}
\end{align}

The state in Eq.~(\ref{cs-3/2-1-1111}) correspond to the multiplet of the $SU(6)_{CS}$ representation for $q^4$, $[1^4]$, not only because it is obvious from  Table~\ref{su6-2} that the color state for $q^4$ is in a triplet state, and at the same time the spin state  for $q^4$ is in a triplet state, but also because these Young-Yamanouchi bases should be represented in the form of Young diagam $[1^4]$ for $q^4$. Therefore, the state in Eq.~(\ref{cs-3/2-1-1111}) is eigenstate of the quadratic Casimir operator of  $SU(6)_{CS}$ for $q^4$, with an eigenvalue of 14/3, as can be seen in  Table~\ref{su6-2}. It should be noted that since the Young-Yamanouchi bases in the left hand side of Eq.~(\ref{cs-3/2-1-1111}) from the coupling scheme means a fully antisymmetric property, each term in the right hand side should be conjugated to each other in terms of the color state and spin state.

Besides, from the standpoint of the permutation group, $S_4$, the state in Eq.~(\ref{cs-3/2-1-1111}) are exactly equivalent to  state in Eq.~(\ref{cs-1/2-1-1111}), since for four quarks the Young tableau of the spin states in Eq.~(\ref{cs-3/2-1-1111}) are the same as those of the spin states in Eq.~(\ref{cs-1/2-1-1111}). In addition,  we can see from the  Young tableau for  $q^3$ in Eq.~(\ref{cs-3/2-1-1111}) that the state  belongs to the multiplet of $SU(6)_{CS}$ representation for $q^3$ represented by  Young diagram $[1^3]$.

\subsection{The Young-Yamanouchi bases of coupling scheme for $S=5/2$}

Finally, we show the Young-Yamanouchi bases for the coupling scheme corresponding to the Young diagram $[21^2]$ with the total spin $S=5/2$ and at the same time $S=2$ among particles 1-4, which come from the coupling scheme in Eq.~(\ref{cs3}):

\begin{align}
&\begin{tabular}{|c|c|}
\hline
           1 &  2    \\
\cline{1-2}
\multicolumn{1}{|c|}{3} \\
\cline{1-1}
\multicolumn{1}{|c|}{4}  \\
\cline{1-1}
\end{tabular}_{CS^{5/2};S^2}
=\vert C_1 \rangle \otimes \vert S^{5/2} \rangle,
\begin{tabular}{|c|c|}
\hline
           1 &  3    \\
\cline{1-2}
\multicolumn{1}{|c|}{2} \\
\cline{1-1}
\multicolumn{1}{|c|}{4}  \\
\cline{1-1}
\end{tabular}_{CS^{5/2};S^2}
=\vert C_2 \rangle \otimes \vert S^{5/2} \rangle,
\nonumber \\
&\begin{tabular}{|c|c|}
\hline
           1 &  4   \\
\cline{1-2}
\multicolumn{1}{|c|}{2} \\
\cline{1-1}
\multicolumn{1}{|c|}{3}  \\
\cline{1-1}
\end{tabular}_{CS^{5/2};S^2}
=\vert C_3 \rangle \otimes \vert S^{5/2} \rangle.
\label{cs-5/2-2-211}
\end{align}

The states in Eq.~(\ref{cs-5/2-2-211}) belong to the multiplet of the  $SU(6)_{CS}$ representation for $q^4$, $[21^2]$, not only because it is obvious from Table~\ref{su6-2} that the color state for $q^4$ is in a triplet state, and at the same time the spin state for $q^4$ is in a quintet state, but also because these Young-Yamanouchi bases should be represented in the form of Young diagram $[21^2]$ for $q^4$. Therefore, the states in Eq.~(\ref{cs-5/2-2-211}) are eigenstates of the quadratic Casimir operator of  $SU(6)_{CS}$ for $q^4$, with an eigenvalue of 26/3, as can be seen in  Table~\ref{su6-2}.

From the standpoint of the permutation group, $S_4$, the state in Eq.~(\ref{cs-5/2-2-211}) are exactly equivalent to state in Eq.~(\ref{cs-3/2-2-211}), since the Young tableau of the spin states in Eq.~(\ref{cs-5/2-2-211}) are the same as those of the spin states in Eq.~(\ref{cs-3/2-2-211}) for the four quarks.

We can also see from the Young tableau for $q^3$ in Eq.~(\ref{cs-5/2-2-211}) that both the first and the second  belong to the multiplet of $SU(6)_{CS}$ representation for $q^3$ represented by  Young diagram $[21]$, while the third  belongs to the multiplet of $SU(6)_{CS}$ representation for $q^3$ represented by  Young diagram $[1^3]$. We note that all of  Young-Yamanouchi bases of the coupling scheme for any total $S$ are orthonormal to each other.

In particular, we now consider the quadratic Casimir operator of $SU(6)_{CS}$ for $q^3$ in order to identify a correspondence between the Young-Yamanouchi bases of the coupling scheme and the multiplets of $SU(6)_{CS}$ representation for $q^3$. Given the quadratic Casimir operator of $SU(6)_{CS}$ for $q^3$, in fact, we can calculate the eigenvalues in Table~\ref{su6-3} by acting the quadratic Casimir operator upon the Young-Yamanouchi bases of the coupling scheme. This can be possible by introducing a formula given by:

\begin{align}
C^{CS}_3=&\frac{1}{4}{\sum}_{i=1}^{2}\lambda_i^c\lambda_3^c{\vec{\sigma}}_i\cdot{\vec{\sigma}}_3
+C^{CS}_2+\frac{1}{2}C^{C}_3-\frac{1}{2}C^{C}_2 +
\nonumber \\
&\frac{1}{3}(\vec{S}\cdot \vec{S})_3-
\frac{1}{3}(\vec{S}\cdot \vec{S})_2+2I,
 \label{Casimir}
 \end{align}
where $C^{CS}$ is the quadratic form of Casimir operator of $SU(6)_{CS}$, $C^C$ the quadratic form of Casimir operator of $SU(3)_C$, $S$ the spin operator, $I$ the identity operator, and the subscript in each Casimir operator indicates the number of the participant quarks. This notation will be used for the subsequent formula associated with  Casimir operator.

For example, we consider the third state in Eq.(\ref{cs-1/2-0-211}) and the first state in Eq.(\ref{cs-1/2-0-31}). From the color and spin state for $q^3$ of the Young-Yamanouchi basis, both states are in a $(\mathbf{8},\mathbf{2})$ state, which is the composition of color state and spin state contained in both $[1^3]$ and $[3]$ of the SU(6)$_{CS}$ representation for $q^3$. However, in spite of being in the same composition of color state and spin state, $(\mathbf{8},\mathbf{2})$, we can distinguish these states immediately from the Young-Yamanouchi basis of the $SU(6)_{CS}$ representation for $q^3$. In fact, we can  calculate their eigenvalues of $C^{CS}_3$ to these states and examine whether these states become eigenstates of $C^{CS}_3$, using the formula in Eq.~(\ref{Casimir}). However, it is not so easy to show this, owing to the first term in Eq.~(\ref{Casimir}). After some complex algebraic calculation, we obtain the following eigenvalue equation:

\begin{align}
&C^{CS}_3
\begin{tabular}{|c|c|}
\hline
           1 &  4   \\
\cline{1-2}
\multicolumn{1}{|c|}{2} \\
\cline{1-1}
\multicolumn{1}{|c|}{3}  \\
\cline{1-1}
\end{tabular}_{CS^{1/2};S^0}
=\frac{21}{4}
\begin{tabular}{|c|c|}
\hline
           1 &  4   \\
\cline{1-2}
\multicolumn{1}{|c|}{2} \\
\cline{1-1}
\multicolumn{1}{|c|}{3}  \\
\cline{1-1}
\end{tabular}_{CS^{1/2};S^0}
,
\nonumber \\
&C^{CS}_3
\begin{tabular}{|c|c|c|}
\hline
           1 &  2 & 3   \\
\cline{1-3}
\multicolumn{1}{|c|}{4} \\
\cline{1-1}
\end{tabular}_{CS^{1/2};S^0}
=\frac{45}{4}
\begin{tabular}{|c|c|c|}
\hline
           1 &  2 & 3   \\
\cline{1-3}
\multicolumn{1}{|c|}{4} \\
\cline{1-1}
\end{tabular}_{CS^{1/2};S^0}
.
\label{Casimir-eigenvalue}
\end{align}
Here, the eigenvalue of $C^{CS}_2$ is manifestly  determined by the composition of the color and spin state for $q^2$, which is either symmetric or antisymmetric. Table~\ref{su6-5} shows the eigenvalues of $C^{CS}_2$ and the composition of color and spin state of the $SU(6)_{CS}$ representation for $q^2$.

\begin{table}[htp]
\caption{ The composition of $SU(3)_C$ and $SU(2)_S$ concerning the $SU(6)_{CS}$ representation
for $q^2$    }
\begin{center}
\begin{tabular}{c|c|c|c}
\hline \hline
$SU(6)_{CS}$  &         &          &           \\
  Young diagram  & $SU(3)_C$ $\otimes$  $SU(2)_S$ & Dimesion  &  Eigenvalue \\
\hline
 $[2]$  &    $(\mathbf{6},\mathbf{3})$,    $(\mathbf{\bar{3}},\mathbf{1})$        &    21         &  20/3        \\
   \hline  
 $[1^2]$  &     $(\mathbf{6},\mathbf{1})$,   $(\mathbf{\bar{3}},\mathbf{3})$             &   15         &  14/3         \\
  \hline \hline        
\end{tabular}
\end{center}
\label{su6-5}
\end{table}

\section{Systematic analysis of $q^4\bar{q}$}

\subsection{ The $SU(6)_{CS}$ representation  of pentaquark}

In this subsection, we deal with a correspondence between the irreducible representations of $SU(6)_{CS}$ for the pentaquark and the color and spin states in the coupling scheme mentioned in a preceding section. For our purpose, we categorize the direct product of the fundamental representation of $SU(6)_{CS}$ for the pentaquark, which leads to useful information in calculating the color-spin interaction. This procedure is achieved by classifying the 7776 dimensional color $\otimes$ spin states of $q^4\bar{Q}$ into the direct sum of the irreducible representations of $SU(6)_{CS}$. This is obtained from multiplying Eq.~(\ref{cs-decom}) by $\mathbf{\bar{6}}_{CS}$, as given in Eq.~(\ref{cs-decomposition}).

In Eq.~(\ref{cs-decomposition}), the subscripts outside of the brackets in the second and third lines indicate the SU(6)$_{CS}$ representation for $q^4$, since the SU(6)$_{CS}$ representations for $q^4\bar{q}$ inside the bracket are originally due to the direct product of the SU(6)$_{CS}$ representation for $q^4$ and $\mathbf{\bar{6}}_{CS}$ for $\bar{q}$ in the first line. Therefore, the SU(6)$_{CS}$ representations for $q^4\bar{q}$ inside the bracket, represented by their corresponding Young diagram in the subscript, have a certain symmetry for $q^4$ which is expressed in terms of the Young diagram of the SU(6)$_{CS}$ representation for $q^4$.  The dimension of each of irreducible SU(6)$_{CS}$ representation for the pentaquark is given in the third column of Table~\ref{su6-1}.

On the one hand, we further decompose the SU(6)$_{CS}$ into the sum of $SU(3)_C \otimes SU(2)_S$ multiplets in order to select out the physically allowed color singlet states among the $SU(6)_{CS}$ representations in Eq.~(\ref{cs-decomposition}).    
Table \ref{colorsinglet-spin} shows the allowed color singlet states with their possible spin states, denoted by $[\mathbf{1}_C,S]$, within each SU(6)$_{CS}$ representation of pentaquark.
Table~\ref{su6-1} shows the composition of $SU(3)_C$ and $SU(2)_S$ multiplets in the $SU(6)_{CS}$ representation, as well as the dimension and the eigenvalue of $C^{CS}$, which is the quadratic form of the Casimir operator of the $SU(6)_{CS}$ representation.

\begin{widetext}

\begin{align}
&\mathbf{6}_{CS}\otimes\mathbf{6}_{CS}\otimes\mathbf{6}_{CS}\otimes\mathbf{6}_{CS}
\otimes\mathbf{\bar{6}}_{CS}=
\Bigg(
\quad
\begin{tabular}{|c|c|c|c|}
\hline
$\quad$ &$\quad$  &$\quad$ & $\quad$   \\
\hline
\end{tabular}
\oplus
3\times
\begin{tabular}{|c|c|}
\hline
      $\quad$        &    $\quad$      \\
\cline{1-2} 
\multicolumn{1}{|c|}{  $\quad$  }  \\
\cline{1-1} 
\multicolumn{1}{|c|}{  $\quad$  }  \\
\cline{1-1} 
\end{tabular}
\oplus
\begin{tabular}{|c|}
\hline
     $\quad$          \\
\hline
    $\quad$          \\
  \hline
     $\quad$          \\
\hline
    $\quad$          \\
\hline         
\end{tabular}
\oplus
3\times
\begin{tabular}{|c|c|c|}
\hline
    $\quad$       &  $\quad$  & $\quad$    \\
\cline{1-3} 
\multicolumn{1}{|c|}{$\quad$}  \\
\cline{1-1} 
\end{tabular}
\oplus
2\times
\begin{tabular}{|c|c|}
\hline
   $\quad$    & $\quad$         \\
\hline
   $\quad$    & $\quad$         \\
\hline
\end{tabular}
\quad
\Bigg)
\otimes
\begin{tabular}{|c|}
\hline
     $\quad$          \\
\hline
    $\quad$          \\
  \hline
     $\quad$          \\
\hline
    $\quad$          \\
\hline  
   $\quad$          \\
\hline           
\end{tabular}
=
\nonumber \\
&\Bigg(
\quad
\begin{tabular}{|c|c|c|c|c|}
\hline
 $\quad$ &  $\quad$  & $\quad$  & $\quad$   & $\quad$   \\
\cline{1-5}
\multicolumn{1}{|c|}{$\quad$}  \\  
\cline{1-1}
\multicolumn{1}{|c|}{$\quad$}  \\  
\cline{1-1}
\multicolumn{1}{|c|}{$\quad$}  \\ 
 \cline{1-1}
\multicolumn{1}{|c|}{$\quad$}  \\  
\cline{1-1}
 \end{tabular}
 \oplus
\begin{tabular}{|c|c|c|}
\hline
 $\quad$ &  $\quad$  & $\quad$  \\
 \hline 
  \end{tabular}
 \quad  \Bigg
 )_{[4]}
 \oplus
 3\Bigg( 
 \quad
\begin{tabular}{|c|c|c|}
\hline
 $\quad$ &  $\quad$  & $\quad$   \\
\cline{1-3}
\multicolumn{1}{|c|}{$\quad$} & \multicolumn{1}{c|}{$\quad$} \\  
\cline{1-2}
\multicolumn{1}{|c|}{$\quad$} & \multicolumn{1}{c|}{$\quad$} \\  
\cline{1-2}
\multicolumn{1}{|c|}{$\quad$}  \\ 
 \cline{1-1}
\multicolumn{1}{|c|}{$\quad$}  \\  
\cline{1-1}
 \end{tabular}
 \oplus
  \begin{tabular}{|c|c|}
\hline
 $\quad$ &  $\quad$    \\
 \cline{1-2}
 \multicolumn{1}{|c|}{$\quad$}  \\  
\cline{1-1}
 \end{tabular}
 \oplus
 \begin{tabular}{|c|}
\hline
 $\quad$     \\
\hline
 $\quad$     \\
  \hline
 $\quad$     \\
  \hline
 \end{tabular}
 \quad
   \Bigg )_{[21^2]} 
    \oplus
\Bigg( 
\quad
 \begin{tabular}{|c|c|}
\hline
 $\quad$ &  $\quad$  \\
 \hline
 $\quad$ &  $\quad$  \\
 \hline
 $\quad$ &  $\quad$  \\
 \hline
 $\quad$ &  $\quad$  \\
\cline{1-2}
\multicolumn{1}{|c|}{$\quad$}  \\  
\cline{1-1}
 \end{tabular}
 \oplus 
   \begin{tabular}{|c|}
\hline
 $\quad$     \\
\hline
 $\quad$     \\
  \hline
 $\quad$     \\
  \hline
 \end{tabular}
 \quad
   \Bigg )_{[1^4]}
 \oplus 
 \nonumber \\
  &3\Bigg( 
  \quad
\begin{tabular}{|c|c|c|c|}
\hline
 $\quad$ &  $\quad$  & $\quad$  & $\quad$     \\
\cline{1-4}
\multicolumn{1}{|c|}{$\quad$} & \multicolumn{1}{c|}{$\quad$} \\  
\cline{1-2}
\multicolumn{1}{|c|}{$\quad$}  \\  
\cline{1-1}
\multicolumn{1}{|c|}{$\quad$}  \\ 
 \cline{1-1}
\multicolumn{1}{|c|}{$\quad$}  \\  
\cline{1-1}
 \end{tabular}
  \oplus
\begin{tabular}{|c|c|c|}
\hline
 $\quad$ &  $\quad$  & $\quad$  \\
 \hline 
  \end{tabular}
    \oplus
  \begin{tabular}{|c|c|}
\hline
 $\quad$ &  $\quad$    \\
 \cline{1-2}
 \multicolumn{1}{|c|}{$\quad$}  \\  
\cline{1-1}
 \end{tabular}
 \quad
  \Bigg )_{[31]} 
  \oplus
  2\Bigg( 
  \quad 
\begin{tabular}{|c|c|c|}
\hline
 $\quad$ &  $\quad$  & $\quad$   \\
\cline{1-3}
\multicolumn{1}{|c|}{$\quad$} & \multicolumn{1}{c|}{$\quad$} & \multicolumn{1}{c|}{$\quad$}  \\  
\cline{1-3}
\multicolumn{1}{|c|}{$\quad$}  \\  
\cline{1-1}
\multicolumn{1}{|c|}{$\quad$}  \\ 
 \cline{1-1}
\multicolumn{1}{|c|}{$\quad$}  \\  
\cline{1-1}
 \end{tabular}
 \oplus
 \begin{tabular}{|c|c|}
\hline
 $\quad$ &  $\quad$    \\
 \cline{1-2}
 \multicolumn{1}{|c|}{$\quad$}  \\  
\cline{1-1}
 \end{tabular}
 \quad
  \Bigg )_{[2^2]}.
\label{cs-decomposition}      
\end{align}

\end{widetext}

In the simplest constituent quark model based on the color spin interaction, the attraction of stabilizing pentaquark depends critically on the expectation value  of the interaction. 
Now that we have completed the necessary classifications and obtained the required results, we are in a position to calculate the color-spin interaction resulting from quantum chromodynamics. This interaction is given by:

\begin{align}
H_{CS}= - {\sum}_{i<j}^{5}\lambda_i^c\lambda_j^c{\vec{\sigma}}_i\cdot{\vec{\sigma}}_j.
\label{formula-1}  
\end{align}

\begin{table}[htp]
\caption{ The  SU(6)$_{CS}$ representations  containing $[\mathbf{1}_C, S]$ multiplet. }
\begin{center}
\begin{tabular}{c|c}
\hline \hline
  $SU(3)_C \otimes SU(2)_S$                 &    SU(6)$_{CS}$ representation    \\
\hline
$[\mathbf{1}_C, 1/2]$  & $[2^41]$, $[32^21^2]$, [21],   $[421^3]$ \\
$[\mathbf{1}_C, 3/2]$  &  $[1^3]$, $[32^21^2]$, $[3^21^3]$,  $[421^3]$ \\
$[\mathbf{1}_C, 5/2]$  & $[32^21^2]$          \\
\hline \hline
\end{tabular}
\end{center}
\label{colorsinglet-spin}
\end{table}
  
One can derive the following elegant formula for Eq.~({\ref{formula-1}) relevant to the pentaquark configuration by introducing the quadratic form of Casimir operator of SU(6)$_{CS}$, which is denoted by $C^{CS}$:

\begin{align}
&- {\sum}_{i<j}^{5}\lambda_i^c\lambda_j^c{\vec{\sigma}}_i\cdot{\vec{\sigma}}_j = 
4C^{CS}_5-8C^{CS}_4-2C^{C}_5+4C^{C}_4-
\nonumber \\
&\frac{4}{3}(\vec{S}\cdot \vec{S})_5 +\frac{8}{3} (\vec{S}\cdot  \vec{S})_4+24I,
\label{formula-2}  
 \end{align}
where the subscript in each Casimir operator indicates the number of the participant quarks, 
the terms $C^{CS}$ and $C^C$ refer to the quadratic form of Casimir operator of $SU(6)_{CS}$ and $SU(3)_C$, respectively. The operator $S$ denotes the spin operator, and $I$ the identity operator. The derivation of the formula is similar to that of the corresponding formula for the dibaryon system described in \cite{Aerts:1977rw}. However, to account for the antiquark, we replace $\lambda_i$ ($\vec{\sigma}_j$) with -$\lambda_i^*$ (-$\vec{\sigma}_j^*$). In  Eq.~(\ref{formula-2}), the eigenvalue of $C^C_4$ is 4/3, because the Young tableau of $q^4$ is $[21^2]$, when the pentaquark is in the color singlet configuration.

We are now in a position to construct $\otimes$ color $\otimes$ spin states in terms of the irreducible $SU(6)_{CS}$ representation of the pentaquark, which is useful for understanding the symmetry property among the four quarks as well as making a good choice of full wave function, involving a flavor and a spatial function.
To do this, we examine all the cases corresponding to the $SU(3)_F$ limit available to Eq.~(\ref{formula-2}).

\begin{widetext}

\begin{table}[htp]
\caption{ The composition of $SU(3)_C$ and $SU(2)_S$ concerning the $SU(6)_{CS}$ representation
for $q^4\bar{q}$    }
\begin{center}
\begin{tabular}{c|c|c|c}
\hline \hline
$SU(6)_{CS}$  &         &          &           \\
  Young tableau & $SU(3)_C$ $\otimes$  $SU(2)_S$ & Dimesion  &  Eigenvalue \\
\hline
$[51^4]$  &    ($\mathbf{35}$,$\mathbf{6}$), ($\mathbf{35}$,$\mathbf{4}$), ($\mathbf{27}$,$\mathbf{4}$), ($\mathbf{27}$,$\mathbf{2}$), ($\mathbf{10}$,$\mathbf{6}$), ($\mathbf{8}$,$\mathbf{4}$),                         &    700        &  81/4        \\
    & ($\mathbf{10}$,$\mathbf{4}$), ($\mathbf{10}$,$\mathbf{2}$), ($\bar{\mathbf{10}}$,$\mathbf{2}$), ($\mathbf{8}$,$\mathbf{2}$)  \qquad   \qquad  \qquad \qquad  \qquad    &           &          \\                                                                           
\hline
$[3]$  &    ($\mathbf{10}$,$\mathbf{4}$), ($\mathbf{8}$,$\mathbf{2}$)                             &   56         &  45/4        \\
\hline
$[421^3]$  &    ($\mathbf{35}$,$\mathbf{4}$), ($\mathbf{35}$,$\mathbf{2}$), ($\mathbf{27}$,$\mathbf{6}$), 2($\mathbf{27}$,$\mathbf{4}$), 2($\mathbf{27}$,$\mathbf{2}$), ($\mathbf{10}$,$\mathbf{6}$),
                                                               &   1134         &    65/4      \\
     &2($\mathbf{10}$,$\mathbf{4}$), 2($\mathbf{10}$,$\mathbf{2}$), ($\bar{\mathbf{10}}$,$\mathbf{4}$), ($\bar{\mathbf{10}}$,$\mathbf{2}$), ($\mathbf{8}$,$\mathbf{6}$), 3($\mathbf{8}$,$\mathbf{4}$),    &       &  \\
 &   3($\mathbf{8}$,$\mathbf{2}$), ($\mathbf{1}$,$\mathbf{4}$), ($\mathbf{1}$,$\mathbf{2}$)  \qquad   \qquad  \qquad \qquad  \qquad  \qquad  \qquad   \qquad    &       &  \\  
 \hline  
 $[21]$  &    ($\mathbf{10}$,$\mathbf{2}$), ($\mathbf{8}$,$\mathbf{4}$), ($\mathbf{8}$,$\mathbf{2}$), ($\mathbf{1}$,$\mathbf{2}$)                            &    70        &  33/4        \\  
 \hline
$[3^21^3]$  &    ($\mathbf{35}$,$\mathbf{2}$), ($\mathbf{27}$,$\mathbf{4}$), ($\mathbf{27}$,$\mathbf{2}$), ($\mathbf{10}$,$\mathbf{4}$), ($\mathbf{10}$,$\mathbf{2}$), ($\bar{\mathbf{10}}$,$\mathbf{6}$),                          &  560          &  57/4        \\
        & ($\bar{\mathbf{10}}$,$\mathbf{4}$), ($\bar{\mathbf{10}}$,$\mathbf{2}$),  ($\mathbf{8}$,$\mathbf{6}$),   2($\mathbf{8}$,$\mathbf{4}$), 2($\mathbf{8}$,$\mathbf{2}$), ($\mathbf{1}$,$\mathbf{4}$)   \qquad    &       &  \\
  \hline
$[32^21^2]$  &    ($\mathbf{27}$,$\mathbf{4}$), 2($\mathbf{27}$,$\mathbf{2}$), ($\mathbf{10}$,$\mathbf{4}$), ($\mathbf{10}$,$\mathbf{2}$), ($\bar{\mathbf{10}}$,$\mathbf{4}$), ($\bar{\mathbf{10}}$,$\mathbf{2}$),                                &  540          &  49/4        \\
 & ($\mathbf{8}$,$\mathbf{6}$), 3($\mathbf{8}$,$\mathbf{4}$),  3($\mathbf{8}$,$\mathbf{2}$),   ($\mathbf{1}$,$\mathbf{6}$), ($\mathbf{1}$,$\mathbf{4}$), ($\mathbf{1}$,$\mathbf{2}$)   \qquad   \qquad  &       &  \\
 \hline  
  $[1^3]$  &    ($\mathbf{8}$,$\mathbf{2}$), ($\mathbf{1}$,$\mathbf{4}$)                            &   20         &    21/4      \\
   \hline  
 $[2^41]$  &  ($\bar{\mathbf{10}}$,$\mathbf{2}$), ($\mathbf{8}$,$\mathbf{4}$), ($\mathbf{8}$,$\mathbf{2}$), ($\mathbf{1}$,$\mathbf{2}$)       &     70       &   33/4       \\
\hline \hline
\end{tabular}
\end{center}
\label{su6-1}
\end{table}

\end{widetext}

\subsection{Color $\otimes$ spin states with $S=1/2$ in terms of the irreducible $SU(6)_{CS}$ representation of the pentaquark}

We first consider flavor $\mathbf{{15}^\prime}$ case with $S=1/2$. In this case, there are two orthonormal flavor $\otimes$ color $\otimes$ spin states which are fully antisymmetryic among the four quarks. Through the coupling scheme and the systematic account for the irreducible $SU(6)_{CS}$ representation of the pentaquark, we can see that in general there are two independent methods by which we can construct the fully antisymmetric flavor $\otimes$ color $\otimes$ spin state among particles 1-4. This construction will be helpful in making a suitable choice for the full wave function involving a flavor and a spatial function.

Before analyzing this in detail, we introduce two formulas for the four quarks, similar to Eq.~(\ref{formula-2}), and make a use of these in calculating the color and spin interaction between a pair of four quarks.
The relevant formula depends upon the flavor state, given as follows:

\begin{align}
 - {\sum}_{i<j}^{4}\lambda_i^c\lambda_j^c{\vec{\sigma}}_i\cdot{\vec{\sigma}}_j=
 &4C^F_4+2C^{C}_4+\frac{4}{3}(\vec{S}\cdot \vec{S})_4-24I,
\label{formula-4}  
\end{align}
where $C^F_4$ is the quadratic Casimir operator of $SU(3)_F$ for $q^4$. There is another formula depending upon the $SU(6)_{CS}$ representation for particles 1-4, given as follows:

\begin{align}
 - {\sum}_{i<j}^{4}\lambda_i^c\lambda_j^c{\vec{\sigma}}_i\cdot{\vec{\sigma}}_j=
 &-4C^{CS}_4+2C^{C}_4+\frac{4}{3}(\vec{S}\cdot \vec{S})_4+32I,
\label{formula-3}  
\end{align}
where $C^{CS}_4$ is a quadratic Casimir operator of $SU(6)_{CS}$ for particles 1-4. The most important aspect of the flavor $\otimes$ color $\otimes$ spin state of the pentaquark under consideration is that it can be well understood in a fresh way convenient for using the eigenstate of any Casimir operator consisting of Eq.~(\ref{formula-4}) and  Eq.~(\ref{formula-3}), or Eq.~(\ref{formula-2}).
In this view, we emphasize that the color $\otimes$ spin state of the pentaquark can also be considered as the eigenstate of the Casimir operator of the $SU(6)_{CS}$ representation of the pentaquark. In this subsection, our main purpose is to investigate a correspondence between the color $\otimes$ spin state in the coupling scheme and the irreducible $SU(6)_{CS}$ representation of the pentaquark, by making a systematic analysis of Eq.~(\ref{cs-decomposition}) and Table~\ref{su6-1}.

Now, let us return to the case of flavor $\mathbf{{15}^\prime}$ for $S=1/2$. As one approach, we can obtain two orthonormal flavor $\otimes$ color $\otimes$ spin states by multiplying the color $\otimes$ spin states coming from the coupling scheme in Eq.(\ref{cs-1/2-0-211}) and Eq.(\ref{cs-1/2-1-211}) by its conjugate flavor $\mathbf{{15}^\prime}$ states, respectively, in order to satisfy the fully antisymmetry, as shown below:

\begin{align}
&\vert \psi_1 \rangle=
\nonumber \\
&\frac{1}{\sqrt{3}}
\bigg(
\begin{tabular}{|c|c|c|}
\hline
1 & 2  & 3    \\
\cline{1-3}
\multicolumn{1}{|c|}{4}  \\
\cline{1-1}
 \end{tabular}_{F}
 \otimes 
 \begin{tabular}{|c|c|}
\hline
1 & 4    \\
\cline{1-2}
\multicolumn{1}{|c|}{2}  \\
\cline{1-1}
\multicolumn{1}{|c|}{3}  \\
\cline{1-1}
 \end{tabular}_{CS^{1/2};S^0}
 -
 \begin{tabular}{|c|c|c|}
\hline
1 & 2  & 4    \\
\cline{1-3}
\multicolumn{1}{|c|}{3}  \\
\cline{1-1}
 \end{tabular}_{F}
 \otimes 
 \begin{tabular}{|c|c|}
\hline
1 & 3    \\
\cline{1-2}
\multicolumn{1}{|c|}{2}  \\
\cline{1-1}
\multicolumn{1}{|c|}{4}  \\
\cline{1-1}
 \end{tabular}_{CS^{1/2};S^0}
 \nonumber \\
& +\begin{tabular}{|c|c|c|}
\hline
1 & 3  & 4    \\
\cline{1-3}
\multicolumn{1}{|c|}{2}  \\
\cline{1-1}
 \end{tabular}_{F}
 \otimes 
 \begin{tabular}{|c|c|}
\hline
1 & 2   \\
\cline{1-2}
\multicolumn{1}{|c|}{3}  \\
\cline{1-1}
\multicolumn{1}{|c|}{4}  \\
\cline{1-1}
 \end{tabular}_{CS^{1/2};S^0}
 \bigg),
 \nonumber
\end{align}
\begin{align}
  &\vert \psi_2 \rangle=
  \nonumber \\
&\frac{1}{\sqrt{3}}
\bigg(
\begin{tabular}{|c|c|c|}
\hline
1 & 2  & 3    \\
\cline{1-3}
\multicolumn{1}{|c|}{4}  \\
\cline{1-1}
 \end{tabular}_{F}
 \otimes 
 \begin{tabular}{|c|c|}
\hline
1 & 4    \\
\cline{1-2}
\multicolumn{1}{|c|}{2}  \\
\cline{1-1}
\multicolumn{1}{|c|}{3}  \\
\cline{1-1}
 \end{tabular}_{CS^{1/2};S^1}
 -
 \begin{tabular}{|c|c|c|}
\hline
1 & 2  & 4    \\
\cline{1-3}
\multicolumn{1}{|c|}{3}  \\
\cline{1-1}
 \end{tabular}_{F}
 \otimes 
 \begin{tabular}{|c|c|}
\hline
1 & 3    \\
\cline{1-2}
\multicolumn{1}{|c|}{2}  \\
\cline{1-1}
\multicolumn{1}{|c|}{4}  \\
\cline{1-1}
 \end{tabular}_{CS^{1/2};S^1}
 \nonumber \\
  &+\begin{tabular}{|c|c|c|}
\hline
1 & 3  & 4    \\
\cline{1-3}
\multicolumn{1}{|c|}{2}  \\
\cline{1-1}
 \end{tabular}_{F}
 \otimes 
 \begin{tabular}{|c|c|}
\hline
1 & 2   \\
\cline{1-2}
\multicolumn{1}{|c|}{3}  \\
\cline{1-1}
\multicolumn{1}{|c|}{4}  \\
\cline{1-1}
 \end{tabular}_{CS^{1/2};S^1}
 \bigg).
\label{cs-1/2-211-1}
\end{align}

Here, as mentioned above, the spin state of  $\vert \psi_1 \rangle$ ($\vert \psi_2 \rangle$) in Eq.~(\ref{cs-1/2-211-1}) for particles 1-4 is 0 (1). Since the two formulas in Eq.~(\ref{formula-4}) and Eq.~(\ref{formula-3}) are made up of only Casimir operators except for the identity operator, it is obvious that the states in Eq.~(\ref{cs-1/2-211-1}) are the eigenstates of both Eq.~(\ref{formula-4}) and Eq.~(\ref{formula-3}). Therefore, the matrix element of Eq.~(\ref{formula-4}) in terms of $\vert \psi_1 \rangle$ and $\vert \psi_2 \rangle$ in Eq.~(\ref{cs-1/2-211-1}) become diagonalized in a 2 by 2 matrix form as follows:

\begin{align}
 \langle -{\sum}_{i<j}^{4}\lambda_i^c\lambda_j^c{\vec{\sigma}}_i\cdot{\vec{\sigma}}_j \rangle=
  \left(\begin{array}{cc}
0 &  0     \\
0  &    \frac{8}{3}     \\
\end{array} \right).     
 \label{cs-1/2-211-matrix-1}  
\end{align}
It is easy to calculate this expectation value from the formula in  Eq.~(\ref{formula-4}) where the eigenvalue of the flavor multiplet $\mathbf{{15}^\prime}$ for $C^F_4$ is 16/3. For the color part, the eigenvalue of $C^{C}_4$ for the four quarks is given as 4/3 because the color state is in a triplet of $SU(3)_C$. Moreover, we can indeed obtain the same result through another formula of Eq.~(\ref{formula-3}), because the Young-Yamanouchi bases in Eq.~(\ref{cs-1/2-211-1}) obtained from the coupling scheme belong to the multiplets of $SU(6)_{CS}$ representation, $[21^2]$ for particles 1-4. This indicates that these states become the eigenstates of the quadratic Casimir operator, $C^{CS}_4$, with an eigenvalue of 26/3, as shown in Table~\ref{su6-2}. These are given as follows:

\begin{align}
 &- {\sum}_{i<j}^{4}\lambda_i^c\lambda_j^c{\vec{\sigma}}_i\cdot{\vec{\sigma}}_j
\vert \psi_1  \rangle =
\nonumber \\
& (-4\times26/3+2\times4/3+4/3\times0+32)\vert \psi_1  \rangle=0\vert \psi_1  \rangle,
 \nonumber \\
 &- {\sum}_{i<j}^{4}\lambda_i^c\lambda_j^c{\vec{\sigma}}_i\cdot{\vec{\sigma}}_j
\vert \psi_2  \rangle =
\nonumber \\
 &(-4\times26/3+2\times4/3+4/3\times1\times2+32)\vert \psi_2  \rangle=8/3\vert \psi_2  \rangle.
\label{cs-1/2-211-eigenvalue-1}  
\end{align}

We can infer another way of constructing the two orthonormal flavor $\otimes$ color $\otimes$ spin states for $S=1/2$ from the $SU(6)_{CS}$ representation of the pentaquark in Eq.~(\ref{cs-decomposition}). This can be accomplished in the same way as those from the coupling scheme in Eq.~(\ref{cs-1/2-211-1}). That is to say, first of all, both the $[21]$ and $[32^21^2]$ of the $SU(6)_{CS}$ representation in Table~\ref{su6-1} contain states which are in the color singlet, and at the same time $S=1/2$ states, that is, $[\mathbf{1}_{C}, \mathbf{2}_{S}]$, as can see in Table~\ref{colorsinglet-spin} and   Table~\ref{su6-1}. Second, both the $[21]$ and $[32^21^2]$ in the  $SU(6)_{CS}$ representation have a symmetry property for $q^4$ corresponding to the Young diagram $[21^2]$, which are conjugate to the flavor $\mathbf{{15}^\prime}$ multiplet, as can be seen in Eq.~(\ref{cs-decomposition}). Then, the two orthonormal flavor $\otimes$ color $\otimes$ spin states for $S=1/2$ based on the $SU(6)_{CS}$ representation of the pentaquark are given as follows:

\begin{align}
&\vert \Psi_1 \rangle =
\nonumber \\
&\frac{1}{\sqrt{3}}
\bigg(
\begin{tabular}{|c|c|c|}
\hline
1 & 2  & 3    \\
\cline{1-3}
\multicolumn{1}{|c|}{4}  \\
\cline{1-1}
 \end{tabular}_{F}
 \otimes 
 \big{\vert}
 \begin{tabular}{|c|c|}
\hline
1 & 4    \\
\cline{1-2}
\multicolumn{1}{|c|}{2}  \\
\cline{1-1}
\multicolumn{1}{|c|}{3}  \\
\cline{1-1}
 \end{tabular},
 [21]
 \big{\rangle}_{CS^{1/2}}
 -
 \nonumber \\
&\begin{tabular}{|c|c|c|}
\hline
1 & 2  & 4    \\
\cline{1-3}
\multicolumn{1}{|c|}{3}  \\
\cline{1-1}
 \end{tabular}_{F}
 \otimes
\big{\vert}  
 \begin{tabular}{|c|c|}
\hline
1 & 3    \\
\cline{1-2}
\multicolumn{1}{|c|}{2}  \\
\cline{1-1}
\multicolumn{1}{|c|}{4}  \\
\cline{1-1}
 \end{tabular},
 [21]
  \big{\rangle}_{CS^{1/2}}
 +
 \begin{tabular}{|c|c|c|}
\hline
1 & 3  & 4    \\
\cline{1-3}
\multicolumn{1}{|c|}{2}  \\
\cline{1-1}
 \end{tabular}_{F}
 \otimes 
 \big{\vert} 
 \begin{tabular}{|c|c|}
\hline
1 & 2   \\
\cline{1-2}
\multicolumn{1}{|c|}{3}  \\
\cline{1-1}
\multicolumn{1}{|c|}{4}  \\
\cline{1-1}
 \end{tabular},
 [21]
  \big{\rangle}_{CS^{1/2}}
 \bigg),
 \nonumber
\\
 &\vert \Psi_2 \rangle =
 \nonumber \\
&\frac{1}{\sqrt{3}}
\bigg(
\begin{tabular}{|c|c|c|}
\hline
1 & 2  & 3    \\
\cline{1-3}
\multicolumn{1}{|c|}{4}  \\
\cline{1-1}
 \end{tabular}_{F}
 \otimes 
 \big{\vert}
 \begin{tabular}{|c|c|}
\hline
1 & 4    \\
\cline{1-2}
\multicolumn{1}{|c|}{2}  \\
\cline{1-1}
\multicolumn{1}{|c|}{3}  \\
\cline{1-1}
 \end{tabular},
 [32^21^2]
 \big{\rangle}_{CS^{1/2}}
 -
 \nonumber \\
 &\begin{tabular}{|c|c|c|}
\hline
1 & 2  & 4    \\
\cline{1-3}
\multicolumn{1}{|c|}{3}  \\
\cline{1-1}
 \end{tabular}_{F}
 \otimes
\big{\vert}  
 \begin{tabular}{|c|c|}
\hline
1 & 3    \\
\cline{1-2}
\multicolumn{1}{|c|}{2}  \\
\cline{1-1}
\multicolumn{1}{|c|}{4}  \\
\cline{1-1}
 \end{tabular},
 [32^21^2]
  \big{\rangle}_{CS^{1/2}}
  +
  \nonumber \\
&\begin{tabular}{|c|c|c|}
\hline
1 & 3  & 4    \\
\cline{1-3}
\multicolumn{1}{|c|}{2}  \\
\cline{1-1}
 \end{tabular}_{F}
 \otimes 
 \big{\vert} 
 \begin{tabular}{|c|c|}
\hline
1 & 2   \\
\cline{1-2}
\multicolumn{1}{|c|}{3}  \\
\cline{1-1}
\multicolumn{1}{|c|}{4}  \\
\cline{1-1}
 \end{tabular},
 [32^21^2]
  \big{\rangle}_{CS^{1/2}}
\bigg).
\label{cs-1/2-211-2}
\end{align}
Here, we introduce new notation for the part of the color and spin state. In the right hand side of Eq.~(\ref{cs-1/2-211-2}), the ket states in the first and second equations express the states belonging to the [21] and $[32^21^2]$ multiplets of the $SU(6)_{CS}$ representation of the pentaquark, respectively. The Young-Yamanouchi bases in the ket states represent the respective symmetry properties for particles 1-4. Also, the subscript $CS^{1/2}$ outside of the ket states means that the states are in the color singlet and $S=1/2$ state for the pentaquark.

Furthermore, it should be noted that each part of the color and spin states in Eq.~(\ref{cs-1/2-211-2}) are the eigenstates of the quadratic Casimir operator, $C^{CS}_5$ of the $SU(6)_{CS}$ representation for the pentaquark. Therefore, we can obtain the eigenvalue equation for the quadratic Casimir operator $C^{CS}_5$, which allows us to construct the $SU(6)_{CS}$ representation of the pentaquark using the states obtained from the coupling scheme. The eigenvalue equation from Table~\ref{su6-1} is as follows:
\begin{align}
&C^{CS}_5
 \big{\vert}
 \begin{tabular}{|c|c|}
\hline
1 & 2    \\
\cline{1-2}
\multicolumn{1}{|c|}{3}  \\
\cline{1-1}
\multicolumn{1}{|c|}{4}  \\
\cline{1-1}
 \end{tabular},
 [21]
 \big{\rangle}_{CS^{1/2}}
 =33/4
  \big{\vert}
 \begin{tabular}{|c|c|}
\hline
1 & 2    \\
\cline{1-2}
\multicolumn{1}{|c|}{3}  \\
\cline{1-1}
\multicolumn{1}{|c|}{4}  \\
\cline{1-1}
 \end{tabular},
 [21]
 \big{\rangle}_{CS^{1/2}}
,
\nonumber \\
&C^{CS}_5
 \big{\vert}
 \begin{tabular}{|c|c|}
\hline
1 & 3    \\
\cline{1-2}
\multicolumn{1}{|c|}{2}  \\
\cline{1-1}
\multicolumn{1}{|c|}{4}  \\
\cline{1-1}
 \end{tabular},
 [21]
 \big{\rangle}_{CS^{1/2}}
 =33/4
  \big{\vert}
 \begin{tabular}{|c|c|}
\hline
1 & 3   \\
\cline{1-2}
\multicolumn{1}{|c|}{2}  \\
\cline{1-1}
\multicolumn{1}{|c|}{4}  \\
\cline{1-1}
 \end{tabular},
 [21]
 \big{\rangle}_{CS^{1/2}}
 ,
 \nonumber \\
 &C^{CS}_5
 \big{\vert}
 \begin{tabular}{|c|c|}
\hline
1 & 4    \\
\cline{1-2}
\multicolumn{1}{|c|}{2}  \\
\cline{1-1}
\multicolumn{1}{|c|}{3}  \\
\cline{1-1}
 \end{tabular},
 [21]
 \big{\rangle}_{CS^{1/2}}
 =33/4
  \big{\vert}
 \begin{tabular}{|c|c|}
\hline
1 & 4    \\
\cline{1-2}
\multicolumn{1}{|c|}{2}  \\
\cline{1-1}
\multicolumn{1}{|c|}{3}  \\
\cline{1-1}
 \end{tabular},
 [21]
 \big{\rangle}_{CS^{1/2}}
 .
\label{cs-1/2-211-eigenvalue-2} 
\end{align}
\begin{align}
&C^{CS}_5
 \big{\vert}
 \begin{tabular}{|c|c|}
\hline
1 & 2    \\
\cline{1-2}
\multicolumn{1}{|c|}{3}  \\
\cline{1-1}
\multicolumn{1}{|c|}{4}  \\
\cline{1-1}
 \end{tabular},
 [32^21^2]
 \big{\rangle}_{CS^{1/2}}
 =49/4
  \big{\vert}
 \begin{tabular}{|c|c|}
\hline
1 & 2    \\
\cline{1-2}
\multicolumn{1}{|c|}{3}  \\
\cline{1-1}
\multicolumn{1}{|c|}{4}  \\
\cline{1-1}
 \end{tabular},
 [32^21^2]
 \big{\rangle}_{CS^{1/2}}
,
\nonumber \\
&C^{CS}_5
 \big{\vert}
 \begin{tabular}{|c|c|}
\hline
1 & 3    \\
\cline{1-2}
\multicolumn{1}{|c|}{2}  \\
\cline{1-1}
\multicolumn{1}{|c|}{4}  \\
\cline{1-1}
 \end{tabular},
 [32^21^2]
 \big{\rangle}_{CS^{1/2}}
 =49/4
  \big{\vert}
 \begin{tabular}{|c|c|}
\hline
1 & 3   \\
\cline{1-2}
\multicolumn{1}{|c|}{2}  \\
\cline{1-1}
\multicolumn{1}{|c|}{4}  \\
\cline{1-1}
 \end{tabular},
 [32^21^2]
 \big{\rangle}_{CS^{1/2}}
 ,
 \nonumber \\
 &C^{CS}_5
 \big{\vert}
 \begin{tabular}{|c|c|}
\hline
1 & 4    \\
\cline{1-2}
\multicolumn{1}{|c|}{2}  \\
\cline{1-1}
\multicolumn{1}{|c|}{3}  \\
\cline{1-1}
 \end{tabular},
 [32^21^2]
 \big{\rangle}_{CS^{1/2}}
 =49/4
  \big{\vert}
 \begin{tabular}{|c|c|}
\hline
1 & 4    \\
\cline{1-2}
\multicolumn{1}{|c|}{2}  \\
\cline{1-1}
\multicolumn{1}{|c|}{3}  \\
\cline{1-1}
 \end{tabular},
 [32^21^2]
 \big{\rangle}_{CS^{1/2}}
 .
\label{cs-1/2-211-eigenvalue-3} 
\end{align}

Since the color and spin states in Eq.~(\ref{cs-1/2-211-2}) become the eigenstates of the $C^{CS}_5$, we can infer from Eq.~(\ref{cs-decomposition}) that the linear sum of the color $\otimes$ spin parts between $\vert \psi_1 \rangle$ and $\vert \psi_2 \rangle$ in Eq.~(\ref{cs-1/2-211-1}) must belong to either the [21] of $SU(6)_{CS}$ representation or the $[32^21^2]$ of $SU(6)_{CS}$ representation in Eq.~(\ref{cs-1/2-211-2}).
The coefficients of this linear combination can be determined by requiring that the linear combination becomes an eigenstate of the Casimir operator $C^{CS}_5$ of the $SU(6)_{CS}$ representation, as well as satisfying the normalization condition for the linear sum.
Therefore, action of $C^{CS}_5$ upon the linear sum that gives the eigenvalues of either 33/4 or 49/4 in Table~\ref{su6-1} leads to an equation involving two variables. The operator $C^{CS}_5$ is
given as follows.

\begin{align}
C^{CS}_5=&-\frac{1}{4}{\sum}_{i=1}^{4}\lambda_i^c\lambda_5^c{\vec{\sigma}}_i\cdot{\vec{\sigma}}_5
+C^{CS}_4+\frac{1}{2}C^{C}_5-\frac{1}{2}C^{C}_4 
+
\nonumber \\
&\frac{1}{3}(\vec{S}\cdot \vec{S})_5-
\frac{1}{3}(\vec{S}\cdot \vec{S})_4+2I.
 \label{Casimir-C6}
 \end{align}

After some algebraic calculations of  a coupled equation with two unknown variables, 
we obtain the following relations:
 \begin{align}
&\big{\vert}
\begin{tabular}{|c|c|}
\hline
1 & 2      \\
\cline{1-2}
\multicolumn{1}{|c|}{3}  \\
\cline{1-1}
\multicolumn{1}{|c|}{4}  \\
\cline{1-1}
 \end{tabular},
 [21]
 \big{\rangle}_{CS^{1/2}}
 =
 \frac{1}{\sqrt{2}}
 \begin{tabular}{|c|c|}
\hline
1 & 2      \\
\cline{1-2}
\multicolumn{1}{|c|}{3}  \\
\cline{1-1}
\multicolumn{1}{|c|}{4}  \\
\cline{1-1}
 \end{tabular}_{CS^{1/2};S^0}
 +\frac{1}{\sqrt{2}}
 \begin{tabular}{|c|c|}
\hline
1 & 2      \\
\cline{1-2}
\multicolumn{1}{|c|}{3}  \\
\cline{1-1}
\multicolumn{1}{|c|}{4}  \\
\cline{1-1}
 \end{tabular}_{CS^{1/2};S^1},
\nonumber
  \end{align}
 \begin{align}
\big{\vert}
\begin{tabular}{|c|c|}
\hline
1 & 2      \\
\cline{1-2}
\multicolumn{1}{|c|}{3}  \\
\cline{1-1}
\multicolumn{1}{|c|}{4}  \\
\cline{1-1}
 \end{tabular},
 [32^21^2]
 \big{\rangle}_{CS^{1/2}}
 =
 -\frac{1}{\sqrt{2}}
 \begin{tabular}{|c|c|}
\hline
1 & 2      \\
\cline{1-2}
\multicolumn{1}{|c|}{3}  \\
\cline{1-1}
\multicolumn{1}{|c|}{4}  \\
\cline{1-1}
 \end{tabular}_{CS^{1/2};S^0}
 +\frac{1}{\sqrt{2}}
 \begin{tabular}{|c|c|}
\hline
1 & 2      \\
\cline{1-2}
\multicolumn{1}{|c|}{3}  \\
\cline{1-1}
\multicolumn{1}{|c|}{4}  \\
\cline{1-1}
 \end{tabular}_{CS^{1/2};S^1}.
  \label{cs-1/2-211-coefficient-1} 
  \end{align}
The same procedure is equally applied to the rest of the Young-Yamanouchi bases of [21] ($[32^21^2]$). We can see obviously that the same relation holds for the others:
 \begin{align}
\big{\vert}
\begin{tabular}{|c|c|}
\hline
1 & 3      \\
\cline{1-2}
\multicolumn{1}{|c|}{2}  \\
\cline{1-1}
\multicolumn{1}{|c|}{4}  \\
\cline{1-1}
 \end{tabular},
 [21]
 \big{\rangle}_{CS^{1/2}}
 =
 \frac{1}{\sqrt{2}}
 \begin{tabular}{|c|c|}
\hline
1 & 3    \\
\cline{1-2}
\multicolumn{1}{|c|}{2}  \\
\cline{1-1}
\multicolumn{1}{|c|}{4}  \\
\cline{1-1}
 \end{tabular}_{CS^{1/2};S^0}
 +\frac{1}{\sqrt{2}}
 \begin{tabular}{|c|c|}
\hline
1 & 3      \\
\cline{1-2}
\multicolumn{1}{|c|}{2}  \\
\cline{1-1}
\multicolumn{1}{|c|}{4}  \\
\cline{1-1}
 \end{tabular}_{CS^{1/2};S^1},
\nonumber 
  \end{align}
 \begin{align}
\big{\vert}
\begin{tabular}{|c|c|}
\hline
1 & 3     \\
\cline{1-2}
\multicolumn{1}{|c|}{2}  \\
\cline{1-1}
\multicolumn{1}{|c|}{4}  \\
\cline{1-1}
 \end{tabular},
 [32^21^2]
 \big{\rangle}_{CS^{1/2}}
 =
 -\frac{1}{\sqrt{2}}
 \begin{tabular}{|c|c|}
\hline
1 & 3     \\
\cline{1-2}
\multicolumn{1}{|c|}{2}  \\
\cline{1-1}
\multicolumn{1}{|c|}{4}  \\
\cline{1-1}
 \end{tabular}_{CS^{1/2};S^0}
 +\frac{1}{\sqrt{2}}
 \begin{tabular}{|c|c|}
\hline
1 & 3     \\
\cline{1-2}
\multicolumn{1}{|c|}{2}  \\
\cline{1-1}
\multicolumn{1}{|c|}{4}  \\
\cline{1-1}
 \end{tabular}_{CS^{1/2};S^1}.
  \label{cs-1/2-211-coefficient-2} 
  \end{align}

 \begin{align}
\big{\vert}
\begin{tabular}{|c|c|}
\hline
1 & 4      \\
\cline{1-2}
\multicolumn{1}{|c|}{2}  \\
\cline{1-1}
\multicolumn{1}{|c|}{3}  \\
\cline{1-1}
 \end{tabular},
 [21]
 \big{\rangle}_{CS^{1/2}}
 =
 \frac{1}{\sqrt{2}}
 \begin{tabular}{|c|c|}
\hline
1 & 4    \\
\cline{1-2}
\multicolumn{1}{|c|}{2}  \\
\cline{1-1}
\multicolumn{1}{|c|}{3}  \\
\cline{1-1}
 \end{tabular}_{CS^{1/2};S^0}
 +\frac{1}{\sqrt{2}}
 \begin{tabular}{|c|c|}
\hline
1 & 4      \\
\cline{1-2}
\multicolumn{1}{|c|}{2}  \\
\cline{1-1}
\multicolumn{1}{|c|}{3}  \\
\cline{1-1}
 \end{tabular}_{CS^{1/2};S^1},
\nonumber 
  \end{align}
 \begin{align}
\big{\vert}
\begin{tabular}{|c|c|}
\hline
1 & 4    \\
\cline{1-2}
\multicolumn{1}{|c|}{2}  \\
\cline{1-1}
\multicolumn{1}{|c|}{3}  \\
\cline{1-1}
 \end{tabular},
 [32^21^2]
 \big{\rangle}_{CS^{1/2}}
 =
 -\frac{1}{\sqrt{2}}
 \begin{tabular}{|c|c|}
\hline
1 & 4    \\
\cline{1-2}
\multicolumn{1}{|c|}{2}  \\
\cline{1-1}
\multicolumn{1}{|c|}{3}  \\
\cline{1-1}
 \end{tabular}_{CS^{1/2};S^0}
 +\frac{1}{\sqrt{2}}
 \begin{tabular}{|c|c|}
\hline
1 & 4    \\
\cline{1-2}
\multicolumn{1}{|c|}{2}  \\
\cline{1-1}
\multicolumn{1}{|c|}{3}  \\
\cline{1-1}
 \end{tabular}_{CS^{1/2};S^1}.
  \label{cs-1/2-211-coefficient-3} 
  \end{align}

On the other hand, there is a general rule of the permutation group that the second Young-Yamanouchi basis of the [21] multiplet in Eq.~(\ref{cs-1/2-211-eigenvalue-2}) (the $[32^21^2]$ in Eq.~(\ref{cs-1/2-211-eigenvalue-3})) can be obtained by acting the permutation operator (23) upon the first Young-Yamanouchi basis of the [21] in Eq.~(\ref{cs-1/2-211-eigenvalue-2}) (the $[32^21^2]$ in Eq.~(\ref{cs-1/2-211-eigenvalue-3})). This process is also applied to the last basis of the [21] multiplet (and the [32$^21^2$] multiplet) by acting the permutation (34) upon the second basis.

From Eqs.~(\ref{cs-1/2-211-coefficient-1}-\ref{cs-1/2-211-coefficient-3}),
we find the following relations between the states in Eq.~(\ref{cs-1/2-211-1}) and the states in Eq.~(\ref{cs-1/2-211-2}):
\begin{align}
\vert \Psi_1 \rangle =\frac{1}{\sqrt{2}}\vert \psi_1 \rangle +
\frac{1}{\sqrt{2}}\vert \psi_2 \rangle,
\quad
 \vert \Psi_2 \rangle =-\frac{1}{\sqrt{2}}\vert \psi_1 \rangle +
\frac{1}{\sqrt{2}}\vert \psi_2 \rangle.
 \label{cs-1/2-211-coefficient-4} 
 \end{align}

We can finally calculate the matrix element of Eq.~(\ref{formula-2})  in terms of $\vert \Psi_1 \rangle$ and $\vert \Psi_2 \rangle$.  Using Eq.~(\ref{cs-1/2-211-coefficient-4}), we find the following 
2 by 2 matrix elements:

\begin{align}
&\langle  \Psi_1 \vert -{\sum}_{i<j}^{5}
\lambda_i^c\lambda_j^c{\vec{\sigma}}_i\cdot{\vec{\sigma}}_j  \vert  \Psi_1  \rangle =
4\times33/4-8\times26/3-
\nonumber \\ 
&2\times0+4\times4/3-4/3\times1/2\times3/2+
8/3\times(1/2\times0+
\nonumber \\
&1/2\times1\times2)+24=-16/3, \nonumber \\
&\langle  \Psi_2 \vert -{\sum}_{i<j}^{5}
\lambda_i^c\lambda_j^c{\vec{\sigma}}_i\cdot{\vec{\sigma}}_j  \vert  \Psi_2  \rangle =
4\times49/4-8\times26/3-
\nonumber \\
&2\times0+4\times4/3-4/3\times1/2\times3/2+
8/3\times(1/2\times0+
\nonumber \\
&1/2\times1\times2)+24=32/3, \nonumber \\
&\langle  \Psi_1 \vert -{\sum}_{i<j}^{5}
\lambda_i^c\lambda_j^c{\vec{\sigma}}_i\cdot{\vec{\sigma}}_j  \vert  \Psi_2  \rangle =
8/3\times(-1/2\times0+
\nonumber \\
 &1/2\times1\times2)=8/3.
\label{cs-1/2-211-eigenvalue-4}
\end{align}
By diagonalizing the matrix in Eq.~(\ref{cs-1/2-211-eigenvalue-4}), we find that the eigenvalue of Eq.~(\ref{formula-2}) is either $-8/3(-1+\sqrt{10})$ or $8/3(1+\sqrt{10})$.

It is instructive to examine the expectation value of  $\lambda_i^c\lambda_j^c{\vec{\sigma}}_i\cdot{\vec{\sigma}}_j$ from the symmetry property.
Furthermore, using Eq.(\ref{formula-3}) and Eq.(\ref{cs-1/2-211-coefficient-4}), we can easily express the expectation value of Eq.(\ref{formula-3}) for particles 1-4 in terms of $\vert \Psi_1 \rangle$ and $\vert \Psi_2 \rangle$ in Eq.(\ref{cs-1/2-211-2}) as follows:
\begin{align}
 \langle -{\sum}_{i<j}^{4}\lambda_i^c\lambda_j^c{\vec{\sigma}}_i\cdot{\vec{\sigma}}_j \rangle=
  \left(\begin{array}{cc}
\frac{4}{3} &  \frac{4}{3}     \\
 \frac{4}{3} &    \frac{4}{3}     \\
\end{array} \right).     
 \label{cs-1/2-211-matrix-2}  
\end{align}
Then, diagonalizing  Eq.~(\ref{cs-1/2-211-matrix-2}) gives either 0, or 8/3 as the same eigenvalues, which are obtained from Eq.~(\ref{cs-1/2-211-matrix-1}). In addition, due to the fact that the antisymmetry property of Eq.~(\ref{cs-1/2-211-2}) and the following relation, $(i k) \lambda^c_i \lambda^c_j {\vec{\sigma}}_i\cdot{\vec{\sigma}}_j (i k) = \lambda^c_k \lambda^c_j {\vec{\sigma}}_k\cdot{\vec{\sigma}}_j$, the expectation value of $\lambda_i^c\lambda_j^c{\vec{\sigma}}_i\cdot{\vec{\sigma}}_j$ ($i<j$ =1, 2, 3, 4) is all the same. Therefore, we find the following:

\begin{align}
 \langle -\lambda_i^c\lambda_j^c{\vec{\sigma}}_i\cdot{\vec{\sigma}}_j \rangle=
  \left(\begin{array}{cc}
\frac{2}{9} &  \frac{2}{9}     \\
 \frac{2}{9} &    \frac{2}{9}     \\
\end{array} \right)    
(i<j =1, 2, 3, 4).
 \label{cs-1/2-211-matrix-3}  
\end{align}
In a similar manner, we can calculate  the expectation value of  $\lambda_i^c\lambda_5^c{\vec{\sigma}}_i\cdot{\vec{\sigma}}_5$ ($i$ =1, 2, 3, 4) in terms of $\vert \Psi_1 \rangle$ and $\vert \Psi_2 \rangle$ in Eq.~(\ref{cs-1/2-211-2}), using Eq.~(\ref{cs-1/2-211-eigenvalue-4}),  Eq.~(\ref{cs-1/2-211-matrix-2}), and the fully antisymmetric property for the four quarks:

\begin{align}
 \langle -\lambda_i^c\lambda_5^c{\vec{\sigma}}_i\cdot{\vec{\sigma}}_5 \rangle=
  \left(\begin{array}{cc}
-\frac{5}{3} &  \frac{1}{3}     \\
 \frac{1}{3} &    \frac{7}{3}     \\
\end{array} \right)    
(i =1, 2, 3, 4).
 \label{cs-1/2-211-matrix-4}  
\end{align}

For the flavor $\mathbf{3}$ multiplet represented by Young diagram $[21^2]$ for $q^4$, there are  two orthonormal flavor $\otimes$ color $\otimes$ spin states, satisfying fully antisymmetry property. These states can be obtained by multiplying the color $\otimes$ spin states coming from the coupling scheme in Eq.~(\ref{cs-1/2-0-31}) and Eq.~(\ref{cs-1/2-1-31}) by its conjugate flavor $\mathbf{3}$ states, respectively:

\begin{align}
&\vert \psi_1 \rangle=
\frac{1}{\sqrt{3}}
\bigg(
\begin{tabular}{|c|c|}
\hline
1 & 2    \\
\cline{1-2}
\multicolumn{1}{|c|}{3}  \\
\cline{1-1}
\multicolumn{1}{|c|}{4}  \\
\cline{1-1}
 \end{tabular}_{F}
\otimes 
\begin{tabular}{|c|c|c|}
\hline
1 & 3  & 4   \\
\cline{1-3}
\multicolumn{1}{|c|}{2}  \\
\cline{1-1}
 \end{tabular}_{CS^{1/2};S^0}
 -
 \nonumber \\
 &\begin{tabular}{|c|c|}
\hline
1 & 3    \\
\cline{1-2}
\multicolumn{1}{|c|}{2}  \\
\cline{1-1}
\multicolumn{1}{|c|}{4}  \\
\cline{1-1}
 \end{tabular}_{F}
\otimes 
\begin{tabular}{|c|c|c|}
\hline
1 & 2  & 4   \\
\cline{1-3}
\multicolumn{1}{|c|}{3}  \\
\cline{1-1}
 \end{tabular}_{CS^{1/2};S^0}+
 \begin{tabular}{|c|c|}
\hline
1 & 4   \\
\cline{1-2}
\multicolumn{1}{|c|}{2}  \\
\cline{1-1}
\multicolumn{1}{|c|}{3}  \\
\cline{1-1}
 \end{tabular}_{F}
\otimes 
\begin{tabular}{|c|c|c|}
\hline
1 & 2  & 3  \\
\cline{1-3}
\multicolumn{1}{|c|}{4}  \\
\cline{1-1}
 \end{tabular}_{CS^{1/2};S^0}
 \bigg),
\nonumber 
 \end{align}
\begin{align}
&\vert \psi_2 \rangle=
\frac{1}{\sqrt{3}}
\bigg(
\begin{tabular}{|c|c|}
\hline
1 & 2    \\
\cline{1-2}
\multicolumn{1}{|c|}{3}  \\
\cline{1-1}
\multicolumn{1}{|c|}{4}  \\
\cline{1-1}
 \end{tabular}_{F}
\otimes 
\begin{tabular}{|c|c|c|}
\hline
1 & 3  & 4   \\
\cline{1-3}
\multicolumn{1}{|c|}{2}  \\
\cline{1-1}
 \end{tabular}_{CS^{1/2};S^1}
 -
 \nonumber \\
 &\begin{tabular}{|c|c|}
\hline
1 & 3    \\
\cline{1-2}
\multicolumn{1}{|c|}{2}  \\
\cline{1-1}
\multicolumn{1}{|c|}{4}  \\
\cline{1-1}
 \end{tabular}_{F}
\otimes 
\begin{tabular}{|c|c|c|}
\hline
1 & 2  & 4   \\
\cline{1-3}
\multicolumn{1}{|c|}{3}  \\
\cline{1-1}
 \end{tabular}_{CS^{1/2};S^1}+
 \begin{tabular}{|c|c|}
\hline
1 & 4   \\
\cline{1-2}
\multicolumn{1}{|c|}{2}  \\
\cline{1-1}
\multicolumn{1}{|c|}{3}  \\
\cline{1-1}
 \end{tabular}_{F}
\otimes 
\begin{tabular}{|c|c|c|}
\hline
1 & 2  & 3  \\
\cline{1-3}
\multicolumn{1}{|c|}{4}  \\
\cline{1-1}
 \end{tabular}_{CS^{1/2};S^1}
\bigg).
  \label{cs-1/2-31-1}
 \end{align}

Obviously, since the states in Eq.~(\ref{cs-1/2-31-1}) are the eigenstates of Eq.~(\ref{formula-4}), the matrix element of Eq.~(\ref{formula-4}) is diagonalized in terms of those states, as follows:

\begin{align}
 \langle -{\sum}_{i<j}^{4}\lambda_i^c\lambda_j^c{\vec{\sigma}}_i\cdot{\vec{\sigma}}_j \rangle=
  \left(\begin{array}{cc}
-16 &  0     \\
0  &    -\frac{40}{3}     \\
\end{array} \right).     
 \label{cs-1/2-31-matrix-1}  
\end{align}

It is easy  to calculate this expectation value from the formula in Eq.~(\ref{formula-4}) where the eigenvalue of the flavor  $\mathbf{3}$ multiplet for  $C_4^F$ is 4/3. Moreover, the application of Eq.~(\ref{formula-3}) to the states in Eq.~(\ref{cs-1/2-31-1}) leads to the same expectation value as that of Eq.~(\ref{cs-1/2-31-matrix-1}) because the Young-Yamanouchi bases in Eq.~(\ref{cs-1/2-31-1}) obtained from the coupling scheme belong to the multiplets of $SU(6)_{CS}$ representation, $[31]$ for particles 1-4. These states have an eigenvalue of 38/3 in Table~\ref{su6-2}, as the eigenstates of the quadratic Casimir operator, $C^{CS}_4$. These are given as follows:

\begin{align}
  &-{\sum}_{i<j}^{4}\lambda_i^c\lambda_j^c{\vec{\sigma}}_i\cdot{\vec{\sigma}}_j
\vert \psi_1  \rangle =
 (-4\times38/3+2\times4/3+
 \nonumber \\
 &4/3\times0+32)\vert \psi_1  \rangle=-16\vert \psi_1  \rangle,
 \nonumber \\
 &- {\sum}_{i<j}^{4}\lambda_i^c\lambda_j^c{\vec{\sigma}}_i\cdot{\vec{\sigma}}_j
\vert \psi_2  \rangle =
 (-4\times38/3+2\times4/3+
 \nonumber \\
& 4/3\times1\times2+32)\vert \psi_2  \rangle=-40/3\vert \psi_2  \rangle.
 \label{cs-1/2-31-matrix-2}  
\end{align}

On the one hand, from the point of view of the $SU(6)_{CS}$ representation of pentaquark, both the $[21]$ and the $[421^3]$ in the $SU(6)_{CS}$ representation conjugate to the flavor $\mathbf{3}$ multiplet for the four quarks contain the state which are in both the color singlet and $S=1/2$ state, as shown in Table~\ref{colorsinglet-spin} and Table~\ref{su6-1} and Eq.~(\ref{cs-decomposition}).
Therefore, another approach based on the $SU(6)_{CS}$ representation in Eq.~(\ref{cs-decomposition}) gives the two orthonormal flavor $\otimes$ color $\otimes$ spin states for $S=1/2$ which satisfy fully antisymmetry in the same way as in the coupling scheme:

\begin{align}
\vert \Psi_1 \rangle =
&\frac{1}{\sqrt{3}}
\bigg(
\begin{tabular}{|c|c|}
\hline
1 & 2    \\
\cline{1-2}
\multicolumn{1}{|c|}{3}  \\
\cline{1-1}
\multicolumn{1}{|c|}{4}  \\
\cline{1-1}
 \end{tabular}_{F}
\otimes 
\big{\vert}
\begin{tabular}{|c|c|c|}
\hline
1 & 3  & 4   \\
\cline{1-3}
\multicolumn{1}{|c|}{2}  \\
\cline{1-1}
 \end{tabular},
 [21]
 \big{\rangle}_{CS^{1/2}}
 -
 \nonumber \\
&\begin{tabular}{|c|c|}
\hline
1 & 3   \\
\cline{1-2}
\multicolumn{1}{|c|}{2}  \\
\cline{1-1}
\multicolumn{1}{|c|}{4}  \\
\cline{1-1}
 \end{tabular}_{F}
\otimes 
\big{\vert}
\begin{tabular}{|c|c|c|}
\hline
1 & 2  & 4   \\
\cline{1-3}
\multicolumn{1}{|c|}{3}  \\
\cline{1-1}
 \end{tabular},
 [21]
 \big{\rangle}_{CS^{1/2}} 
 +
 \nonumber \\
&\begin{tabular}{|c|c|}
\hline
1 & 4  \\
\cline{1-2}
\multicolumn{1}{|c|}{2}  \\
\cline{1-1}
\multicolumn{1}{|c|}{3}  \\
\cline{1-1}
 \end{tabular}_{F}
\otimes 
\big{\vert}
\begin{tabular}{|c|c|c|}
\hline
1 & 2  & 3   \\
\cline{1-3}
\multicolumn{1}{|c|}{4}  \\
\cline{1-1}
 \end{tabular},
 [21]
 \big{\rangle}_{CS^{1/2}}
 \bigg),
 \nonumber
\end{align}
\begin{align}
\vert \Psi_2 \rangle =
&\frac{1}{\sqrt{3}}
\bigg(
\begin{tabular}{|c|c|}
\hline
1 & 2    \\
\cline{1-2}
\multicolumn{1}{|c|}{3}  \\
\cline{1-1}
\multicolumn{1}{|c|}{4}  \\
\cline{1-1}
 \end{tabular}_{F}
\otimes 
\big{\vert}
\begin{tabular}{|c|c|c|}
\hline
1 & 3  & 4   \\
\cline{1-3}
\multicolumn{1}{|c|}{2}  \\
\cline{1-1}
 \end{tabular},
 [421^3]
 \big{\rangle}_{CS^{1/2}}
 -
 \nonumber \\
&\begin{tabular}{|c|c|}
\hline
1 & 3   \\
\cline{1-2}
\multicolumn{1}{|c|}{2}  \\
\cline{1-1}
\multicolumn{1}{|c|}{4}  \\
\cline{1-1}
 \end{tabular}_{F}
\otimes 
\big{\vert}
\begin{tabular}{|c|c|c|}
\hline
1 & 2  & 4   \\
\cline{1-3}
\multicolumn{1}{|c|}{3}  \\
\cline{1-1}
 \end{tabular},
 [421^3]
 \big{\rangle}_{CS^{1/2}} 
 +
 \nonumber \\
 &\begin{tabular}{|c|c|}
\hline
1 & 4  \\
\cline{1-2}
\multicolumn{1}{|c|}{2}  \\
\cline{1-1}
\multicolumn{1}{|c|}{3}  \\
\cline{1-1}
 \end{tabular}_{F}
\otimes 
\big{\vert}
\begin{tabular}{|c|c|c|}
\hline
1 & 2  & 3   \\
\cline{1-3}
\multicolumn{1}{|c|}{4}  \\
\cline{1-1}
 \end{tabular},
 [421^3]
 \big{\rangle}_{CS^{1/2}}
 \bigg).
  \label{cs-1/2-31-2}
\end{align}
We again use the same notation introduced in the case of flavor $\mathbf{{15}^\prime}$ to express the states belonging to the multiplets of the [21] and $[421^3]$ of the $SU(6)_{CS}$ representation of the pentaquark.

It should be noted that both the $[21]$ and $[421^3]$ multiplets in Eq.~(\ref{cs-1/2-31-2}) are eigenstates of $C^{CS}_5$ of the $SU(6)_{CS}$ representation. Therefore, we can see from Table~\ref{su6-1} that the following relations hold:

\begin{align}
 C^{CS}_5\vert \Psi_1 \rangle =33/4\vert \Psi_1 \rangle,
\quad \quad
 C^{CS}_5\vert \Psi_2 \rangle =65/4\vert \Psi_2 \rangle.
 \label{cs-1/2-31-eigenvalue} 
\end{align}

As in the case of flavor $\mathbf{{15}^\prime}$, for the specific the Young-Yamanouchi basis, it is found that the linear sum of the color and spin parts between $\vert \psi_1 \rangle$ and $\vert \psi_2 \rangle$ in Eq.~(\ref{cs-1/2-31-1}) coming from the coupling scheme must belong to either the [21] of $SU(6)_{CS}$ representation or the $[421^3]$ of $SU(6)_{CS}$ representation in Eq.~(\ref{cs-1/2-31-2}). Therefore, we can find the coefficients of the linear sum by solving a coupled equation that satisfies two requirements, as we have already done. Thus, we obtain the following relation:

\begin{align}
\big{\vert}
\begin{tabular}{|c|c|c|}
\hline
1 & 3  & 4   \\
\cline{1-3}
\multicolumn{1}{|c|}{2}  \\
\cline{1-1}
 \end{tabular},
 [21]
 \big{\rangle}_{CS^{1/2}}
 =\frac{1}{2}
 \begin{tabular}{|c|c|c|}
\hline
1 & 3  & 4   \\
\cline{1-3}
\multicolumn{1}{|c|}{2}  \\
\cline{1-1}
 \end{tabular}_{CS^{1/2};S^0}
 -\frac{\sqrt{3}}{2}
 \begin{tabular}{|c|c|c|}
\hline
1 & 3  & 4   \\
\cline{1-3}
\multicolumn{1}{|c|}{2}  \\
\cline{1-1}
 \end{tabular}_{CS^{1/2};S^1}.
\label{cs-1/2-31-coefficient-0}
\end{align}
For the multiplet [$421^3$], we obtain the following relation in the same method:
\begin{align}
\big{\vert}
\begin{tabular}{|c|c|c|}
\hline
1 & 3  & 4   \\
\cline{1-3}
\multicolumn{1}{|c|}{2}  \\
\cline{1-1}
 \end{tabular},
 [421^3]
 \big{\rangle}_{CS^{1/2}}
 =\frac{\sqrt{3}}{2}
 \begin{tabular}{|c|c|c|}
\hline
1 & 3  & 4   \\
\cline{1-3}
\multicolumn{1}{|c|}{2}  \\
\cline{1-1}
 \end{tabular}_{CS^{1/2};S^0}
\nonumber
\\
 +\frac{1}{2}
 \begin{tabular}{|c|c|c|}
\hline
1 & 3  & 4   \\
\cline{1-3}
\multicolumn{1}{|c|}{2}  \\
\cline{1-1}
 \end{tabular}_{CS^{1/2};S^1}.
\label{cs-1/2-31-coefficient-1} 
\end{align}
In addition to these, we obtain the same relations applicable to the rest of Young-Yamanouchi bases of [21] ($[421^3]$), as we have already done in the case of flavor $\mathbf{{15}^\prime}$. Then, we obtain the following relation from a correspondence between Eq.~(\ref{cs-1/2-31-1}) and Eq.~(\ref{cs-1/2-31-2}):

\begin{align}
&\vert \Psi_1 \rangle =\frac{1}{2}\vert \psi_1 \rangle -
\frac{\sqrt{3}}{2}\vert \psi_2 \rangle,
 \vert \Psi_2 \rangle =\frac{\sqrt{3}}{2}\vert \psi_1 \rangle +
\frac{1}{2}\vert \psi_2 \rangle.
 \label{cs-1/2-31-coefficient} 
 \end{align}

We can now straightforwardly calculate the matrix element of Eq.~(\ref{formula-2}) in terms of $\vert \Psi_1 \rangle$ and $\vert \Psi_2 \rangle$ in Eq.~(\ref{cs-1/2-31-2}). Using Eq.~(\ref{cs-1/2-31-coefficient}), we find the following 2 by 2 matrix elements:

\begin{align}
&\langle  \Psi_1 \vert -{\sum}_{i<j}^{5}
\lambda_i^c\lambda_j^c{\vec{\sigma}}_i\cdot{\vec{\sigma}}_j  \vert  \Psi_1  \rangle =
4\times33/4-8\times38/3-
\nonumber \\
&2\times0+4\times4/3-4/3\times1/2\times3/2+
8/3\times(1/4\times0+
\nonumber \\
&3/4\times1\times2)+24=-36, \nonumber \\
&\langle  \Psi_2 \vert -{\sum}_{i<j}^{5}
\lambda_i^c\lambda_j^c{\vec{\sigma}}_i\cdot{\vec{\sigma}}_j  \vert  \Psi_2  \rangle =
4\times65/4-8\times38/3-
\nonumber \\
&2\times0+4\times4/3-4/3\times1/2\times3/2+
8/3\times(3/4\times0+
\nonumber \\
&1/4\times1\times2)+24=-20/3, 
\nonumber \\
&\langle  \Psi_1 \vert -{\sum}_{i<j}^{5}
\lambda_i^c\lambda_j^c{\vec{\sigma}}_i\cdot{\vec{\sigma}}_j  \vert  \Psi_2  \rangle =
8/3\times(\sqrt{3}/4\times0-
\nonumber \\
 &\sqrt{3}/4\times1\times2)=-4/\sqrt{3}.
\label{CS-1/2-F[3]}
\end{align}
By diagonalizing the matrix in Eq.~(\ref{CS-1/2-F[3]}), we obtain that
the eigenvalue of Eq.~(\ref{formula-2}) is either $-8/3(8+\sqrt{31})$ or $8/3(-8+\sqrt{31})$.

We also examine the expectation value of  $\lambda_i^c\lambda_j^c{\vec{\sigma}}_i\cdot{\vec{\sigma}}_j$ from the symmetry property. Using Eq.~(\ref{formula-3}) and Eq.~(\ref{cs-1/2-31-coefficient}), one can easily show that the expectation value of Eq.~(\ref{formula-3}) for particles 1-4 is given in terms of $\vert \Psi_1 \rangle$ and $\vert \Psi_2 \rangle$ in Eq.~(\ref{cs-1/2-31-2}) as follows.

\begin{align}
 \langle -{\sum}_{i<j}^{4}\lambda_i^c\lambda_j^c{\vec{\sigma}}_i\cdot{\vec{\sigma}}_j \rangle=
  \left(\begin{array}{cc}
-14 &  -\frac{2}{\sqrt{3}}     \\
-\frac{2}{\sqrt{3}}  &    -\frac{46}{3}     \\
\end{array} \right).     
 \label{cs-1/2-31-matrix-2}  
\end{align}
Then, diagonalizing Eq.~(\ref{cs-1/2-31-matrix-2}) gives either -16 or -40/3, which are the same eigenvalues obtained from Eq.~(\ref{cs-1/2-31-matrix-1}). In addition, due to the fact that the antisymmetry property of Eq.~(\ref{cs-1/2-31-2}) and the following relation, $(i k) \lambda^c_i \lambda^c_j {\vec{\sigma}}_i\cdot{\vec{\sigma}}_j (i k) = \lambda^c_k \lambda^c_j {\vec{\sigma}}_k\cdot{\vec{\sigma}}_j$, the expectation value of $\lambda_i^c\lambda_j^c{\vec{\sigma}}_i\cdot{\vec{\sigma}}_j$ ($i<j$ =1, 2, 3, 4) is all the same. Therefore, we find the following:

\begin{align}
 \langle -\lambda_i^c\lambda_j^c{\vec{\sigma}}_i\cdot{\vec{\sigma}}_j \rangle=
  \left(\begin{array}{cc}
-\frac{7}{3} &  -\frac{1}{3\sqrt{3}}     \\
-\frac{1}{3\sqrt{3}}    &    -\frac{23}{9}     \\
\end{array} \right)    
(i<j =1, 2, 3, 4).
 \label{cs-1/2-31-matrix-3}  
\end{align}
In a similar manner, we can calculate  the expectation value of  $\lambda_i^c\lambda_5^c{\vec{\sigma}}_i\cdot{\vec{\sigma}}_5$ 
($i$ =1, 2, 3, 4) in terms of $\vert \Psi_1 \rangle$ and $\vert \Psi_2 \rangle$ in Eq.~(\ref{cs-1/2-31-2}), using Eq.~(\ref{CS-1/2-F[3]}),  Eq.~(\ref{cs-1/2-31-matrix-2}), and the fully antisymmetric property for the four quarks:

\begin{align}
 \langle -\lambda_i^c\lambda_5^c{\vec{\sigma}}_i\cdot{\vec{\sigma}}_5 \rangle=
  \left(\begin{array}{cc}
-\frac{11}{2} &  -\frac{1}{2\sqrt{3}}     \\
 -\frac{1}{2\sqrt{3}}  &    \frac{13}{6}     \\
\end{array} \right)    
(i =1, 2, 3, 4).
 \label{cs-1/2-31-matrix-4}  
\end{align}

For the flavor $\mathbf{\bar{6}}$ multiplet represented by Young diagram $[2^2]$ for $q^4$, there is one fully antisymmetric  flavor $\otimes$ color $\otimes$ spin state coming from the coupling scheme in Eq.~(\ref{cs-1/2-1-22}) as follows:

\begin{align}
&\vert \psi \rangle =
\frac{1}{\sqrt{2}} \bigg{(}
\begin{tabular}{|c|c|}
\hline
1 & 2    \\
\hline
3 &  4  \\
\hline
 \end{tabular}_{F}
 \otimes 
 \begin{tabular}{|c|c|}
\hline
1 & 3    \\
\hline
3 &  4  \\
\hline
 \end{tabular}_{CS^{1/2};S^1}
- 
 \begin{tabular}{|c|c|}
\hline
1 & 3    \\
\hline
2 &  4  \\
\hline
 \end{tabular}_{F}
 \otimes 
\begin{tabular}{|c|c|}
\hline
1 & 2   \\
\hline
3 &  4  \\
\hline
 \end{tabular}_{CS^{1/2};S^1}
 \bigg{)}
 .
  \label{cs-1/2-22-1} 
 \end{align}
Obviously, since the state in Eq.~(\ref{cs-1/2-22-1}) must be the eigenstate of  Eq.~(\ref{formula-4}), we obtain the following relation:
\begin{align}
& -{\sum}_{i<j}^{4}
\lambda_i^c\lambda_j^c{\vec{\sigma}}_i\cdot{\vec{\sigma}}_j  \vert  \psi \rangle =-16/3 \vert  \psi \rangle,
\label{S-1/2-F[6bar]-1}
\end{align}
where the eigenvalue of the flavor $\mathbf{\bar{6}}$ for $C_4^F$ is 10/3.

We can also obtain the same result of Eq.~(\ref{S-1/2-F[6bar]-1}) by using Eq.~(\ref{formula-3}) and the Young-Yamanouchi bases  in color and spin part of Eq.~(\ref{cs-1/2-22-1}) obtained from the coupling scheme. This is so because these Young-Yamanouchi bases in Eq.~(\ref{cs-1/2-22-1}) belong to the multiplets of $SU(6)_{CS}$ representation $[2^2]$ for particles 1-4 and the states become the eigenstates of the quadratic Casimir operator, $C^{CS}_4$, whose eigenvalue is 32/3.
This is given as follows:
\begin{align}
 -& {\sum}_{i<j}^{4}\lambda_i^c\lambda_j^c{\vec{\sigma}}_i\cdot{\vec{\sigma}}_j
\vert \psi  \rangle =
 (-4\times32/3+2\times4/3+
 \nonumber \\
 &4/3\times1\times2+32)\vert \psi  \rangle=-16/3\vert \psi \rangle.
\label{S-1/2-F[6bar]-2}  
\end{align}

On the one hand, from the point of view of the $SU(6)_{CS}$ representation of pentaquark,  the $[21]$ in the $SU(6)_{CS}$ representation conjugate to the flavor $\mathbf{\bar{6}}$ multiplet for the four quarks contains the state which is in both the color singlet and $S=1/2$ state, as shown in Table~\ref{colorsinglet-spin} and Table~\ref{su6-1} and Eq.~(\ref{cs-decomposition}). Also, it has the symmetry property for $q^4$ corresponding to Young diagram $[2^2]$.
Therefore, another approach based on the $SU(6)_{CS}$ representation in Eq.~(\ref{cs-decomposition}) gives a flavor $\otimes$ color $\otimes$ spin state for $S=1/2$ which satisfies fully antisymmetry in the same way as in the coupling scheme:
\begin{align}
\vert \Psi \rangle =
&\frac{1}{\sqrt{2}} \bigg{(}
\begin{tabular}{|c|c|}
\hline
1 & 2    \\
\hline
3 &  4  \\
\hline
 \end{tabular}_{F}
 \otimes 
 \big{\vert}
 \begin{tabular}{|c|c|}
\hline
1 & 3    \\
\hline
2 &  4  \\
\hline
 \end{tabular}
,
 [21]
 \big{\rangle}_{CS^{1/2}} 
- 
\nonumber \\
 &\begin{tabular}{|c|c|}
\hline
1 & 3    \\
\hline
2 &  4  \\
\hline
 \end{tabular}_{F}
 \otimes 
 \big{\vert}
 \begin{tabular}{|c|c|}
\hline
1 & 2    \\
\hline
3 &  4  \\
\hline
 \end{tabular}
,
 [21]
 \big{\rangle}_{CS^{1/2}} 
 \bigg{)}.
  \label{cs-1/2-22-2} 
 \end{align}

It should be noted that the Young-Yamanouchi bases of the color and spin part of Eq.~(\ref{cs-1/2-22-1}) are equivalent to those of Eq.~(\ref{cs-1/2-22-2}).

Since the Young-Yamanouchi bases in
Eq.~(\ref{cs-1/2-22-2}) are the eigenstates of $C^{CS}_5$ of $SU(6)_{CS}$ representation, we obtain from Table~\ref{su6-1} the following eigenvalue equation valid for the Casimir operator of the  $SU(6)_{CS}$:

\begin{align}
&C^{CS}_5
\vert \Psi \rangle
 =33/4\vert \Psi \rangle.
\label{cs-1/2-22-eigenvalue} 
\end{align}

In addition to this, $\vert \Psi  \rangle$ in Eq.~(\ref{cs-1/2-22-2}) becomes the eigenstate of Eq.~(\ref{formula-2}), by using Eq.~(\ref{cs-1/2-22-eigenvalue}). Thus, it is straightforward to calculate the following:
\begin{align}
& -{\sum}_{i<j}^{5}
\lambda_i^c\lambda_j^c{\vec{\sigma}}_i\cdot{\vec{\sigma}}_j  \vert  \Psi \rangle =(
4\times33/4-8\times32/3- 2\times0
\nonumber \\
&+4\times4/3-4/3\times1/2\times3/2+
8/3\times1\times2+24) \vert  \Psi \rangle =
\nonumber \\
&-56/3 \vert  \Psi \rangle.
\label{S-1/2-F[6bar]-2}
\end{align}
   
On the other hand, the antisymmetry property of $\vert \Psi  \rangle$ in Eq.~(\ref{cs-1/2-22-2}) leads to the expectation value of $\lambda_i^c\lambda_j^c{\vec{\sigma}}_i\cdot{\vec{\sigma}}_j$ ($i<j$ =1, 2, 3, 4), which are all the same, by using  Eq.~(\ref{S-1/2-F[6bar]-1}):
\begin{align}
&\langle  \Psi \vert
-\lambda_i^c\lambda_j^c{\vec{\sigma}}_i\cdot{\vec{\sigma}}_j  \vert  \Psi \rangle =-8/9.
\quad  (i<j =1, 2, 3, 4)
\label{S-1/2-F[6bar]-3}
\end{align}

In a similar manner as we have already done, we can calculate the expectation value of  $\lambda_i^c\lambda_5^c{\vec{\sigma}}_i\cdot{\vec{\sigma}}_5$
($i$ =1, 2, 3, 4), by using  Eq.~(\ref{S-1/2-F[6bar]-2}), Eq.~(\ref{S-1/2-F[6bar]-1}), and the fully antisymmetric property for the four quarks:
\begin{align}
&\langle  \Psi \vert
-\lambda_i^c\lambda_5^c{\vec{\sigma}}_i\cdot{\vec{\sigma}}_5  \vert  \Psi \rangle =-10/3.
\quad  (i =1, 2, 3, 4)
\label{S-1/2-F[6bar]-4}
\end{align}

For the fully symmetric flavor $\mathbf{15}$ multiplet represented by Young diagram [4], there is one fully antisymmetric flavor $\otimes$ color $\otimes$ spin state coming from the coupling scheme in  Eq.~(\ref{cs-1/2-1-1111}) as follows:
\begin{align}
\vert \psi \rangle=
\begin{tabular}{|c|c|c|c|}
\hline
1 & 2  & 3  &4   \\
\hline
 \end{tabular}_{F}
\otimes
\begin{tabular}{|c|}
\hline
           1     \\
\hline
          2   \\
 \hline 
        3  \\
 \hline 
 4     \\        
\hline
\end{tabular}_{CS^{1/2};S^1}
,
\label{cs-1/2-1111-1}
\end{align}
where the spin state for particles 1-4 is 1.

On the one hand, from the point of view of the $SU(6)_{CS}$ representation of pentaquark,  the $[2^41]$ of  $SU(6)_{CS}$ representation conjugate to the flavor $\mathbf{15}$ multiplet for the four quarks contains the state which is in both the color singlet and $S=1/2$ state, as shown in Table~\ref{colorsinglet-spin} and  Table~\ref{su6-1}. Also, it has the symmetry property for $q^4$ corresponding to Young diagram $[1^4]$. Therefore, another approach based on the $SU(6)_{CS}$ representation in Eq.~(\ref{cs-decomposition}) gives a flavor $\otimes$ color $\otimes$ spin state for $S=1/2$ which satisfies fully antisymmetry in the same way as in the coupling scheme:
\begin{align}
\vert \Psi \rangle=
\begin{tabular}{|c|c|c|c|}
\hline
1 & 2  & 3  &4   \\
\hline
 \end{tabular}_{F}
\otimes
\big{\vert}
 \begin{tabular}{|c|}
\hline
1 \\
\hline
2 \\
\hline
3 \\
\hline
4 \\
\hline
 \end{tabular},
 [2^41]
 \big{\rangle}_{CS^{1/2}}
 .
 \label{cs-1/2-1111-2}
\end{align}

It should be noted that the Young-Yamanouchi basis of the color and spin part of Eq.~(\ref{cs-1/2-1111-1}) is equivalent to that of Eq.~(\ref{cs-1/2-1111-2}). We can see from these properties that this state becomes the eigenstate of Eq.~(\ref{formula-4}) as well as Eq.~(\ref{formula-2}). Then, by using the fact that the eigenvalue of the flavor $\mathbf{15}$ multiplet to  the $C^F_4$ is 28/3, we obtain from Eq.~(\ref{formula-4}) the following:
\begin{align}
 -& {\sum}_{i<j}^{4}\lambda_i^c\lambda_j^c{\vec{\sigma}}_i\cdot{\vec{\sigma}}_j
\vert \psi  \rangle =
 (4\times28/3+2\times4/3+
 \nonumber \\
 &4/3\times1\times2-24)=56/3\vert \psi  \rangle.
 \label{cs-1/2-1111-eigenvalue-1}  
\end{align}

We can also obtain the same result of Eq.~(\ref{cs-1/2-1111-eigenvalue-1}) by using  Eq.~(\ref{formula-3}) and the Young-Yamanouchi basis in color and spin part of Eq.~(\ref{cs-1/2-1111-2}) obtained from the coupling scheme. This is so because these Young-Yamanouchi basis in Eq.~(\ref{cs-1/2-1111-2}) belong to the multiplets of $SU(6)_{CS}$ representation, $[1^4]$ for particles 1-4 and the states become the eigenstate of the quadratic Casimir operator, $C^{CS}_4$, whose eigenvalue is 14/3. This is given as follows:
\begin{align}
 - {\sum}_{i<j}^{4}\lambda_i^c\lambda_j^c{\vec{\sigma}}_i\cdot{\vec{\sigma}}_j
\vert \Psi  \rangle =
 56/3\vert \Psi  \rangle.
 \label{cs-1/2-1111-eigenvalue-2}  
\end{align}

Since the Young-Yamanouchi basis in Eq.~(\ref{cs-1/2-1111-2}) is the eigenstate of $C^{CS}_5$ of $SU(6)_{CS}$ representation of the pentaquark, we obtain from Table~\ref{su6-1} the following eigenvalue equation valid for the Casimir operator of the  $SU(6)_{CS}$:
\begin{align}
&C^{CS}_5
\vert \Psi \rangle
 =33/4\vert \Psi \rangle.
\label{cs-1/2-1111-eigenvalue-3} 
\end{align}

In addition to this, $\vert \Psi \rangle$ in Eq.~(\ref{cs-1/2-1111-2}) becomes the eigenstate of Eq.~(\ref{formula-2}), by using Eq.~(\ref{cs-1/2-1111-eigenvalue-3}). Thus, it is straightforward to calculate the followings:
\begin{align}
 &-{\sum}_{i<j}^{5}
\lambda_i^c\lambda_j^c{\vec{\sigma}}_i\cdot{\vec{\sigma}}_j  \vert  \Psi \rangle =(
4\times33/4-8\times14/3- 
2\times0
\nonumber \\
&+4\times4/3-4/3\times1/2\times3/2+
8/3\times1\times2+24) \vert  \Psi \rangle =
\nonumber \\
&88/3 \vert  \Psi \rangle.
\label{cs-1/2-1111-eigenvalue-4}
\end{align}

On the other hand, the antisymmetry property of $\vert \Psi \rangle$ in Eq.~(\ref{cs-1/2-1111-2}) leads to the expectation value of $\lambda_i^c\lambda_j^c{\vec{\sigma}}_i\cdot{\vec{\sigma}}_j$ ($i<j$ =1, 2, 3, 4), which are all the same, by using Eq.~(\ref{cs-1/2-1111-eigenvalue-1}):
\begin{align}
\langle  \Psi \vert
-\lambda_i^c\lambda_j^c{\vec{\sigma}}_i\cdot{\vec{\sigma}}_j  \vert  \Psi \rangle =28/9.  \quad  \quad  (i<j =1, 2, 3, 4) 
\label{cs-1/2-1111-eigenvalue-5}
\end{align}
Finally, we can calculate the expectation value of $\lambda_i^c\lambda_5^c{\vec{\sigma}}_i\cdot{\vec{\sigma}}_5$ ($i$ =1, 2, 3, 4), by using Eq.~(\ref{cs-1/2-1111-eigenvalue-1}), Eq.~(\ref{cs-1/2-1111-eigenvalue-4}), and the fully antisymmetric property for the four quarks:
\begin{align}
\langle  \Psi \vert
-\lambda_i^c\lambda_5^c{\vec{\sigma}}_i\cdot{\vec{\sigma}}_5  \vert  \Psi \rangle =8/3.   \quad  \quad  (i =1, 2, 3, 4)
\label{cs-1/2-1111-eigenvalue-6}
\end{align}

 \subsection{Color $\otimes$ spin states with $S=3/2$ in terms of the
irreducible $SU(6)_{CS}$ representation of pentaquark}

In this subsection, we examine the $S=3/2$ case with respect to the flavor states among the particles 1-4. For the fully symmetric flavor $\mathbf{15}$ multiplet, there is one fully antisymmetric color $\otimes$ spin state conjugate to the flavor $\mathbf{15}$ multiplet coming from the coupling scheme in Eq.~(\ref{cs-3/2-1-1111}). This state is exactly equivalent to that obtained from Young diagram $[1^3]$ of the $SU(6)_{CS}$ representation of the pentaquark, since the multiplet of $[1^3]$ of
the $SU(6)_{CS}$ representation of the pentaquark contains the state which is in both the color singlet and $S=3/2$ state, as shown in Table~\ref{colorsinglet-spin},  Table~\ref{su6-1} and Eq.~(\ref{cs-decomposition}). The state is given as,
\begin{align}
\vert \Psi \rangle&=
\begin{tabular}{|c|c|c|c|}
\hline
1 & 2  & 3  &4   \\
\hline
 \end{tabular}_{F}
\otimes
\begin{tabular}{|c|}
\hline
           1     \\
\hline
          2   \\
 \hline 
        3  \\
 \hline 
 4     \\        
\hline
\end{tabular}_{CS^{3/2};S^1}
\nonumber \\
&=
\begin{tabular}{|c|c|c|c|}
\hline
1 & 2  & 3  &4   \\
\hline
 \end{tabular}_{F}
\otimes
\big{\vert}
 \begin{tabular}{|c|}
\hline
1 \\
\hline
2 \\
\hline
3 \\
\hline
4 \\
\hline
 \end{tabular},
 [1^3]
 \big{\rangle}_{CS^{3/2}}
 ,
\label{cs-3/2-1111}
\end{align}
where the state is in $S=1 $ for the four quarks. Then, we can see that the state in Eq.~(\ref{cs-3/2-1111}) becomes not only the eigenstate of  Eq.~(\ref{formula-4}) for particles 1-4, but also the eigenstate of  Eq.~(\ref{formula-2}), because this state can be viewed as consisting of the eigenstate of any Casimir operator that are involved in  Eq.~(\ref{formula-4}) and  Eq.~(\ref{formula-2}).
 
For  Eq.~(\ref{formula-4}), we obtain the following:

 \begin{align}
& -{\sum}_{i<j}^{4}
\lambda_i^c\lambda_j^c{\vec{\sigma}}_i\cdot{\vec{\sigma}}_j  \vert  \Psi \rangle =(
4\times28/3+2\times4/3+
\nonumber \\
&4/3\times1\times2-24)
 \vert  \Psi \rangle =56/3 \vert  \Psi \rangle.
\label{S-3/2-F[15]-2}
\end{align}
Here, we remind the fact that the eigenvalue of the flavor $\mathbf{15}$ multiplet to the $C^F_4$ is 28/3. We can also see that another calculation obtained from Eq.~(\ref{formula-3}) is the same as that of Eq.~(\ref{S-3/2-F[15]-2}), since the color and spin part for the four quarks in  Eq.~(\ref{cs-3/2-1111}) belongs to the multiplet of  $SU(6)_{CS}$ representation, $[1^4]$, and the eigenvalue of this state to the Casimir operator, $C^{CS}_4$ is 14/3, as shown in Table~\ref{su6-2}.
  
For  Eq.~(\ref{formula-2}), we obtain the following:  
\begin{align}
&-{\sum}_{i<j}^{5}
\lambda_i^c\lambda_j^c{\vec{\sigma}}_i\cdot{\vec{\sigma}}_j  \vert  \Psi \rangle =(
4\times21/4-8\times14/3- 2\times0+
\nonumber \\
&4\times4/3-4/3\times3/2\times5/2+
8/3\times1\times2+24) \vert  \Psi \rangle =
\nonumber \\
&40/3 \vert  \Psi \rangle.
\label{S-3/2-F[15]-1}
\end{align}
In this calculation performed with the second expression in  Eq.~(\ref{cs-3/2-1111}), we use the eigenvalue of $SU(6)_{CS}$ representation, $[1^3]$, of 21/4 for the Casimir operator, $C^{CS}_5$.

Besides, the antisymmetry property for particles 1-4 in Eq.~(\ref{cs-3/2-1111}) makes it possible to calculate the expectation value of $\lambda_i^c\lambda_j^c{\vec{\sigma}}_i\cdot{\vec{\sigma}}_j$ ($i<j$=1, 2, 3, 4), immediately resulting from  Eq.~(\ref{S-3/2-F[15]-2}), as follows:

\begin{align}
&\langle  \Psi \vert
-\lambda_i^c\lambda_j^c{\vec{\sigma}}_i\cdot{\vec{\sigma}}_j  \vert  \Psi \rangle =28/9.
\quad  (i<j=1, 2, 3, 4)
\label{S-3/2-F[15]-3}
\end{align}
In a similar method, we can calculate the expectation value of $\lambda_i^c\lambda_5^c{\vec{\sigma}}_i\cdot{\vec{\sigma}}_5$ ($i$ =1, 2, 3, 4), by using Eq.~(\ref{S-3/2-F[15]-1}), Eq.~(\ref{S-3/2-F[15]-2}), and the fully antisymmetric property for the four quarks:
\begin{align}
&\langle  \Psi \vert
-\lambda_i^c\lambda_5^c{\vec{\sigma}}_i\cdot{\vec{\sigma}}_5  \vert  \Psi \rangle =-4/3.
\quad  (i =1, 2, 3, 4)
\label{S-3/2-F[15]-4}
\end{align}

For the flavor $\mathbf{{15}^\prime}$ multiplet represented by Young diagram [31] for $q^4$, there are two orthonormal flavor $\otimes$ color $\otimes$ spin states, satisfying fully antisymmetry property. These states can be obtained by multiplying the color $\otimes$ spin states coming from the coupling scheme in Eq.~(\ref{cs-3/2-1-211}) and Eq.~(\ref{cs-3/2-2-211}) by its conjugate flavor $\mathbf{{15}^\prime}$ states, respectively:

\begin{align}
\vert \psi_1 \rangle=
&\frac{1}{\sqrt{3}}
\bigg(
\begin{tabular}{|c|c|c|}
\hline
1 & 2  & 3    \\
\cline{1-3}
\multicolumn{1}{|c|}{4}  \\
\cline{1-1}
 \end{tabular}_{F}
 \otimes 
 \begin{tabular}{|c|c|}
\hline
1 & 4    \\
\cline{1-2}
\multicolumn{1}{|c|}{2}  \\
\cline{1-1}
\multicolumn{1}{|c|}{3}  \\
\cline{1-1}
 \end{tabular}_{CS^{3/2};S^1}
 -
 \nonumber \\
 &\begin{tabular}{|c|c|c|}
\hline
1 & 2  & 4    \\
\cline{1-3}
\multicolumn{1}{|c|}{3}  \\
\cline{1-1}
 \end{tabular}_{F}
 \otimes 
 \begin{tabular}{|c|c|}
\hline
1 & 3    \\
\cline{1-2}
\multicolumn{1}{|c|}{2}  \\
\cline{1-1}
\multicolumn{1}{|c|}{4}  \\
\cline{1-1}
 \end{tabular}_{CS^{3/2};S^1}
  +
  \begin{tabular}{|c|c|c|}
\hline
1 & 3  & 4    \\
\cline{1-3}
\multicolumn{1}{|c|}{2}  \\
\cline{1-1}
 \end{tabular}_{F}
 \otimes 
 \begin{tabular}{|c|c|}
\hline
1 & 2   \\
\cline{1-2}
\multicolumn{1}{|c|}{3}  \\
\cline{1-1}
\multicolumn{1}{|c|}{4}  \\
\cline{1-1}
 \end{tabular}_{CS^{3/2};S^1}
 \bigg),
 \nonumber
 \end{align}
\begin{align}
  \vert \psi_2 \rangle=
&\frac{1}{\sqrt{3}}
\bigg(
\begin{tabular}{|c|c|c|}
\hline
1 & 2  & 3    \\
\cline{1-3}
\multicolumn{1}{|c|}{4}  \\
\cline{1-1}
 \end{tabular}_{F}
 \otimes 
 \begin{tabular}{|c|c|}
\hline
1 & 4    \\
\cline{1-2}
\multicolumn{1}{|c|}{2}  \\
\cline{1-1}
\multicolumn{1}{|c|}{3}  \\
\cline{1-1}
 \end{tabular}_{CS^{3/2};S^2}
 -
 \nonumber \\
 &\begin{tabular}{|c|c|c|}
\hline
1 & 2  & 4    \\
\cline{1-3}
\multicolumn{1}{|c|}{3}  \\
\cline{1-1}
 \end{tabular}_{F}
 \otimes 
 \begin{tabular}{|c|c|}
\hline
1 & 3    \\
\cline{1-2}
\multicolumn{1}{|c|}{2}  \\
\cline{1-1}
\multicolumn{1}{|c|}{4}  \\
\cline{1-1}
 \end{tabular}_{CS^{3/2};S^2}
 +
  \begin{tabular}{|c|c|c|}
\hline
1 & 3  & 4    \\
\cline{1-3}
\multicolumn{1}{|c|}{2}  \\
\cline{1-1}
 \end{tabular}_{F}
 \otimes 
 \begin{tabular}{|c|c|}
\hline
1 & 2   \\
\cline{1-2}
\multicolumn{1}{|c|}{3}  \\
\cline{1-1}
\multicolumn{1}{|c|}{4}  \\
\cline{1-1}
 \end{tabular}_{CS^{3/2};S^2}
 \bigg). 
 \label{cs-3/2-211-1}
 \end{align}
Obviously, since the states in Eq.~(\ref{cs-3/2-211-1}) are the eigenstates of Eq.~(\ref{formula-4}), the matrix element of Eq.~(\ref{formula-4}) is diagonalized in terms of those states, as follows:
\begin{align}
 \langle -{\sum}_{i<j}^{4}\lambda_i^c\lambda_j^c{\vec{\sigma}}_i\cdot{\vec{\sigma}}_j \rangle=
  \left(\begin{array}{cc}
\frac{8}{3} &  0     \\
0  &    8     \\
\end{array} \right).    
 \label{cs-3/2-211-matrix-1}  
\end{align}
Here, the eigenvalue of the flavor $\mathbf{{15}^\prime}$ multiplet to the $C^F_4$ is 16/3. We can also obtain the same values as those of  Eq.~(\ref{cs-3/2-211-matrix-1}) through  Eq.~(\ref{formula-3}), because the Young-Yamanouchi bases in Eq.~(\ref{cs-3/2-211-1}) obtained from the coupling scheme belong to the multiplets of $SU(6)_{CS}$ representation, $[21^2]$ for particles 1-4. These states have an eigenvalue of 26/3 in Table~\ref{su6-2}, as the eigenstates of the quadratic Casimir operator, $C^{CS}_4$. These are given as follows:

\begin{align}
 &- {\sum}_{i<j}^{4}\lambda_i^c\lambda_j^c{\vec{\sigma}}_i\cdot{\vec{\sigma}}_j
\vert \psi_1  \rangle =
 (-4\times26/3+2\times4/3+
 \nonumber \\
& 4/3\times1\times2+32)\vert \psi_1  \rangle=8/3\vert \psi_1 \rangle,
 \nonumber \\
  &- {\sum}_{i<j}^{4}\lambda_i^c\lambda_j^c{\vec{\sigma}}_i\cdot{\vec{\sigma}}_j
\vert \psi_2  \rangle =
 (-4\times26/3+2\times4/3+
 \nonumber \\ 
& 4/3\times2\times3+32)\vert \psi_2  \rangle=8\vert \psi_2 \rangle.
\label{cs-3/2-211-matrix-2}  
\end{align}

On the one hand, from the point of view of the $SU(6)_{CS}$ representation of pentaquark, both the $[32^21^2]$ and $[1^3]$ in the $SU(6)_{CS}$ representation conjugate to the flavor $\mathbf{{15}^\prime}$ multiplet for the four quarks contain the state which are in both the color singlet and $S=3/2$ state, as shown in Table~\ref{colorsinglet-spin} and Table~\ref{su6-1} and Eq.~(\ref{cs-decomposition}).
Therefore, another approach based on the $SU(6)_{CS}$ representation in Eq.~(\ref{cs-decomposition}) gives the two orthonormal flavor $\otimes$ color $\otimes$ spin states for $S=3/2$ which satisfy fully antisymmetry in the same way as in the coupling scheme:
\begin{align}
\vert \Psi_1 \rangle =
&\frac{1}{\sqrt{3}}
\bigg(
\begin{tabular}{|c|c|c|}
\hline
1 & 2  & 3    \\
\cline{1-3}
\multicolumn{1}{|c|}{4}  \\
\cline{1-1}
 \end{tabular}_{F}
 \otimes 
 \big{\vert}
 \begin{tabular}{|c|c|}
\hline
1 & 4    \\
\cline{1-2}
\multicolumn{1}{|c|}{2}  \\
\cline{1-1}
\multicolumn{1}{|c|}{3}  \\
\cline{1-1}
 \end{tabular},
 [32^21^2]
 \big{\rangle}_{CS^{3/2}}
 -
 \nonumber \\
 &\begin{tabular}{|c|c|c|}
\hline
1 & 2  & 4    \\
\cline{1-3}
\multicolumn{1}{|c|}{3}  \\
\cline{1-1}
 \end{tabular}_{F}
 \otimes
\big{\vert}  
 \begin{tabular}{|c|c|}
\hline
1 & 3    \\
\cline{1-2}
\multicolumn{1}{|c|}{2}  \\
\cline{1-1}
\multicolumn{1}{|c|}{4}  \\
\cline{1-1}
 \end{tabular},
 [32^21^2]
  \big{\rangle}_{CS^{3/2}}
 +
 \nonumber \\
 &\begin{tabular}{|c|c|c|}
\hline
1 & 3  & 4    \\
\cline{1-3}
\multicolumn{1}{|c|}{2}  \\
\cline{1-1}
 \end{tabular}_{F}
 \otimes 
 \big{\vert} 
 \begin{tabular}{|c|c|}
\hline
1 & 2   \\
\cline{1-2}
\multicolumn{1}{|c|}{3}  \\
\cline{1-1}
\multicolumn{1}{|c|}{4}  \\
\cline{1-1}
 \end{tabular},
 [32^21^2]
  \big{\rangle}_{CS^{3/2}}
 \bigg),
\nonumber 
 \end{align}
\begin{align} 
 &\vert \Psi_2 \rangle =
\frac{1}{\sqrt{3}}
\bigg(
\begin{tabular}{|c|c|c|}
\hline
1 & 2  & 3    \\
\cline{1-3}
\multicolumn{1}{|c|}{4}  \\
\cline{1-1}
 \end{tabular}_{F}
 \otimes 
 \big{\vert}
 \begin{tabular}{|c|c|}
\hline
1 & 4    \\
\cline{1-2}
\multicolumn{1}{|c|}{2}  \\
\cline{1-1}
\multicolumn{1}{|c|}{3}  \\
\cline{1-1}
 \end{tabular},
 [1^3]
 \big{\rangle}_{CS^{3/2}}
 -
 \nonumber \\
 &\begin{tabular}{|c|c|c|}
\hline
1 & 2  & 4    \\
\cline{1-3}
\multicolumn{1}{|c|}{3}  \\
\cline{1-1}
 \end{tabular}_{F}
 \otimes
\big{\vert}  
 \begin{tabular}{|c|c|}
\hline
1 & 3    \\
\cline{1-2}
\multicolumn{1}{|c|}{2}  \\
\cline{1-1}
\multicolumn{1}{|c|}{4}  \\
\cline{1-1}
 \end{tabular},
 [1^3]
  \big{\rangle}_{CS^{3/2}}
 +
  \begin{tabular}{|c|c|c|}
\hline
1 & 3  & 4    \\
\cline{1-3}
\multicolumn{1}{|c|}{2}  \\
\cline{1-1}
 \end{tabular}_{F}
 \otimes 
 \big{\vert} 
 \begin{tabular}{|c|c|}
\hline
1 & 2   \\
\cline{1-2}
\multicolumn{1}{|c|}{3}  \\
\cline{1-1}
\multicolumn{1}{|c|}{4}  \\
\cline{1-1}
 \end{tabular},
 [1^3]
  \big{\rangle}_{CS^{3/2}}
 \bigg).
 \label{cs-3/2-211-2}
 \end{align}
It should be noted that both $[32^21^2]$ and $[1^3]$ multiplets in Eq.~(\ref{cs-3/2-211-2}) are eigenstates of $C^{CS}_5$ of the $SU(6)_{CS}$ representation. Therefore, we can see from Table~\ref{su6-1} that the following relations hold:
\begin{align}
 C^{CS}_5\vert \Psi_1 \rangle =49/4\vert \Psi_1 \rangle,
\quad \quad
 C^{CS}_5\vert \Psi_2 \rangle =21/4\vert \Psi_2 \rangle.
 \label{cs-3/2-211-eigenvalue} 
\end{align}

Due to the situation of a similar kind that occurs in the case of $S=1/2$, the approach used in the previous subsection can also be applied to this case to find a correspondence between Eq.(\ref{cs-3/2-211-1}) and Eq.(\ref{cs-3/2-211-2}). This can be achieved by solving an algebraic problem involving a coupled equation with two unknown variables. Specifically, we obtain the following relation:
 \begin{align}
&\big{\vert}
\begin{tabular}{|c|c|}
\hline
1 & 2      \\
\cline{1-2}
\multicolumn{1}{|c|}{3}  \\
\cline{1-1}
\multicolumn{1}{|c|}{4}  \\
\cline{1-1}
 \end{tabular},
 [32^21^2]
 \big{\rangle}_{CS^{3/2}}
 =
 \sqrt{\frac{5}{7}}
 \begin{tabular}{|c|c|}
\hline
1 & 2      \\
\cline{1-2}
\multicolumn{1}{|c|}{3}  \\
\cline{1-1}
\multicolumn{1}{|c|}{4}  \\
\cline{1-1}
 \end{tabular}_{CS^{3/2};S^1}
 +\sqrt{\frac{2}{7}}
 \begin{tabular}{|c|c|}
\hline
1 & 2      \\
\cline{1-2}
\multicolumn{1}{|c|}{3}  \\
\cline{1-1}
\multicolumn{1}{|c|}{4}  \\
\cline{1-1}
 \end{tabular}_{CS^{3/2};S^2},
\nonumber
\end{align}
\begin{align}
\big{\vert}
\begin{tabular}{|c|c|}
\hline
1 & 2      \\
\cline{1-2}
\multicolumn{1}{|c|}{3}  \\
\cline{1-1}
\multicolumn{1}{|c|}{4}  \\
\cline{1-1}
 \end{tabular},
 [1^3]
 \big{\rangle}_{CS^{3/2}}
 =
 -\sqrt{\frac{2}{7}}
 \begin{tabular}{|c|c|}
\hline
1 & 2      \\
\cline{1-2}
\multicolumn{1}{|c|}{3}  \\
\cline{1-1}
\multicolumn{1}{|c|}{4}  \\
\cline{1-1}
 \end{tabular}_{CS^{3/2};S^1}
 +\sqrt{\frac{5}{7}}
 \begin{tabular}{|c|c|}
\hline
1 & 2      \\
\cline{1-2}
\multicolumn{1}{|c|}{3}  \\
\cline{1-1}
\multicolumn{1}{|c|}{4}  \\
\cline{1-1}
\end{tabular}_{CS^{3/2};S^1}.
\label{cs-3/2-211-coefficient-0} 
\end{align}
The same procedure holds good for the rest of Young-Yamanouchi bases of $[32^21^2]$ ($[1^3]$). We can then find at once the relation between the states in Eq.~(\ref{cs-3/2-211-1}) and Eq.~(\ref{cs-3/2-211-2}), by using Eq.~(\ref{cs-3/2-211-coefficient-0}), as the following:

\begin{align}
&\vert \Psi_1 \rangle =\sqrt{\frac{5}{7}}\vert \psi_1 \rangle +
\sqrt{\frac{2}{7}}\vert \psi_2 \rangle,
 \vert \Psi_2 \rangle =-\sqrt{\frac{2}{7}}\vert \psi_1 \rangle +
\sqrt{\frac{5}{7}}\vert \psi_2 \rangle.
 \label{cs-3/2-211-coefficient} 
 \end{align}

We can now straightforwardly calculate the matrix element of Eq.~(\ref{formula-2}) in terms of $\vert \Psi_1 \rangle$ and $\vert \Psi_2 \rangle$ in Eq.~(\ref{cs-3/2-211-2}). Using Eq.~(\ref{cs-3/2-211-coefficient}), we find the following 2 by 2 matrix elements:
\begin{align}
.&\langle  \Psi_1 \vert -{\sum}_{i<j}^{5}
\lambda_i^c\lambda_j^c{\vec{\sigma}}_i\cdot{\vec{\sigma}}_j  \vert  \Psi_1  \rangle =
4\times49/4-8\times26/3-
\nonumber \\
&2\times0+4\times4/3-4/3\times3/2\times5/2+
8/3(5/7\times2+
\nonumber \\
&2/7\times2\times3)+24=260/21, 
\nonumber \\
&\langle  \Psi_2 \vert -{\sum}_{i<j}^{5}
\lambda_i^c\lambda_j^c{\vec{\sigma}}_i\cdot{\vec{\sigma}}_j  \vert  \Psi_2  \rangle =
4\times21/4-8\times26/3-
\nonumber \\
  &2\times0+4\times4/3-4/3\times3/2\times5/2+
8/3(2/7\times2+
\nonumber \\
&5/7\times2\times3)+24=-232/21, 
\nonumber \\
&\langle  \Psi_1 \vert -{\sum}_{i<j}^{5}
\lambda_i^c\lambda_j^c{\vec{\sigma}}_i\cdot{\vec{\sigma}}_j  \vert  \Psi_2  \rangle =
8/3\times(-\sqrt{10}/\sqrt{49}\times2
\nonumber \\
&+\sqrt{10}/\sqrt{49}\times2\times3)=32\sqrt{10}/21
\label{CS-3/2-F[15`]}
\end{align}
By diagonalizing the matrix, we find that the eigenvalue of Eq.~(\ref{formula-2}) is either -12 or 40/3.

We also examine the expectation value of  $\lambda_i^c\lambda_j^c{\vec{\sigma}}_i\cdot{\vec{\sigma}}_j$ from the symmetry property. Using Eq.~(\ref{formula-3}) and Eq.~(\ref{cs-3/2-211-coefficient}), one can easily show that the expectation value of Eq.~(\ref{formula-3}) for particles 1-4 is given in terms of $\vert \Psi_1 \rangle$ and $\vert \Psi_2 \rangle$ in Eq.~(\ref{cs-3/2-211-2}) as follows.
\begin{align}
 \langle -{\sum}_{i<j}^{4}\lambda_i^c\lambda_j^c{\vec{\sigma}}_i\cdot{\vec{\sigma}}_j \rangle=
  \left(\begin{array}{cc}
  \frac{88}{21} &  \frac{16\sqrt{10}}{21}     \\
 \frac{16\sqrt{10}}{21}  &    \frac{136}{21}     \\
\end{array} \right).     
 \label{cs-3/2-211-matrix-2}  
\end{align}
Then, the matrix in Eq.~(\ref{cs-3/2-211-matrix-2}) is diagonalized to give the eigenvalues of 8 and 8/3, which are the same as those obtained from Eq.~(\ref{cs-3/2-211-matrix-1}). The antisymmetry property  for particles 1-4 makes it possible to calculate the expectation value of $\lambda_i^c\lambda_j^c{\vec{\sigma}}_i\cdot{\vec{\sigma}}_j$ ($i<j$ =1, 2, 3, 4), immediately resulting from Eq.~(\ref{cs-3/2-211-matrix-2}), as follows:

\begin{align}
 \langle -\lambda_i^c\lambda_j^c{\vec{\sigma}}_i\cdot{\vec{\sigma}}_j \rangle=
  \left(\begin{array}{cc}
  \frac{44}{63} &  \frac{8\sqrt{10}}{63}     \\
 \frac{8\sqrt{10}}{63}  &    \frac{68}{63}     \\
\end{array} \right).  
\quad
(i<j =1, 2, 3, 4).
 \label{cs-3/2-211-matrix-3}  
\end{align}

In a similar manner, we can calculate  the expectation value of  $\lambda_i^c\lambda_5^c{\vec{\sigma}}_i\cdot{\vec{\sigma}}_5$ 
($i$ =1, 2, 3, 4) in terms of $\vert \Psi_1 \rangle$ and $\vert \Psi_2 \rangle$ in Eq.~(\ref{cs-3/2-211-2}), using Eq.~(\ref{CS-3/2-F[15`]}) and Eq.~(\ref{cs-3/2-211-matrix-2}), and the fully antisymmetric property for the four quarks:
\begin{align}
 \langle -\lambda_i^c\lambda_5^c{\vec{\sigma}}_i\cdot{\vec{\sigma}}_5 \rangle=
  \left(\begin{array}{cc}
  \frac{43}{21} &  \frac{4\sqrt{10}}{21}     \\
 \frac{4\sqrt{10}}{21}  &    -\frac{92}{21}     \\
\end{array} \right).  
\quad
(i=1, 2, 3, 4).
 \label{cs-3/2-211-matrix-4}  
\end{align}

For the the flavor $\mathbf{3}$ multiplet, there is one fully antisymmetric flavor $\otimes$ color $\otimes$ spin state coming from the coupling scheme in Eq.~(\ref{cs-3/2-1-31}), which is given as:

\begin{align}
\vert \psi \rangle=
\frac{1}{\sqrt{3}}
\bigg(
&\begin{tabular}{|c|c|}
\hline
1 & 2    \\
\cline{1-2}
\multicolumn{1}{|c|}{3}  \\
\cline{1-1}
\multicolumn{1}{|c|}{4}  \\
\cline{1-1}
 \end{tabular}_{F}
\otimes 
\begin{tabular}{|c|c|c|}
\hline
1 & 3  & 4   \\
\cline{1-3}
\multicolumn{1}{|c|}{2}  \\
\cline{1-1}
 \end{tabular}_{CS^{3/2};S^1}
 \nonumber \\
 &-
\begin{tabular}{|c|c|}
\hline
1 & 3    \\
\cline{1-2}
\multicolumn{1}{|c|}{2}  \\
\cline{1-1}
\multicolumn{1}{|c|}{4}  \\
\cline{1-1}
 \end{tabular}_{F}
\otimes 
\begin{tabular}{|c|c|c|}
\hline
1 & 2  & 4   \\
\cline{1-3}
\multicolumn{1}{|c|}{3}  \\
\cline{1-1}
 \end{tabular}_{CS^{3/2};S^1}
 \nonumber \\
 &+
 \begin{tabular}{|c|c|}
\hline
1 & 4   \\
\cline{1-2}
\multicolumn{1}{|c|}{2}  \\
\cline{1-1}
\multicolumn{1}{|c|}{3}  \\
\cline{1-1}
 \end{tabular}_{F}
\otimes 
\begin{tabular}{|c|c|c|}
\hline
1 & 2  & 3  \\
\cline{1-3}
\multicolumn{1}{|c|}{4}  \\
\cline{1-1}
 \end{tabular}_{CS^{3/2};S^1}
 \bigg),
  \label{cs-3/2-31-1}
 \end{align}
where the spin state for particles 1-4 is 1. Since the state in Eq.~(\ref{cs-3/2-31-1}) is an eigenstate of Eq.~(\ref{formula-4}) for particles 1-4, we find  from Eq.~(\ref{formula-4}) the following equation:

\begin{align}
& -{\sum}_{i<j}^{4}
\lambda_i^c\lambda_j^c{\vec{\sigma}}_i\cdot{\vec{\sigma}}_j  \vert  \psi \rangle =-40/3 \vert  \psi \rangle,
\label{S-3/2-F[3]-1}
\end{align}
where the eigenvalue of the flavor $\mathbf{3}$ to $C_4^F$ is 4/3. 
We can also obtain the same  result of Eq.~(\ref{S-3/2-F[3]-1}) through  Eq.~(\ref{formula-3}), because the Young-Yamanouchi bases in color and spin part of Eq.~(\ref{cs-3/2-31-1}) obtained from the coupling scheme belong to the multiplets of $SU(6)_{CS}$ representation, $[31]$ for particles 1-4. Then, by using Eq.~(\ref{formula-3}), we obtain the following:

\begin{align}
 - &{\sum}_{i<j}^{4}\lambda_i^c\lambda_j^c{\vec{\sigma}}_i\cdot{\vec{\sigma}}_j
\vert \psi  \rangle =
 (-4\times38/3+2\times4/3+
 \nonumber \\
& 4/3\times1\times2+32)\vert \psi  \rangle
 =-40/3\vert \psi \rangle,
\label{S-3/2-F[3]-2}  
\end{align}

On the one hand, from the point of view of the $SU(6)_{CS}$ representation of pentaquark,  the $[421^3]$ in the $SU(6)_{CS}$ representation conjugate to the flavor $\mathbf{3}$ multiplet for the four quarks contains the state which is in both the color singlet and $S=3/2$ state, as shown in Table~\ref{colorsinglet-spin} and Table~\ref{su6-1} and Eq.~(\ref{cs-decomposition}). Also, it has the symmetry property for $q^4$ corresponding to Young diagram $[21^2]$.
Therefore, another approach based on the $SU(6)_{CS}$ representation in Eq.~(\ref{cs-decomposition}) gives a flavor $\otimes$ color $\otimes$ spin state for $S=3/2$ which satisfies fully antisymmetry in the same way as in the coupling scheme:
\begin{align}
\vert \Psi \rangle =
\frac{1}{\sqrt{3}}
\bigg(
&\begin{tabular}{|c|c|}
\hline
1 & 2    \\
\cline{1-2}
\multicolumn{1}{|c|}{3}  \\
\cline{1-1}
\multicolumn{1}{|c|}{4}  \\
\cline{1-1}
 \end{tabular}_{F}
\otimes 
\big{\vert}
\begin{tabular}{|c|c|c|}
\hline
1 & 3  & 4   \\
\cline{1-3}
\multicolumn{1}{|c|}{2}  \\
\cline{1-1}
 \end{tabular},
 [421^3]
 \big{\rangle}_{CS^{3/2}}
 -
 \nonumber \\
&\begin{tabular}{|c|c|}
\hline
1 & 3   \\
\cline{1-2}
\multicolumn{1}{|c|}{2}  \\
\cline{1-1}
\multicolumn{1}{|c|}{4}  \\
\cline{1-1}
 \end{tabular}_{F}
\otimes 
\big{\vert}
\begin{tabular}{|c|c|c|}
\hline
1 & 2  & 4   \\
\cline{1-3}
\multicolumn{1}{|c|}{3}  \\
\cline{1-1}
 \end{tabular},
 [421^3]
 \big{\rangle}_{CS^{3/2}} 
 +
 \nonumber \\
 &\begin{tabular}{|c|c|}
\hline
1 & 4  \\
\cline{1-2}
\multicolumn{1}{|c|}{2}  \\
\cline{1-1}
\multicolumn{1}{|c|}{3}  \\
\cline{1-1}
 \end{tabular}_{F}
\otimes 
\big{\vert}
\begin{tabular}{|c|c|c|}
\hline
1 & 2  & 3   \\
\cline{1-3}
\multicolumn{1}{|c|}{4}  \\
\cline{1-1}
 \end{tabular},
 [421^3]
 \big{\rangle}_{CS^{3/2}}
 \bigg).
\label{cs-3/2-31-2}
\end{align}
Since the color-spin parts of Eq.~(\ref{cs-3/2-31-2}) are the eigenstates with an eigenvalue of 65/4 for $C^{CS}_5$, using  Eq.~(\ref{formula-2}), we can obtain the following equation:

\begin{align}
 -&{\sum}_{i<j}^{5}
\lambda_i^c\lambda_j^c{\vec{\sigma}}_i\cdot{\vec{\sigma}}_j  \vert  \Psi \rangle =(
4\times65/4-8\times38/3-2\times0+
\nonumber \\
&4\times4/3-4/3\times3/2\times5/2+
8/3\times1\times2+24) \vert  \Psi \rangle 
\nonumber \\
&=-20/3 \vert  \Psi \rangle.
\label{S-3/2-F[3]-3}
\end{align}

For the the flavor $\mathbf{\bar{6}}$ multiplet, there is one fully antisymmetric flavor $\otimes$ color $\otimes$ spin state coming from the coupling scheme in Eq.~(\ref{cs-3/2-1-22}). This state must be exactly equivalent to that coming from Young diagram $[3^21^3]$ of the $SU(6)_{CS}$ representation of pentaquark, which is\ conjugate to the flavor $\mathbf{\bar{6}}$ multiplet, as can be seen from Table~\ref{colorsinglet-spin}, Table~\ref{su6-1}, and Eq.~(\ref{cs-decomposition}). It has to be so for the reason that the multiplet of the $[3^21^3]$ representation contains the state in the color singlet and at the same time $S=3/2$ state. The fully antisymmetric flavor $\otimes$ color $\otimes$ spin state for $S=3/2$ is given as follows:

\begin{align}
\vert \Psi \rangle =
&\frac{1}{\sqrt{2}}
 \bigg{(}
\begin{tabular}{|c|c|}
\hline
1 & 2    \\
\hline
3 &  4  \\
\hline
 \end{tabular}_{F}
 \otimes 
 \begin{tabular}{|c|c|}
\hline
1 & 3    \\
\hline
3 &  4  \\
\hline
 \end{tabular}_{CS^{3/2};S^1}
- 
 \begin{tabular}{|c|c|}
\hline
1 & 3    \\
\hline
2 &  4  \\
\hline
 \end{tabular}_{F}
 \otimes 
\begin{tabular}{|c|c|}
\hline
1 & 2   \\
\hline
3 &  4  \\
\hline
 \end{tabular}_{CS^{3/2};S^1}
 \bigg{)}
 \nonumber \\
= 
&\frac{1}{\sqrt{2}} 
\bigg{(}
\begin{tabular}{|c|c|}
\hline
1 & 2    \\
\hline
3 &  4  \\
\hline
 \end{tabular}_{F}
 \otimes 
 \big{\vert}
 \begin{tabular}{|c|c|}
\hline
1 & 3    \\
\hline
2 &  4  \\
\hline
 \end{tabular}
,
 [3^21^3]
 \big{\rangle}_{CS^{3/2}}
 \nonumber \\
&- 
 \begin{tabular}{|c|c|}
\hline
1 & 3    \\
\hline
2 &  4  \\
\hline
 \end{tabular}_{F}
 \otimes 
 \big{\vert}
 \begin{tabular}{|c|c|}
\hline
1 & 2    \\
\hline
3 &  4  \\
\hline
 \end{tabular}
,
 [3^21^3]
 \big{\rangle}_{CS^{3/2}}
 \bigg{)}.
  \label{cs-3/2-22} 
 \end{align}
It should be noted that the color $\otimes$ spin states in Eq.~(\ref{cs-3/2-22}) become the eigenstates of $C^{CS}_4$ as well as $C^{CS}_5$. From either Eq.~(\ref{formula-3}) or Eq.~(\ref{formula-4}), we find the eigenvalue equation for particles 1-4 as follows:

\begin{align}
& -{\sum}_{i<j}^{4}
\lambda_i^c\lambda_j^c{\vec{\sigma}}_i\cdot{\vec{\sigma}}_j  \vert  \Psi \rangle =-16/3 \vert  \Psi \rangle.
\label{S-3/2-F[6]-1}
\end{align}
Moreover, we straightforwardly find the eigenvalue equation of Eq.~(\ref{formula-2}) as follows:
\begin{align}
 &-{\sum}_{i<j}^{5}
\lambda_i^c\lambda_j^c{\vec{\sigma}}_i\cdot{\vec{\sigma}}_j  \vert  \Psi \rangle =(
4\times57/4-8\times32/3- 2\times0
\nonumber \\
&+4\times4/3-4/3\times3/2\times5/2+
8/3\times1\times2+24) \vert  \Psi \rangle 
\nonumber \\
&=4/3 \vert  \Psi \rangle.
\label{S-3/2-F[6]-2}
\end{align}

\subsection{the flavor $\otimes$ color $\otimes$ spin states in the case of $S=5/2$}

In the case of $S=5/2$, we have only to consider the flavor $\mathbf{{15}^\prime}$ multiplet, because color $\otimes$ spin states coming from the coupling scheme are given by Eq.~(\ref{cs-5/2-2-211}) conjugate to the flavor $\mathbf{{15}^\prime}$ states.
With the color $\otimes$ spin states of the coupling scheme, the fully antisymmetric flavor $\otimes$ color $\otimes$ spin state for $S=5/2$ can be constructed, as follows:

\begin{align}
\vert \psi \rangle=
&\frac{1}{\sqrt{3}}
\bigg(
\begin{tabular}{|c|c|c|}
\hline
1 & 2  & 3    \\
\cline{1-3}
\multicolumn{1}{|c|}{4}  \\
\cline{1-1}
 \end{tabular}_{F}
 \otimes 
 \begin{tabular}{|c|c|}
\hline
1 & 4    \\
\cline{1-2}
\multicolumn{1}{|c|}{2}  \\
\cline{1-1}
\multicolumn{1}{|c|}{3}  \\
\cline{1-1}
 \end{tabular}_{CS^{5/2};S^2}
 -
 \nonumber \\
 &\begin{tabular}{|c|c|c|}
\hline
1 & 2  & 4    \\
\cline{1-3}
\multicolumn{1}{|c|}{3}  \\
\cline{1-1}
 \end{tabular}_{F}
 \otimes 
 \begin{tabular}{|c|c|}
\hline
1 & 3    \\
\cline{1-2}
\multicolumn{1}{|c|}{2}  \\
\cline{1-1}
\multicolumn{1}{|c|}{4}  \\
\cline{1-1}
 \end{tabular}_{CS^{5/2};S^2}
  +
  \begin{tabular}{|c|c|c|}
\hline
1 & 3  & 4    \\
\cline{1-3}
\multicolumn{1}{|c|}{2}  \\
\cline{1-1}
 \end{tabular}_{F}
 \otimes 
 \begin{tabular}{|c|c|}
\hline
1 & 2   \\
\cline{1-2}
\multicolumn{1}{|c|}{3}  \\
\cline{1-1}
\multicolumn{1}{|c|}{4}  \\
\cline{1-1}
 \end{tabular}_{CS^{5/2};S^2}
 \bigg).
\label{cs-5/2-211-1}
\end{align}
This state in Eq.~(\ref{cs-5/2-211-1}) must be exactly equivalent to that coming from Young diagram $[32^21^2]$ of the $SU(6)_{CS}$ representation of the pentaquark, involving the state which are in the color singlet and $S=5/2$, as can be seen from Table~\ref{colorsinglet-spin}, Table~\ref{su6-1}, and Eq.~(\ref{cs-decomposition}).
The fully antisymmetric flavor $\otimes$ color $\otimes$ spin state for $S=5/2$ coming from the Young diagram $[32^21^2]$ of the $SU(6)_{CS}$ representation of the pentaquark is given as:

\begin{align}
\vert \Psi \rangle =
\frac{1}{\sqrt{3}}
\bigg(
&\begin{tabular}{|c|c|c|}
\hline
1 & 2  & 3    \\
\cline{1-3}
\multicolumn{1}{|c|}{4}  \\
\cline{1-1}
 \end{tabular}_{F}
 \otimes 
 \big{\vert}
 \begin{tabular}{|c|c|}
\hline
1 & 4    \\
\cline{1-2}
\multicolumn{1}{|c|}{2}  \\
\cline{1-1}
\multicolumn{1}{|c|}{3}  \\
\cline{1-1}
 \end{tabular},
 [32^21^2]
 \big{\rangle}_{CS^{5/2}}
 -
 \nonumber \\
 &\begin{tabular}{|c|c|c|}
\hline
1 & 2  & 4    \\
\cline{1-3}
\multicolumn{1}{|c|}{3}  \\
\cline{1-1}
 \end{tabular}_{F}
 \otimes
\big{\vert}  
 \begin{tabular}{|c|c|}
\hline
1 & 3    \\
\cline{1-2}
\multicolumn{1}{|c|}{2}  \\
\cline{1-1}
\multicolumn{1}{|c|}{4}  \\
\cline{1-1}
 \end{tabular},
 [32^21^2]
  \big{\rangle}_{CS^{5/2}}
 +
 \nonumber \\
 & \begin{tabular}{|c|c|c|}
\hline
1 & 3  & 4    \\
\cline{1-3}
\multicolumn{1}{|c|}{2}  \\
\cline{1-1}
 \end{tabular}_{F}
 \otimes 
 \big{\vert} 
 \begin{tabular}{|c|c|}
\hline
1 & 2   \\
\cline{1-2}
\multicolumn{1}{|c|}{3}  \\
\cline{1-1}
\multicolumn{1}{|c|}{4}  \\
\cline{1-1}
 \end{tabular},
 [32^21^2]
  \big{\rangle}_{CS^{5/2}}
\bigg).
\label{cs-5/2-211-2}
\end{align}
The color $\otimes$ spin states coming from the coupling scheme in Eq.(\ref{cs-5/2-211-1}) become the eigenstates of $C^{CS}_4$. Also, those coming from the Young diagram $[32^21^2]$ of the $SU(6)_{CS}$ representation of pentaquark become the eigenstates of $C^{CS}_5$. These indicate that these color $\otimes$ spin states are the eigenstates of both Eq.(\ref{formula-3}) and Eq.(\ref{formula-2}).
From either Eq.~(\ref{formula-3}) or Eq.~(\ref{formula-4}), we find the eigenvalue equation for particles 1-4 as follows:
\begin{align}
& -{\sum}_{i<j}^{4}
\lambda_i^c\lambda_j^c{\vec{\sigma}}_i\cdot{\vec{\sigma}}_j  \vert  \Psi \rangle =8 \vert  \Psi \rangle.
\label{S-5/2-1}
\end{align}
Moreover, we straightforwardly find the eigenvalue equation of Eq.~(\ref{formula-2}) as follows:
\begin{align}
 -&{\sum}_{i<j}^{5}
\lambda_i^c\lambda_j^c{\vec{\sigma}}_i\cdot{\vec{\sigma}}_j  \vert  \Psi \rangle =(
4\times49/4-8\times26/3-2\times0
\nonumber \\
&+4\times4/3-4/3\times5/2\times7/2+
8/3\times2\times3+24) \vert  \Psi \rangle
\nonumber \\
  &=40/3 \vert  \Psi \rangle.
\label{S-3/2-2}
\end{align}

\section{Hamiltonian}

For the purpose of investigating the stability of a pentaquark system against the strong decay, we present a nonrelativistic  Hamiltonian in a constituent quark model, which includes one gluon exchange potential. This model is an effective tool for describing a multiquark configuration involving the short-range interaction of the potentials. The Hamiltonian is given by
\begin{align}
H &=& \sum^{5}_{i=1} \left( m_i+\frac{{\mathbf p}^{2}_i}{2 m_i} \right)-\frac{3}{4}\sum^{5}_{i<j}\frac{\lambda^{c}_{i}}{2} \,\, \frac{\lambda^{c}_{j}}{2} \left( V^{C}_{ij} + V^{CS}_{ij} \right), \qquad
\label{Hamiltonian}
\end{align}
where $m_i$ is the quark mass and $\lambda^c_{i}/2$ is the SU(3) color operator for the $i$-th quark.
In Hamiltonian in Eq.~(\ref{Hamiltonian}), $V^{C}_{ij}$ and $V^{CS}_{ij}$ are a confinement and a hyperfine potential, respectively. $V^{C}_{ij}$ consist of a linear-rising and a coulomb in a potential form, and $V^{CS}_{ij}$ is chosen to be entirely a delta function in the limit of two  extremely heavy quarks. These are given as the follows:
\begin{eqnarray}
V^{C}_{ij} &=& - \frac{\kappa}{r_{ij}} + \frac{r_{ij}}{a^2_0} - D,
\label{ConfineP}
\\
V^{CS}_{ij} &=& \frac{\hbar^2 c^2 \kappa'}{m_i m_j c^4} \frac{e^{- \left( r_{ij} \right)^2 / \left( r_{0ij} \right)^2}}{(r_{0ij}) r_{ij}} \boldsymbol{\sigma}_i \cdot \boldsymbol{\sigma}_j.
\label{CSP}
\end{eqnarray}
Here
\begin{eqnarray}
r_{0ij} &=& 1/ \left( \alpha + \beta \frac{m_i m_j}{m_i + m_j} \right)	,	
\label{Parameter1}
\\
\kappa' &=& \kappa_0 \left( 1+ \gamma \frac{m_i m_j}{m_i + m_j} \right)	,	
\label{Parameter2}
\end{eqnarray}
where $r_{ij}=|{\mathbf r}_i - {\mathbf r}_j |$ is the relative distance between the $i$ and $j$ quarks, and $\boldsymbol{\sigma}_i$ is the spin operator. The parameters appearing in Eqs.~(\ref{ConfineP})-(\ref{Parameter2}) are determined by fitting them to the experimental value of both mesons and baryons, demanding the requirement that ${\chi}^2$ should be minimized for a number of set of fitting parameters\cite{Noh:Prd2021}. These  fitting parameters are given as follows.
\begin{eqnarray}
&\kappa=120.0 \, \textrm{MeV fm}, \quad a_0=0.0334066 \, \textrm{(MeV$^{-1}$fm)$^{1/2}$},& \nonumber \\
&D=917  \, \textrm{MeV}, & \nonumber \\
&m_{u}=342 \, \textrm{MeV}, \qquad m_{s}=642 \, \textrm{MeV}, &\nonumber \\
&m_{c}=1922 \, \textrm{MeV}, \qquad m_{b}=5337 \, \textrm{MeV},	&\nonumber \\
&\alpha = 1.0749 \, \textrm{fm$^{-1}$}, \,\, \beta = 0.0008014 \, \textrm{(MeV fm)$^{-1}$}, &	\nonumber \\
&\gamma = 0.001380 \, \textrm{MeV$^{-1}$}, \,\, \kappa_0=197.144 \, \textrm{MeV}. 	 &
\label{fitparameters}
\end{eqnarray}

We use a complete set of harmonic oscillator basis as a spatial function to calculate the masses of mesons and baryons within our quark model. We employ the variational method to obtain their exact ground states as accurately as possible. In fact, the variational method for the calculation of the Hamiltonian  in Eq.~(\ref{Hamiltonian}) by the complete set of harmonic oscillator is a tool of great significance, in a sense that it provides a uniform convergence of the ground state mass in terms of ideal methodology. The direct consequence is that in doing so, an exact solution on the part of variational method in a given quark model enables us to understand the property of the structure of multi-quark system, such as tetraquark, pentaquark, and so on.
 
Indeed, one can attain ultimately its convergence of the value, only if a sufficient number of harmonic oscillator basis is considered as a spatial function with which the variational method is concerned. To carry out this with a higher level of precision than in the previous works\cite{Noh:Prd2021,Noh:2023fdy}, it is necessary to consider additionally higher quanta of harmonic oscillator bases. The result is that the ground state becomes very close to a convergent state to the system. The detailed explanation about the quanta will be added in the next section.

In our present work, we fit the masses of mesons and baryons using the variational method and a complete set of harmonic oscillator basis up to the 6th quanta. The fitted values are presented in Tables~\ref{mesons} and \ref{baryons}, respectively, which show slight differences in the fitted values compared to our previous work\cite{Noh:Prd2021}. The overall fitted value to baryons has improved, resulting in a very slight reduction of the ${\chi}^2$ value in this case.
 
\begin{table}[t]

\caption{The masses of mesons obtained (Column 3) with the fitting parameters set given in  Eq.~(\ref{fitparameters}).  Column 4 shows the variational parameter $a$.}	

\centering

\begin{tabular}{cccc}
\hline
\hline	\multirow{2}{*}{Particle}	&	Experimental	&	Mass	&	Variational		\\
							&	Value (MeV)	&	(MeV)	&	Parameter (${\rm fm}^{-2}$)\\
\hline 
\\
$D$			&	1864.8		&	1853.83		&	$a$ = 7.5		\\
$D^*$		&	2007.0		&	2006.22		&	$a$ = 5.7		\\
$\eta_{c}$	&	2983.6		&	2985.97		&	$a$ = 25.2		\\
$J/\Psi$		&	3096.9		&	3118.36		&	$a$ = 19.7	\\
$D_s$		&	1968.3		&	1963.58		&	$a$ = 12.1		\\
$D^*_s$		&	2112.1		&	2109.16		&	$a$ = 9.3		\\
$K$			&	493.68		&	498.318		&	$a$ = 7.7		\\
$K^*$		&	891.66	 	&	874.664		&	$a$ = 4.1		\\

$B$			&	5279.3	 	&	5301.22		&	$a$ = 7.3		\\
$B^*$		&	5325.2		&	5360.45		&	$a$ = 6.5		\\
$\eta_b$		&	9398.0		&	9327.06		&	$a$ = 100.2	\\
$\Upsilon$	&	9460.3		&	9456.56		&	$a$ = 81.9		\\
$B_s$		&	5366.8		&	5375.25		&	$a$ = 13.0		\\
$B_s^*$		&	5415.4		&	5439.34		&	$a$ = 11.5		\\
$B_c$		&	6275.6		&	6268.41		&	$a$ = 38.7		\\
$B_c^*$		&		-		&	6361.84		&	$a$ = 32.6		\\
\\
\hline 
\hline
\label{mesons}
\end{tabular}
\end{table}

\begin{table}[t]
\caption{Same as Table~\ref{mesons} but for baryons. In column 4, $a_1$ and $a_2$ are the variational parameters.}

\centering

\begin{tabular}{cccc}
\hline
\hline	\multirow{2}{*}{Particle}	&	Experimental	&	Mass		&	\quad Variational\quad	\\
								&	Value (MeV)	&	(MeV)	&	\quad Parameters (${\rm fm}^{-2}$)\quad	\\
\hline  
\\
$\Lambda$			&	1115.7	&	1111.21	&	\quad$a_1$ = 4.3, $a_2$ = 3.9\quad	\\
$\Lambda_{c}$		&	2286.5	&	2269.04	&	\quad$a_1$ = 4.4, $a_2$ = 5.0\quad	\\
$\Xi_{cc}$		&	3621.4	&	3620.81	&	\quad$a_1$ = 11.8, $a_2$ = 4.6\quad		\\
$\Lambda_b$		&	5619.4	&	5634.20	&	\quad$a_1$ = 4.5, $a_2$ = 5.6\quad		\\
$\Sigma_{c}$		&	2452.9	&	2438.13	&	\quad$a_1$ = 3.0, $a_2$ = 5.7\quad	\\
$\Sigma_{c}^*$	&	2517.5	&	2523.1	&	\quad$a_1$ = 2.7, $a_2$ = 4.9\quad		\\
$\Sigma_{b}$		&	5811.3	&	5841.38	&	\quad$a_1$ = 2.9, $a_2$ = 6.1\quad	\\
$\Sigma_{b}^*$	&	5832.1	&	5874.89	&	\quad$a_1$ = 2.8, $a_2$ = 5.7\quad		\\
$\Sigma$			&	1192.6	&	1191.78	&	\quad$a_1$ = 3.0, $a_2$ = 5.0\quad	\\
$\Sigma^*$		&	1383.7	&	1395.70	&	\quad$a_1$ = 2.4, $a_2$ = 3.4\quad		\\
$\Xi$			&	1314.9	&	1326.58	&	\quad$a_1$ = 4.6, $a_2$ = 4.6\quad		\\
$\Xi^*$			&	1531.8	&	1540.28	&	\quad$a_1$ = 4.3, $a_2$ = 2.9\quad		\\
$\Xi_{c}$			&	2467.8	&	2471.60	&	\quad$a_1$ = 4.9, $a_2$ = 6.6\quad	\\
$\Xi_{c}^*$		&	2645.9	&	2649.58	&	\quad$a_1$ = 3.5, $a_2$ = 6.3\quad		\\
$\Xi_{b}$			&	5787.8	&	5823.74	&	\quad$a_1$ = 5.0, $a_2$ = 7.9\quad	\\
$\Xi_{b}^*$		&	5945.5	&	5988.91	&	\quad$a_1$ = 3.6, $a_2$ = 7.8\quad		\\

$p$				&	938.27	&	936.67	&	\quad$a_1$ = 2.8, $a_2$ = 2.8\quad		\\
$\Delta$			&	1232		&	1242.1	&	\quad$a_1$ = 1.9, $a_2$ = 1.9\quad	\\
\\
\hline 
\hline
\label{baryons}
\end{tabular}
\end{table}

\section{Wave Function}

In this section, we discuss about the wave function of the pentaquark, primarily involving the spatial function. In our present paper, we deal with a pentaquark system consisting of  $u(1)d(2)c(3)c(4)\bar{s}(5)$, which is in $S$ wave for the total angular momentum, $L$, $S=1/2$ for the total spin, and $I=0$ for the isospin. Generally, in setting up the Hamiltonian by means of a spatial function in the form of a Gaussian, we adopt Jacobian coordinates in the center of mass frame. For the pentaquark system, there are several sets of four Jacobi coordinates available, depending on the method of connecting the position vectors of the constituent quarks. In our study, we choose a specific set of Jacobi coordinates that provides an effective description for our purpose to investigate the stability of the pentaquark configuration. These are given as follows:
\begin{eqnarray}
	& \mathbf{x}_1 = \frac{1}{\sqrt{2}}({\mathbf r}_1 - {\mathbf r}_2), \qquad \mathbf{x}_2 = \frac{1}{\sqrt{2}}({\mathbf r}_3 - {\mathbf r}_4), \ &	\nonumber
	\\
	& \mathbf{x}_3 =\sqrt {\frac{2}{3}} \left(  {\mathbf r}_5 - \frac{1}{2}{\mathbf r}_1-  \frac{1}{2}{\mathbf r}_2   \right), \, &  \nonumber
   \\  
   & \mathbf{x}_4 =	\mu \left(   \frac{m_u {\mathbf r}_1 +m_u {\mathbf r}_2 + m_s {\mathbf r}_5}{m_u+m_u+m_s} -   \frac{m_c {\mathbf r}_3 +m_c {\mathbf r}_4 }{m_c+m_c}         \right),
   \nonumber \\
& \mu = \frac{\sqrt{2} m_c(m_s+2m_u)}{m_c \sqrt{3{m_s}^2+4m_sm_u+8{m_u}^2}}.  
   \label{Jacobian}
\end{eqnarray}

In order to obtain a more precise value for a solution of the Hamiltonian in  Eq.~(\ref{Hamiltonian}), it is essential to understand a three dimensional harmonic oscillator system by using the variational method. One of the most important features which are found in a complicated system like our own is that one should manifestly use the wave function of three dimensional harmonic oscillator system for describing the exact solution, and for investigating the property of its structure.

The wave function of a three-dimensional harmonic oscillator system is well known as consisting of the associated Laguerre polynomials for a radial part and the spherical harmonics for the angular part:

\begin{align}
&\psi(r, \theta, \phi) = R(r)_{n,l} Y^m_l (\theta, \phi)
\nonumber \\
&= \sqrt{\frac{2 \, \Gamma (n+1)}{\Gamma \left( n+l+\frac{3}{2} \right)}} \, r^l \exp \left[-\frac{r^2}{2} \right] L^{l+\frac{1}{2}}_n \left( r^2 \right) Y^m_l (\theta, \phi).
\label{harmonic oscillator}
\end{align}

However, in applying this to the variational method, it is necessary to rescale the  argument of the exponential in the radial part to become the wave function appropriate for our scheme. This is achieved by replacing the radial distance $r$ with $\sqrt{2 a} \, x$, where $x$ is the magnitude of the position vector similar to those in Eq.~(\ref{Jacobian}), and $a$ is the variational parameter. Unlike the wave function in Eq.~(\ref{harmonic oscillator}), the  modified wave function in a three dimensional harmonic oscillator system is useful to deal with a multiquark system such as $T_{cc}$ through  the variational method. In addition, introducing several variational parameters corresponding to the Jacobian coordinates is the best optimum tool for making such a system converge faster. Finally, for the main  purpose of carrying out the variational method, we search for the variational parameter which gives a minimal value of the diagonalized Hamiltonian represented by the complete set of harmonic oscillator bases.

In particular, concerning the convergence, it is convenient to use the concept of quanta, which appears in the expectation value of kinetic term with respect to the complete set of harmonic oscillator bases. In the extreme case of which the constituent quarks and all the variational parameters are identical, the expectation value of kinetic term increases directly as the value of quanta increases. The diagonal term, denoted by $\langle T_c \rangle$, is given by:

\begin{small}
\begin{eqnarray}
\langle T_c \rangle
&=&
\frac{\hbar^2 c^2 a}{m} \bigg[ \left(2 n_1 + l_1 + \frac{3}{2} \right)
+ \left(2 n_2 + l_2 + \frac{3}{2} \right) +
\nonumber \\
 && \left(2 n_3 + l_3 + \frac{3}{2} \right) + \left(2 n_4 + l_4 + \frac{3}{2} \right) \bigg]
\nonumber \\
&=&
\frac{\hbar^2 c^2 a}{m} \bigg[ 2 n_1 + l_1 + 2 n_2 + l_2 + 2 n_3 + l_3 +
\nonumber \\
&& 2 n_4 + l_4 
  +6\bigg].
\label{kinetic}
\end{eqnarray}
\end{small}
{}\
\\
Here, $m_1=m_2=m_3=m_4=m_5 \equiv m$, and $a_1=a_2=a_3=a_4 \equiv a$. For a given set of quantum numbers, $(n_1, n_2, n_3, n_4, l_1, l_2, l_3, l_4)$ that specifies the spatial functions of the pentaquark, the quanta defined in our previous work\cite{Noh:Prd2021} can be expressed as:
\begin{align}
&Q\equiv2 n_1  + 2 n_2 + 2 n_3 +2 n_4+ l_1 + l_2 + l_3 + l_4
\label{quanta}
\end{align}
It shows that the diagonal term in Eq.~(\ref{kinetic}) remains unchanged for the same quanta, even though there are enumerable sets of the quantum numbers that comprise the same quanta.
Moreover, in this concept of quanta, it is remarkable that the value of the Hamiltonian tends to decrease markedly with each step of increasing quanta.
Thus, instead of counting up all the spatial functions, using quanta as a criterion to gauge the extent of convergence is more convenient.
 
In our investigation of $T_{cc}$, composed of two light quarks and two heavy antiquarks, we observed a noticeable decrease in the mass of the system with each quanta, leading to its convergence of the ground state. This finding highlights the usefulness of the quanta as a standard for analyzing the convergence process. Therefore, in our present work, we employ the concept of quanta to facilitate the analysis of the entire convergence process.

Now, for the first time, to investigate the existence of the pentaquark and enhance the precision of calculating the Hamiltonian in a variational method, we use the complete set of harmonic oscillator basis, which can be generally expanded in terms of three Jacobian coordinates in Eq.(\ref{Jacobian}). In particular, in the absence of any  internal orbital angular momentum, the complete set of harmonic oscillator for $S$ wave can be generated by Laguerre polynomials in the form of the square of a single Jacobian coordinate. In this case, the simplest spatial function of the pentaquark is given by
\begin{align}
&\frac{8 {a_1}^{3/4}{a_2}^{3/4}{a_3}^{3/4}{a_4}^{3/4}}{{\pi}^3} \times
\nonumber \\
&{\rm exp} \big[-( a_1{\mathbf{x}_1}^2 +a_2{\mathbf{x}_2}^2+a_3{\mathbf{x}_3}^2+a_4{\mathbf{x}_4}^2  ) \big],
 \label{spatial-1}
\end{align}
where $a_i$($i$=1, 2, 3, 4) is the variational parameter. This spatial function in Eq.~(\ref{spatial-1}), where each quantum number $n_i=0$ in the Laguerre polynomials, gives relatively the largest contribution to the value of the mass of the pentaquark. This only represents the lowest quanta.

The next highest contribution to the mass of the pentaquark comes from a spatial function obtained by increasing the quantum number, $n_i$, of the Laguerre polynomials in any one of the four by one. In this case, the value of quanta $Q$ in Eq.~(\ref{quanta}) is equal to 2.
 
On the other hand, given two non-zero internal orbital angular momentum associated with any two Jacobian coordinates, it is possible to construct the spatial function of the pentaquark for $S$ wave from a general theory of angular momentum addition. Such a spatial function plays an important role in inducing the value of the Hamiltonian to converge uniformly. We also consider the possible combination of $L=0$ of the pentaquark among any non-zero four internal orbital angular momenta, which in some case have complicated patterns to establish.

Nonetheless, it is easy to build the spatial function of the pentaquark for the $S$ wave, as these wave functions are invariant forms under a rotation group whose types are scalar, such as $\mathbf{x}_1 \cdot \mathbf{x}_2$, $\mathbf{x}_1 \cdot \mathbf{x}_3$,
$\mathbf{x}_2 \cdot \mathbf{x}_3$,  $\mathbf{x}_1 \cdot (\mathbf{x}_2 \times \mathbf{x}_3)$, and etc., in a three dimensional vector space. For instance, in the case of $\mathbf{x}_1 \cdot \mathbf{x}_2$, where both $l_1$ and $l_2$ are equal to 1, the value of quanta $Q$ in Eq.~(\ref{quanta}) is equal to 2. Apart from such scalars as the inner product between two Jacobian coordinates, there remains another kind of scalar, $\mathbf{x}_1 \cdot (\mathbf{x}_2 \times \mathbf{x}_3)$, where  $l_1$,  $l_2$, and $l_3$ are all equal to 1, respectively. In this case, the value of quanta $Q$ in Eq.~(\ref{quanta}) is equal to 3. However, since such scalars do contribute little to the value of the off-diagonal Hamiltonian matrix element, the kind of spatial function which contains those types in the angular part is completely neglected. Thus, the quanta only takes on even values.

In an analogy with the case of color $\otimes$ spin states represented by the $SU(6)_{CS}$ representation, the spatial function of the pentaquark for the $S$ wave has a certain symmetry between two light quarks and two heavy quarks, respectively. From the Jacobian coordinates in  Eq.~(\ref{Jacobian}), we find the following symmetry properties for the Jacobian coordinates: $\mathbf{x}_1$ is antisymmetric for the two light quarks and $\mathbf{x}_2$ is antisymmetric for the two heavy quarks. $\mathbf{x}_3$ is symmetric for the two light quarks and $\mathbf{x}_4$ is symmetric for the two light quarks and for the two heavy quarks, respectively. From the standpoint of the quantum number, all the possible spatial functions characterized by only a quantum number, $n_i$ ($i$=1, 2, 3, 4), are symmetric between two light quarks and two heavy quarks, respectively, because the power of the Jacobian coordinate corresponding to the quantum number $n_i$ in the Laguerre polynomials is always even. However, for the spatial functions of the pentaquark involved in the quantum number, $l_i$ ($i$=1, 2, 3, 4), the symmetry properties depend entirely upon whether $l_i$ is even or odd. This is because the increase in $l_i$ is attended by the same increase in the number of the power of the corresponding Jacobian coordinate. In general, we can easily deduce this rule from the condition where the spatial function should be a scalar form for a given relative orbital angular momentum quantum number.

We are now in a position to organize the full wave function for the pentaquark consisting of $u(1)d(2)c(3)c(4)\bar{s}(5)$, which is used to calculate the Hamiltonian in Eq.~(\ref{Hamiltonian}) using the variational method. Since the Pauli principle restricts the full wave function to be antisymmetric between the two light quarks and the two heavy quarks, respectively, we have to choose the full wave function that satisfies these symmetries. 
Such full wave functions can be constructed straightforwardly by combining the symmetry properties of both the color $\otimes$ spin states and the spatial functions.

In particular, for the symmetry between the third and fourth quarks, there are some considerations regarding the color $\otimes$ spin states obtained from either the coupling scheme or the $SU(6)_{CS}$ representation. One can determine definitely the symmetry property between the first and second quarks, but not that between the third and fourth quarks: for examples, as can be seen in Eq.~(\ref{cs-1/2-0-211}), we cannot determine the symmetry between the third and fourth quarks for the second and third Young-Yamanouchi bases in Eq.~(\ref{cs-1/2-0-211}) since they are not neighboring. However, the symmetry between the third and fourth quarks can be directly determined for the first Young-Yamanouchi basis since they are neighboring.

Fortunately, it is well-known that the method for constructing either symmetry or antisymmetry between two non-neighboring particles is to combine two related Young-Yamanouchi bases. For example, the symmetries between the third and fourth quarks depend entirely on how to combine the first and second states in Eq.~(\ref{CS1/2}).

\section{Numerical Analysis}

In this section, we analyze the pentaquark system consisting of $u(1)d(2)c(3)c(4)\bar{s}(5)$ with a total spin of $S=1/2$ and isospin $I=0$ in the $S$ wave. We investigate the convergence of the expectation value of the Hamiltonian with each quanta and also discuss the numerical results regarding the stability of the system against strong decay into a baryon and a meson.

Calculating the pentaquark system is technically more challenging than a tetraquark due to the vast number of numerical inputs required. Our approach method makes it possible to achieve a precise assessment of the calculation despite these technical difficulties. However, this technical method may still be a decisive factor in obtaining the exact solution for the pentaquark system.

In addition, according to our work\cite{Noh:2023fdy}, it should be noted that the study for the tetraquarks shows that the structure depends critically upon the functional form of the hyperfine potential. In particular, we found that a Yukawa type, rather than a Gaussian type, is necessary to describe the compact bound state in $T_{cc}$. Therefore, the former potential should be adopted to investigate the stability as well as the structure even in the pentaquark system.

In the pentaquark system, the precise calculation of the Hamiltonian requires a vast number of the spatial functions of the pentaquark within the 6th quanta. Specifically, for the first six quanta, the number of spatial functions in the pentaquark system is estimated to be more than 1800. However, due to the enormous time-consuming model calculation, we limit it to only 1356 in the actual calculation. This limitation on the number of spatial functions may lead us to overlook the possibility of at least 3 MeV drop in the mass of the pentaquark.

On the other hand, in carrying out the expectation value of the Hamiltonian using the variational method, we have to search for exact four variational parameters in such a way that minimize the eigenvalue of the diagonalized Hamiltonian represented by the full wave functions. However, due to the complexity of the system, it is challenging to achieve this perfectly for higher quanta. For this reason, we find the exact four variational parameters only up to the 4th quanta.

In our practical evaluation, we first find the exact four variational parameters which give the diagonalized Hamiltonian its minimum eigenvalue with the 4th quanta of the spatial bases. Subsequently, we calculate the eigenvalue with the 5th quanta and 6th quanta simply by putting the fixed values of the four variational parameters obtained from the evaluation with the 4th quanta. However, up to the 5th quanta, we consider all possible spatial functions, whereas for the 6th quanta, we only select a limited number of spatial functions that are expected to relatively contribute to the mass value. This selection of spatial functions allows for reducing the computational burden.

As can be seen in Table~\ref{pentaquarkmass}, we find that there is a tendency for the value of the mass to decrease noticeably in accordance with each quanta. The result in Table~\ref{pentaquarkmass} shows that the bound pentaquark configuration is less likely to be found within our quark model, because the mass of the pentaquark has +18.5 MeV above its lowest threshold.

However, as pointed out above, it is highly probable that the value of the mass could be varied within a range of 3-6 MeV, if our technical limitation in the evaluation process is entirely eliminated. Nonetheless, the possibility of the existence of the pentaquark system is not still high due to a repulsion in the energy of at least 12 MeV.

Besides, in the light of the result in Table~\ref{pentaquarkmass}, it is reasonable to expect that the system will converge beyond the 6th quanta, perhaps in the 7th quanta or beyond. In this case, the mass of the pentaquark system would be very close to its threshold value. However, it is important to note that achieving convergence in higher quanta may pose additional technical challenges, and it remains to be seen whether the pentaquark system can be experimentally observed.

Here, it is necessary to return to the problem of how a Yukawa type hyperfine potential affects the organization of $T_{cc}$ more attractively than a Gaussian type hyperfine potential at short range. In the previous work\cite{Noh:Prd2021} that used a Gaussian type hyperfine potential, similar to our present work, there was a significant discrepancy in the value of the binding energy compared to experimental results, as the mass of $T_{cc}$ was found to have about 13 MeV above its threshold. However, there is a remarkable finding from the use of the Yukawa type hyperfine potential. The $T_{cc}$ in our study\cite{Noh:2023fdy} forms a compact and bound configuration, with a mass that is almost identical to the experimentally observed value. Therefore, it is crucial to consider the use of a Yukawa type potential to obtain more accurate results.

Since a similar situation will arise in the pentaquark system, what is at least as certain is that the choice of the Yukawa type potential is most preferable to that of the Gaussian type potential to achieve our purpose. Therefore, it is probable that the significant characteristics of the Yukawa type potential in our quark model lead to a different result from what has been obtained in this work.

\begin{widetext}

\begin{table}[t]
\caption{Mass of the pentaquark $udcc\bar{s}$ in each step of quanta. The lowest threshold of the pentaquark configuration for $S$=1/2 and $I$=0 is $\Xi_{cc} K$ with a mass of 4119.1 MeV. The masses of $\Xi_{cc}$ and $K$ are obtained from the quark model calculation with the variational method. In each step, the value of Q is obtained from $Q=2 n_1  + 2 n_2 + 2 n_3 + 2 n_4 + l_1 + l_2 + l_3+ l_4 $.}
\begin{tabular}{ccccccccccc}
\hline
\hline
Q					&&		Mass(MeV)	&&	Variational Parameters(fm$^{-2}$)	 &&	The number of     &&        
The  number of   &&   Binding  \\
                       	&&               &&          &&	     Spatial Wave function   &&   Total Wave function    
              && energy  	  \\ 
\hline 
$Q_1=0$	&&		4406.4	&&	$a_{1}= 2.5 , \, a_{2}= 7.5 , \, a_{3}= 4.0 , \, a_{4}= 3.2 $	&& 1 && 4  &&	\\
$Q_2=2$	&&		4266.9	&&	$a_{1}= 2.4 , \, a_{2}= 7.3 , \, a_{3}= 4.0 , \, a_{4}= 3.0 $ && 11 && 42   &&\\
$Q_3=4$	&&	4196.3	&&	$a_{1}= 2.5 , \, a_{2}= 7.8 , \, a_{3}= 4.4 ,  \, a_{4}= 2.8 $&& 66	&& 250   && \\
$Q_4=6$	&&		4164.9	&&	$a_{1}= 2.5, \, a_{2}= 8.0, \, a_{3}= 4.6 , \, a_{4}= 2.7$	&& 282	&& 1064  &&\\
$Q_5=8$	&&		4145.9	&&	$a_{1}= 2.5, \, a_{2}= 8.0, \, a_{3}= 4.6 , \, a_{4}= 2.7$	&& 916	&& 3452  && \\
$Q_6=10$	&&		4137.6	&&	$a_{1}= 2.5, \, a_{2}= 8.0, \, a_{3}= 4.6 , \, a_{4}= 2.7$	&& 1356	&& 5179 &&  +18.5 MeV \\
\hline
\hline 
\end{tabular}
\label{pentaquarkmass}
\end{table}

\end{widetext}

\section*{Acknowledgments}
This work was supported by the Korea National Research Foundation under the grant number 2021R1A2C1009486(NRF).

\section*{Appendix}
\numberwithin{equation}{subsection}

In this appendix, we present the color $\otimes$ spin states obtained from a correspondence between the coupling scheme and the $SU(6)_{CS}$ representation of the pentaquark with respect to the total spin $S$. These states are represented by a specific Young-Yamanouchi basis for particles 1-4 corresponding to the Young diagram of the $SU(6)_{CS}$ irreducible representation of the pentaquark. In particular, the color part indicates the color singlet of the pentaquark. Subsequently, we will examine the method of calculating the expectation values of $\lambda_i^c\lambda_j^c{\vec{\sigma}}_i\cdot{\vec{\sigma}}_j$.

\begin{widetext}

\subsection{Color $\otimes$ spin states with $S=1/2$ in terms of the
irreducible $SU(6)_{CS}$ representation of the pentaquark}

There are a total of 15 color $\otimes$ spin states that are in a color singlet and have $S=1/2$, as shown in Eq.(\ref{color2}) and Eq.(\ref{spin1/2}). These states correspond to 15 orthonormal states obtained from the systematic analysis of Eq.~(\ref{cs-decomposition}), each of which is associated with a specific Young-Yamanouchi basis for particles 1-4 and has a certain symmetry. These states arise from the $SU(6)_{CS}$ representation of the pentaquark, and we will examine the method of calculating the expectation values of $\lambda_i^c\lambda_j^c{\vec{\sigma}}_i\cdot{\vec{\sigma}}_j$ using these states:
\begin{align}
&\big{\vert}
 \begin{tabular}{|c|c|c|}
\hline
1 & 2 & 3     \\
\cline{1-3}
\multicolumn{1}{|c|}{4}  \\
\cline{1-1}
 \end{tabular},
 [421^3]
 \big{\rangle}_{CS^{1/2}}
 ,
 \quad
\big{\vert}
 \begin{tabular}{|c|c|c|}
\hline
1 & 2 & 4     \\
\cline{1-3}
\multicolumn{1}{|c|}{3}  \\
\cline{1-1}
 \end{tabular},
 [421^3]
 \big{\rangle}_{CS^{1/2}}
 ,
 \quad
 \big{\vert}
 \begin{tabular}{|c|c|c|}
\hline
1 & 3 & 4     \\
\cline{1-3}
\multicolumn{1}{|c|}{2}  \\
\cline{1-1}
 \end{tabular},
 [421^3]
 \big{\rangle}_{CS^{1/2}}
 ,
 \quad
  \big{\vert}
 \begin{tabular}{|c|c|c|}
\hline
1 & 2 & 3     \\
\cline{1-3}
\multicolumn{1}{|c|}{4}  \\
\cline{1-1}
 \end{tabular},
 [21]
 \big{\rangle}_{CS^{1/2}}
 ,
 \quad
\big{\vert}
 \begin{tabular}{|c|c|c|}
\hline
1 & 2 & 4     \\
\cline{1-3}
\multicolumn{1}{|c|}{3}  \\
\cline{1-1}
 \end{tabular},
 [21]
 \big{\rangle}_{CS^{1/2}}
 ,
 \nonumber \\
 &\big{\vert}
 \begin{tabular}{|c|c|c|}
\hline
1 & 3 & 4     \\
\cline{1-3}
\multicolumn{1}{|c|}{2}  \\
\cline{1-1}
 \end{tabular},
 [21]
 \big{\rangle}_{CS^{1/2}}
 ,
 \quad
 \big{\vert}
 \begin{tabular}{|c|c|}
\hline
1 & 2      \\
\cline{1-2}
\multicolumn{1}{|c|}{3}  \\
\cline{1-1}
\multicolumn{1}{|c|}{4}  \\
\cline{1-1}
 \end{tabular},
 [32^21^2]
 \big{\rangle}_{CS^{1/2}}
 ,
 \quad
 \big{\vert}
 \begin{tabular}{|c|c|}
\hline
1 & 3    \\
\cline{1-2}
\multicolumn{1}{|c|}{2}  \\
\cline{1-1}
\multicolumn{1}{|c|}{4}  \\
\cline{1-1}
 \end{tabular},
 [32^21^2]
 \big{\rangle}_{CS^{1/2}}
 ,
 \quad
\big{\vert}
 \begin{tabular}{|c|c|}
\hline
1 & 4    \\
\cline{1-2}
\multicolumn{1}{|c|}{2}  \\
\cline{1-1}
\multicolumn{1}{|c|}{3}  \\
\cline{1-1}
 \end{tabular},
 [32^21^2]
 \big{\rangle}_{CS^{1/2}}
 ,
 \quad
 \big{\vert}
 \begin{tabular}{|c|c|}
\hline
1 & 2      \\
\cline{1-2}
\multicolumn{1}{|c|}{3}  \\
\cline{1-1}
\multicolumn{1}{|c|}{4}  \\
\cline{1-1}
 \end{tabular},
 [21]
 \big{\rangle}_{CS^{1/2}}
 ,
 \nonumber \\
 &\big{\vert}
 \begin{tabular}{|c|c|}
\hline
1 & 3    \\
\cline{1-2}
\multicolumn{1}{|c|}{2}  \\
\cline{1-1}
\multicolumn{1}{|c|}{4}  \\
\cline{1-1}
 \end{tabular},
 [21]
 \big{\rangle}_{CS^{1/2}}
 , 
 \quad
 \big{\vert}
 \begin{tabular}{|c|c|}
\hline
1 & 4    \\
\cline{1-2}
\multicolumn{1}{|c|}{2}  \\
\cline{1-1}
\multicolumn{1}{|c|}{3}  \\
\cline{1-1}
 \end{tabular},
 [21]
 \big{\rangle}_{CS^{1/2}}
 ,
 \quad
\big{\vert}
 \begin{tabular}{|c|c|}
\hline
1 & 2    \\
\hline
3 & 4    \\
\hline
 \end{tabular},
 [21]
 \big{\rangle}_{CS^{1/2}}
 ,
 \quad
 \big{\vert}
 \begin{tabular}{|c|c|}
\hline
1 & 3    \\
\hline
2 & 4    \\
\hline
 \end{tabular},
 [21]
 \big{\rangle}_{CS^{1/2}}
 ,
 \quad
 \big{\vert}
 \begin{tabular}{|c|}
\hline
1 \\
\hline
2 \\
\hline
3 \\
\hline
4 \\
\hline
 \end{tabular},
 [2^41]
 \big{\rangle}_{CS^{1/2}}
 .
\label{CS1/2}
\end{align}

\subsection{Color $\otimes$ spin states with $S=3/2$ in terms of the
irreducible $SU(6)_{CS}$ representation of the pentaquark}

There are a total of 12 color $\otimes$ spin states that are in a color singlet and have $S=3/2$, as shown in Eq.~(\ref{color2}) and  Eq.~(\ref{spin3/2}). These states correspond to 12 orthonormal states obtained from the systematic analysis of Eq.~(\ref{cs-decomposition}), each of which is associated with a specific Young-Yamanouchi basis for particles 1-4 and has a certain symmetry. These states arise from the $SU(6)_{CS}$ representation of the pentaquark, and we will examine the method of calculating the expectation values of $\lambda_i^c\lambda_j^c{\vec{\sigma}}_i\cdot{\vec{\sigma}}_j$ using these states:

\begin{align}
&\big{\vert}
 \begin{tabular}{|c|c|c|}
\hline
1 & 2 & 3     \\
\cline{1-3}
\multicolumn{1}{|c|}{4}  \\
\cline{1-1}
 \end{tabular},
 [421^3]
 \big{\rangle}_{CS^{3/2}}
 ,
 \quad
\big{\vert}
 \begin{tabular}{|c|c|c|}
\hline
1 & 2 & 4     \\
\cline{1-3}
\multicolumn{1}{|c|}{3}  \\
\cline{1-1}
 \end{tabular},
 [421^3]
 \big{\rangle}_{CS^{3/2}}
 ,
 \quad
 \big{\vert}
 \begin{tabular}{|c|c|c|}
\hline
1 & 3 & 4     \\
\cline{1-3}
\multicolumn{1}{|c|}{2}  \\
\cline{1-1}
 \end{tabular},
 [421^3]
 \big{\rangle}_{CS^{3/2}}
 ,
 \quad
\big{\vert}
 \begin{tabular}{|c|c|}
\hline
1 & 2      \\
\cline{1-2}
\multicolumn{1}{|c|}{3}  \\
\cline{1-1}
\multicolumn{1}{|c|}{4}  \\
\cline{1-1}
 \end{tabular},
 [32^21^2]
 \big{\rangle}_{CS^{3/2}}
 ,
 \quad
 \big{\vert}
 \begin{tabular}{|c|c|}
\hline
1 & 3    \\
\cline{1-2}
\multicolumn{1}{|c|}{2}  \\
\cline{1-1}
\multicolumn{1}{|c|}{4}  \\
\cline{1-1}
 \end{tabular},
 [32^21^2]
 \big{\rangle}_{CS^{3/2}}
 ,
  \nonumber \\
 &\big{\vert}
 \begin{tabular}{|c|c|}
\hline
1 & 4    \\
\cline{1-2}
\multicolumn{1}{|c|}{2}  \\
\cline{1-1}
\multicolumn{1}{|c|}{3}  \\
\cline{1-1}
 \end{tabular},
 [32^21^2]
 \big{\rangle}_{CS^{3/2}}
 ,
 \quad
 \big{\vert}
 \begin{tabular}{|c|c|}
\hline
1 & 2      \\
\cline{1-2}
\multicolumn{1}{|c|}{3}  \\
\cline{1-1}
\multicolumn{1}{|c|}{4}  \\
\cline{1-1}
 \end{tabular},
 [1^3]
 \big{\rangle}_{CS^{3/2}}
 ,
 \quad
\big{\vert}
 \begin{tabular}{|c|c|}
\hline
1 & 3    \\
\cline{1-2}
\multicolumn{1}{|c|}{2}  \\
\cline{1-1}
\multicolumn{1}{|c|}{4}  \\
\cline{1-1}
 \end{tabular},
 [1^3]
 \big{\rangle}_{CS^{3/2}}
 , 
 \quad
 \big{\vert}
 \begin{tabular}{|c|c|}
\hline
1 & 4    \\
\cline{1-2}
\multicolumn{1}{|c|}{2}  \\
\cline{1-1}
\multicolumn{1}{|c|}{3}  \\
\cline{1-1}
 \end{tabular},
 [1^3]
 \big{\rangle}_{CS^{3/2}}
 ,
 \quad
 \big{\vert}
 \begin{tabular}{|c|c|}
\hline
1 & 2    \\
\hline
3 & 4    \\
\hline
 \end{tabular},
 [3^21^3]
 \big{\rangle}_{CS^{3/2}}
 ,
 \nonumber \\
&\big{\vert}
 \begin{tabular}{|c|c|}
\hline
1 & 3    \\
\hline
2 & 4    \\
\hline
 \end{tabular},
 [3^21^3]
 \big{\rangle}_{CS^{3/2}}
 ,
 \quad
 \big{\vert}
 \begin{tabular}{|c|}
\hline
1 \\
\hline
2 \\
\hline
3 \\
\hline
4 \\
\hline
 \end{tabular},
 [1^3]
 \big{\rangle}
 .
\label{CS3/2}
\end{align}

\subsection{Color $\otimes$ spin states with $S=5/2$ in terms of the
irreducible $SU(6)_{CS}$ representation of the pentaquark}

According to Eq.(\ref{color2}) and Eq.(\ref{spin5/2}), there are 3 color $\otimes$ spin states that are in a color singlet and have $S=5/2$. From the systematic analysis of Eq.~(\ref{cs-decomposition}), we find that these states correspond to the multiplet $[32^21^2]$ representation of the pentaquark. Consequently, these 3 orthonormal states can be obtained only from this representation:

\begin{align}
\big{\vert}
 \begin{tabular}{|c|c|}
\hline
1 & 2      \\
\cline{1-2}
\multicolumn{1}{|c|}{3}  \\
\cline{1-1}
\multicolumn{1}{|c|}{4}  \\
\cline{1-1}
 \end{tabular},
 [32^21^2]
 \big{\rangle}
 ,
 \quad
 \quad
 \big{\vert}
 \begin{tabular}{|c|c|}
\hline
1 & 3    \\
\cline{1-2}
\multicolumn{1}{|c|}{2}  \\
\cline{1-1}
\multicolumn{1}{|c|}{4}  \\
\cline{1-1}
 \end{tabular},
 [32^21^2]
 \big{\rangle}
 ,
 \quad
 \quad
 \big{\vert}
 \begin{tabular}{|c|c|}
\hline
1 & 4    \\
\cline{1-2}
\multicolumn{1}{|c|}{2}  \\
\cline{1-1}
\multicolumn{1}{|c|}{3}  \\
\cline{1-1}
 \end{tabular},
 [32^21^2]
 \big{\rangle}
 .
\label{CS5/2}
\end{align}

\subsection{The expectation value of $\lambda_i^c\lambda_j^c{\vec{\sigma}}_i\cdot{\vec{\sigma}}_j$}

In this section, we examine the expectation values of $\lambda_i^c\lambda_j^c{\vec{\sigma}}_i\cdot{\vec{\sigma}}_j$ in terms of color $\otimes$ spin states in Eq.~(\ref{CS1/2}), Eq.~(\ref{CS3/2}) and Eq.~(\ref{CS5/2}).
Though this problem can be completely understood through a systematic analysis in Section IV, the process is not easy, but very complicated. However, there is a simpler way to approach this calculation by taking advantage of the symmetry properties between the first and second quarks, since all the states in Eq.~(\ref{CS1/2}), Eq.~(\ref{CS3/2}) and Eq.~(\ref{CS5/2}) are made up of either symmetric or antisymmetric combinations of color and spin states. Then, it is easy to show that the expectation values of $\lambda_1^c\lambda_2^c{\vec{\sigma}}_1\cdot{\vec{\sigma}}_2$ can be calculated directly through the following:

\begin{align}
&\lambda_1^c\lambda_2^c
\begin{tabular}{|c|c|}
\hline
1 & 2      \\
\hline
 \end{tabular}_{C}
=4/3
\begin{tabular}{|c|c|}
\hline
1 & 2      \\
\hline
 \end{tabular}_{C}
 ,
 \quad  \quad
 \lambda_1^c\lambda_2^c
\begin{tabular}{|c|}
\hline
1     \\
\hline
2     \\
\hline
 \end{tabular}_{C}
=-8/3
\begin{tabular}{|c|}
\hline
1      \\
\hline
2     \\
\hline
 \end{tabular}_{C},
 \quad  \quad
 {\vec{\sigma}}_1\cdot{\vec{\sigma}}_2
\begin{tabular}{|c|c|}
\hline
1 & 2      \\
\hline
 \end{tabular}_{S}
=
\begin{tabular}{|c|c|}
\hline
1 & 2      \\
\hline
 \end{tabular}_{S},
 \quad  \quad \quad
 {\vec{\sigma}}_1\cdot{\vec{\sigma}}_2
\begin{tabular}{|c|}
\hline
1     \\
\hline
2     \\
\hline
 \end{tabular}_{S}
=-3
\begin{tabular}{|c|}
\hline
1      \\
\hline
2     \\
\hline
 \end{tabular}_{S}.
\label{CS-color-spin}
\end{align}
In Eq.~(\ref{CS-color-spin}), the subscript $C$ indicates the color state, and $S$ indicates the spin state. In the case of $S=1/2$, we can calculate the expectation values of $\lambda_i^c\lambda_j^c{\vec{\sigma}}_i\cdot{\vec{\sigma}}_j$ in terms of color $\otimes$ spin states in Eq.~(\ref{CS1/2}), resulting in a 15 by 15 matrix form denoted by $\langle \lambda_1^c\lambda_2^c{\vec{\sigma}}_1\cdot{\vec{\sigma}}_2 \rangle$.

In order to calculate the other expectation values, such as $\langle \lambda_1^c\lambda_3^c{\vec{\sigma}}_1\cdot{\vec{\sigma}}_3 \rangle$, it is necessary to consider a transposition operator, $(ij)$ to take an important role in this situation. Here, for the purpose, we use the following formula:

\begin{align}
(ij)\lambda_1^c\lambda_i^c{\vec{\sigma}}_1\cdot{\vec{\sigma}}_i(ij)=
\lambda_1^c\lambda_j^c{\vec{\sigma}}_1\cdot{\vec{\sigma}}_j,
\label{formula-color-spin}  
 \end{align}
where $(ij)$ is the transposition operator of the permutation group, $S_4$, which acts on the Young-Yamanouchi states corresponding to a given Young diagram of $q^4$. As presented in Ref.~\cite{Park:2015nha}, by the use of the generator of $SU(3)_C$ and $SU(2)_S$, the $(ij)$ operator ($i<j$ = 1, 2, 3, 4) can be replaced by $(1/3I+1/2\lambda_i^c\lambda_j^c)$ $\otimes$ $(1/2I+1/2{\vec{\sigma}}_i\cdot{\vec{\sigma}}_j)$, acting on the color and spine space, respectively. Furthermore, when the transposition operator is represented by the Young-Yamanouchi bases of the color $\otimes$ spin states of the pentaquark in Eq.~(\ref{CS1/2}), it becomes a block diagonal matrix because the Young-Yamanouchi bases of a specific $SU(6)_{CS}$ representation form an invariant subspace under the transposition operator.

Now, we can calculate the expectation value of $\lambda_1^c\lambda_3^c{\vec{\sigma}}_1\cdot{\vec{\sigma}}_3$ by means of $\langle \lambda_1^c\lambda_3^c{\vec{\sigma}}_1\cdot{\vec{\sigma}}_3 \rangle = (23) \langle \lambda_1^c\lambda_2^c{\vec{\sigma}}_1\cdot{\vec{\sigma}}_2  \rangle (23)$.
Here, we present the transposition operator, $(23)$, which is represented in a block diagonal matrix with respect to the color $\otimes$ spin states of the pentaquark in Eq.~(\ref{CS1/2}):

\begin{align}
 &(23)=
 \left(\begin{array}{ccccccccccccccc}
  1   & 0    & 0       & 0 & 0 & 0 & 0 & 0 & 0 & 0& 0 & 0 & 0 &0 & 0 \\
  0   &   -\frac{1}{2}      & \frac{\sqrt{3}}{2}      & 0 & 0 & 0 & 0 & 0 & 0 & 0& 0 & 0 & 0 &0 & 0 \\
  0     &   \frac{\sqrt{3}}{2}   &   \frac{1}{2}    & 0 & 0 & 0 & 0 & 0 & 0 & 0& 0 & 0 & 0 &0 & 0 \\
  0    & 0       & 0       &  1  & 0 & 0 & 0 & 0 & 0 & 0& 0 & 0 & 0 &0 & 0   \\
  0     & 0    & 0       & 0   &  -\frac{1}{2} &  \frac{\sqrt{3}}{2} & 0 & 0 & 0 & 0& 0 & 0 & 0 &0 & 0   \\
  0     & 0    & 0        & 0  &  \frac{\sqrt{3}}{2} & \frac{1}{2}   & 0 & 0 & 0 & 0& 0 & 0 & 0 &0 & 0               \\
  0     & 0    & 0       & 0   & 0  & 0 &-\frac{1}{2}  &  \frac{\sqrt{3}}{2}  & 0 & 0& 0 & 0 & 0 &0 & 0 \\
  0    & 0    & 0       & 0   & 0  & 0  &   \frac{\sqrt{3}}{2}    &  \frac{1}{2} & 0 & 0& 0 & 0 & 0 &0 & 0 \\
  0   & 0    & 0       & 0    & 0  & 0  & 0    & 0   & -1  &  0 & 0 & 0 & 0 &0 & 0 \\
  0   & 0    & 0       & 0   & 0   & 0  & 0   & 0    &  0    & -\frac{1}{2} &  \frac{\sqrt{3}}{2}   & 0 & 0 &0 & 0 \\
  0  & 0    & 0       & 0    & 0   & 0  & 0   & 0   & 0    &  \frac{\sqrt{3}}{2}   &  \frac{1}{2}    & 0 & 0 &0 & 0 \\
  0   & 0    & 0       & 0    & 0  & 0  & 0   & 0   & 0     & 0    & 0    &-1  &  0  &0 & 0      \\
0   & 0    & 0       & 0 & 0 & 0 & 0 & 0 & 0 & 0& 0 & 0 &  -\frac{1}{2}  & \frac{\sqrt{3}}{2}  & 0 \\
0  & 0    & 0       & 0 & 0 & 0 & 0 & 0 & 0 & 0& 0 & 0 & \frac{\sqrt{3}}{2} & \frac{1}{2}   & 0 \\
0 & 0    & 0       & 0 & 0 & 0 & 0 & 0 & 0 & 0& 0 & 0 & 0 &0 & -1 \\
\end{array} \right).
\label{CS-1/2-15-permutation-(23)}  
\end{align}

In a similar manner, we calculate the remaining expectation values of $\lambda_i^c\lambda_j^c{\vec{\sigma}}_i\cdot{\vec{\sigma}}_j$ ($i<j$ = 1, 2, 3, 4) by using other transposition operators.

Additionally, we can obtain the expectation values of $\lambda_i^c\lambda_5^c{\vec{\sigma}}_i\cdot{\vec{\sigma}}_5$ ($i$ = 1, 2, 3, 4) between a quark and an antiquark by deriving the following formula:

\begin{align}
  \lambda_4^c\lambda_5^c{\vec{\sigma}}_4\cdot{\vec{\sigma}}_5=-
 &4C^{CS}_{(4+5)}+2C^{C}_{(4+5)}+\frac{4}{3}(\vec{S}\cdot \vec{S})_{(4+5)}+16I.
\label{last-formula}  
\end{align}
In order to calculate the expectation value of $\lambda_4^c\lambda_5^c{\vec{\sigma}}_4\cdot{\vec{\sigma}}_5$, it is necessary to decompose $\mathbf{6}_{CS}\otimes\mathbf{\bar{6}}_{CS}$ into the direct sum of the composition of $SU(3)_C$ and $SU(2)_S$: $\mathbf{6}_{CS}\otimes\mathbf{\bar{6}}_{CS} = \mathbf{35}_{CS} \oplus \mathbf{1}_{CS}$. The dimensions of the SU(6)$_{CS}$ representation corresponding to the Young diagrams $[21^4]$ and $[1^6]$ are 35 and 1, respectively. Table IX shows the composition of $SU(3)_C$ and $SU(2)_S$ concerning the $SU(6)_{CS}$ representation, and its eigenvalue of the quadratic Casimir operator, $C^{CS}$ for $q\bar{q}$.

\begin{table}[htp]
\caption{The composition of $SU(3)_C$ and $SU(2)_S$ concerning the $SU(6)_{CS}$ representation for $q\bar{q}$.}
\begin{center}
\begin{tabular}{c|c|c|c}
\hline \hline
$SU(6)_{CS}$  &         &          &           \\
  Young digram  & $SU(3)_C$ $\otimes$  $SU(2)_S$ & Dimesion  &  Eigenvalue \\
\hline
 $[21^4]$  &    $(\mathbf{8},\mathbf{3})$,    $(\mathbf{8},\mathbf{1})$,     $(\mathbf{1},\mathbf{3})$        &    35         &  6        \\
   \hline  
 $[1^6]$  &       $(\mathbf{1},\mathbf{1})$                     &   1       &  0        \\
  \hline \hline        
\end{tabular}
\end{center}
\label{last-table}
\end{table}

On the other hand, it should be noted that, in Eq.~(\ref{color2}), the color states between the 4th and 5th quarks are octet in the $|C_1\rangle$ and $|C_2\rangle$, and singlet in the $|C_3\rangle$. For the spin part, since most of the spin states between the 4th and 5th quarks in Eq.~(\ref{spin1/2}) and Eq.~(\ref{spin3/2}) cannot be directly determined, these Young-Yamanouchi spin states should be transformed into those associated with the decay mode.
With the availability of these eigenvalues of such Casimir operators, we can calculate the expectation values of $\lambda_4^c\lambda_5^c{\vec{\sigma}}_4\cdot{\vec{\sigma}}_5$, and the $\lambda_i^c\lambda_5^c{\vec{\sigma}}_i\cdot{\vec{\sigma}}_5$ ($i$ = 1, 2, 3) through the relevant permutations.

\end{widetext}



\begin{thebibliography}{99}

\bibitem{Noh:2023fdy}
Sungsik Noh and Woosung Park,
Observation of $T_{cc}$ and a quark model,
[arXiv:2303.03285 [hep-ph]].



\bibitem{Choi:2003ue} 
  S.~K.~Choi {\it et al.} [Belle Collaboration],
  Phys.\ Rev.\ Lett.\  {\bf 91}, 262001 (2003).





\bibitem{LHCbTcc}
R.~Aaij \textit{et al.} [LHCb],
Nature Phys. \textbf{18}, no.7, 751-754 (2022)
doi:10.1038/s41567-022-01614-y

\bibitem{Ballot:1983iv}
J.~l.~Ballot and J.~M.~Richard,
Phys. Lett. B \textbf{123}, 449-451 (1983)
doi:10.1016/0370-2693(83)90991-7


\bibitem{Zouzou:1986qh}
S.~Zouzou, B.~Silvestre-Brac, C.~Gignoux and J.~M.~Richard,
Z. Phys. C \textbf{30}, 457 (1986)
doi:10.1007/BF01557611


\bibitem{R1} 
  J. Carlson, L. Heller, and J.A. Tjon
  Stability of Dimesons,
  Phys.\ Rev.\ D\  {\bf 37}, 744 (1988).

\bibitem{R2} 
  B. Silvestre-Brac and C. Semay,
  Systematics of $L=0$ $q^2 \bar{q}^2$ systems,
  Z.\ Phys.\ C\  {\bf 57}, 273-282 (1993).





\bibitem{R3} 
  C. Semay and B. Silvestre-Brac
  Diquonia and potential models,
  Z.\ Phys.\ C\  {\bf 61}, 271-275 (1994).

\bibitem{R4} 
  S. Pepin, Fl. Stancu, M. Genovese, and J.-M. Richard,
  Tetraquarks with colour-blind forces in chiral quark models,
  Phys.\ Lett.\ B\  {\bf 393}, 119-123 (1997).

\bibitem{R5} 
  Boris A. Gelman and Shmuel Nussinov,
  Does a narrow tetraquark cc anti-u anti-d state exist?,
  Phys.\ Lett.\ B\  {\bf 551}, 296-304 (2003).

\bibitem{R6} 
  J. Vijande, F. Fernandez, A. Valcarce, and B. Silvestre-Brac,
  Tetraquarks in a chiral constituent-quark model,
  Eur.\ Phys. \ J. \ A\  {\bf 19}, 383-389 (2004).

\bibitem{R7} 
  D. Janc and M. Rosina,
  The $T_{cc} = DD^*$ Molecular State,
  Few Body Syst.\ {\bf 35}, 175-196 (2004).



\bibitem{R9} 
  D. Ebert, R. N. Faustov, V. O. Galkin, and W. Lucha,
  Masses of tetraquarks with two heavy quarks in the relativistic quark model,
  Phys.\ Rev.\ D\  {\bf 76}, 114015 (2007).


\bibitem{R11} 
  Youchang Yang, Chengrong Deng, Jailun Ping, and T. Goldman,
  $S$-wave $QQ\bar{q}\bar{q}$ state in the constituent quark model,
  Phys.\ Rev.\ D\  {\bf 80}, 114023 (2009).


\bibitem{R13} 
  Yoichi Ikeda, Bruno Charron, Sinya Aoki, Takumi Doi, Tetsuo Hatsuda, Takashi Inoue, Noriyoshi Ishii, Keiko Murano, Hidekatsu Nemura, and Kenji Sasaki,
  Charmed tetraquarks $T_{cc}$ and $T_{cs}$ from dynamical lattice QCD simulations,
  Phys.\ Lett.\ B\  {\bf 729}, 85-90 (2014).

\bibitem{Park:2013fda}
W.~Park and S.~H.~Lee,
Nucl. Phys. A \textbf{925} (2014), 161-184
doi:10.1016/j.nuclphysa.2014.02.008
[arXiv:1311.5330 [nucl-th]].


\bibitem{R15} 
  Marek Karliner and Jonathan L. Rosner,
  Discovery of the Doubly Charmed $\Xi_{cc}$ Baryon Implies a Stable $bb\bar{u}\bar{d}$ Tetraquark,
  Phys.\ Rev.\ Lett.\ {\bf 119}, 202001 (2017).

\bibitem{R16} 
  Estia J. Eichten and Chris Quiqq,
  Heavy-Quark Symmetry Implies Stable Heavy Tetraquark Mesons $Q_i Q_j \bar{q}_k \bar{q}_l$,
  Phys.\ Rev.\ Lett.\  {\bf 119}, 202002 (2017).

\bibitem{R14} 
  Si-Qiang Luo, Kan Chen, Xiang Liu, Yan-Rui Liu, and Shi-Lin Zhu,
  Exotic tetraquark states with the $qq\bar{Q}\bar{Q}$ configuration,
  Eur.\ Phys.\ J. \ C\  {\bf 77}, 709 (2017).


\bibitem{R18} 
  Hadron Spectrum Collaboration, Gavin K. C. Cheung, Christopher E. Thomas, Jozef J. Dudek, and Robert G. Edwards,
  Tetraquark operators in lattice QCD and exotic flavour states in the charm sector,
  JHEP\  {\bf 11}, 033 (2017).

\bibitem{R17} 
  Zhi-Gang Wang,
  Analysis of the axialvector doubly heavy tetraquark states with QCD sum rules,
  Acta Phys.\ Polon.\ B\  {\bf 49}, 1781 (2018).

\bibitem{Woosung:NPA2019} 
  Woosung Park, Sungsik Noh, and Su Houng Lee,
  Masses of the doubly heavy tetraquarks in a constituent quark model,
  Nucl.\ Phys.\ A\  {\bf 983}, 1-19 (2019).
  
  
\bibitem{R20} 
  Parikshit Junnarkar, Nilmani Mathur, and M. Padmanath,
  Study of doubly heavy tetraquarks in lattice QCD,
  Phys.\ Rev.\ D\  {\bf 99}, 034507 (2019).
  
  
\bibitem{R19} 
  Anthony Francis, Renwick J. Hudspith, Randy Lewis, and Kim Maltman,
  Evidence for charm-bottom tetraquarks and the mass dependence of heavy-light tetraquark states from lattice QCD,
  Phys.\ Rev.\ D\  {\bf 99}, 054505 (2019).


\bibitem{R22} 
  Ming-Zhu Liu, Tian-Wei Wu, Manuel Pavon Valderrama, Ju-Jun Xie, and Li-Sheng Geng,
  Heavy-quark spin and flavor symmetry partners of the $X(3872)$ revisited: What can we learn from the one boson exchange model?,
  Phys.\ Rev.\ D\  {\bf 99}, 094018 (2019).

\bibitem{R21} 
  Chengrong Deng, Hong Chen, and Jialun Ping,
  Systematical investigation on the stability of doubly heavy tetraquark states,
  Eur.\ Phys.\ J.\ A\  {\bf 56}, 9 (2020).

\bibitem{R23} 
  Gang Yang, Jialun Ping, and Jorge Segovia,
  Doubly-heavy tetraquarks,
  Phys.\ Rev.\ D\  {\bf 101}, 014001 (2020).


\bibitem{R24} 
  Qi-Fang L$\ddot{\rm u}$, Dian-Yong Chen, and Yu-Bing Dong,
  Masses of doubly heavy tetraquarks $T_{QQ'}$ in a relativized quark model,
  Phys.\ Rev.\ D\  {\bf 102}, 034012 (2020).
 

\bibitem{R25} 
  Eric Braaten, Li-Ping He, and Abhishek Mohapatra,
  Masses of doubly heavy tetraquarks with error bars,
  Phys.\ Rev.\ D\  {\bf 103}, 016001 (2021).



\bibitem{R27} 
  Jian-Bo Cheng, Shi-Yuan Li, Yan-Rui Liu, Zong-Guo Si, and Tao Yao
  Double-heavy tetraquark states with heavy diquark-antiquark symmetry,
  Chin.\ Phys.\ C\  {\bf 45}, 043102 (2021).


\bibitem{Noh:Prd2021}
Sungsik Noh, Woosung Park, and Su Houng Lee, Doubly heavy tetraquarks, $qq' \bar{Q} \bar{Q'}$, in a nonrelativistic quark model with a complete set of harmonic oscillator bases,
Phys. Rev. D \textbf{103}, 114009 (2021)


\bibitem{R28} 
  Rudolf N. Faustov, Vladimir O. Galkin, and Elena M. Savchenko,
  Heavy Tetraquarks in the Reltativistic Quark Model,
  Universe\  {\bf 7}, 94 (2021).




\bibitem{Park:2018oib}
W.~Park, S.~Cho and S.~H.~Lee,
Phys. Rev. D \textbf{99}, no.9, 094023 (2019)
doi:10.1103/PhysRevD.99.094023
[arXiv:1811.10911 [hep-ph]].




\bibitem{Park:2017jbn}
W.~Park, A.~Park, S.~Cho and S.~H.~Lee,
Phys. Rev. D \textbf{95}, no.5, 054027 (2017)
doi:10.1103/PhysRevD.95.054027
[arXiv:1702.00381 [hep-ph]].



\bibitem{Stancu:1999qr}
  F.~Stancu and S.~Pepin,
  Few Body Syst.\  {\bf 26}, 113 (1999).



\bibitem{Jaffe:1976ih}
R.~L.~Jaffe,
Phys. Rev. D \textbf{15}, 281 (1977)
doi:10.1103/PhysRevD.15.281





\bibitem{Aerts:1977rw}
A.~T.~M.~Aerts, P.~J.~G.~Mulders and J.~J.~de Swart,
Phys. Rev. D \textbf{17}, 260 (1978)
doi:10.1103/PhysRevD.17.260






\bibitem{Park:2015nha}
W.~Park, A.~Park and S.~H.~Lee,
Phys. Rev. D \textbf{92}, no.1, 014037 (2015)
doi:10.1103/PhysRevD.92.014037
[arXiv:1506.01123 [nucl-th]].













\bibitem{Chen:2002gd}
  J.~Q.~Chen, J.~L.~Ping and F.~Wang,
  ``Group representation theory for physicists,''
  River Edge, USA: World Scientific (2002)   



\bibitem{ComRef9} 
  J. Vijande, E. Weissman, A. Valcarce, and N. Barnea,
  Are there compact heavy four-quark bound states?,
  Phys.\ Rev.\ D\  {\bf 76}, 094027 (2007).







\end{thebibliography}
\end{document}